\documentclass[12pt]{spieman}  
\usepackage{amsmath,amsfonts,amssymb}
\usepackage{graphicx}
\usepackage{setspace}
\usepackage{tocloft}
\usepackage{lineno}
\usepackage{booktabs}
\usepackage{hyperref}
\usepackage[version=4]{mhchem} 

\newcommand{\escape}{ESCAPE}
\newcommand{\euve}{EUVE}
\newcommand{\xmm}{XMM-Newton}
\newcommand{\hst}{HST}
\newcommand\adeg{\mbox{$^\circ$}}%
\newcommand\amin{\mbox{$^\prime$}}%
\newcommand\asec{\mbox{$^{\prime\prime}$}}%

\usepackage{lipsum}
\usepackage{xcolor}
\usepackage{color}

\title{\escape: a small explorer mission to study the stellar drivers of exoplanet evolution}

\author[a,*]{Allison Youngblood}
\author[b,c,d]{Kevin France}
\author[b]{Brian Fleming}
\author[b]{Tim H. Hellickson}
\author[b]{Alan C. Hoskins}
\author[e]{James Paul Mason}
\author[b]{Dolon Bhattacharyya}
\author[f]{Hannah Diamond-Lowe}
\author[g]{Girish M. Duvvuri}
\author[h]{Cynthia S. Froning}
\author[i]{Vinay L. Kashyap}
\author[j]{Tommi T. Koskinen}
\author[b,k]{Yuta Notsu}
\author[l]{Seth Redfield}
\author[b]{David J. Wilson}
\author[m]{Ute V. Amerstorfer}
\author[h,n]{Tracy M. Becker}
\author[o]{Francesco Borsa}
\author[b,c]{David A. Brain}
\author[p]{Luca Fossati}
\author[b]{Briana Indahl}
\author[q]{Meng Jin}
\author[r]{Graham S. Kerr}
\author[c,k,b]{Adam F. Kowalski}
\author[f,s]{Rachel A. Osten}
\author[b]{Tom L. Patton}
\author[b]{Steven V. Penton}
\author[t]{Ignazio Francesco Pillitteri}
\author[h,n]{Kurt D. Retherford}
\author[u,v]{Astrid M. Veronig}
\author[w]{Aline A. Vidotto}
\author[x,i]{Jennifer G. Winters}
\author[y]{Adina D. Feinstein}
\author[z,aa]{Guillaume P. Gronoff}
\author[ab]{Jeffrey L. Linsky}
\author[ac]{R. O. Parke Loyd}
\author[f]{Susan E. Mullally}
\author[ad]{Jorge Sanz-Forcada}
\author[ae,af]{Peter J. Wheatley}
\author[ag,ah]{Amber V. Young}
 
\affil[a]{Exoplanets and Stellar Astrophysics Laboratory, NASA Goddard Space Flight Center, Greenbelt, MD 20771, USA}
\affil[b]{Laboratory for Atmospheric and Space Physics, University of Colorado Boulder, 3665 Discovery Drive, Boulder, CO 80303, USA}
\affil[c]{Department of Astrophysical and Planetary Sciences, University of Colorado Boulder, Boulder, CO 80309, USA}
\affil[d]{Center for Astrophysics and Space Astronomy, University of Colorado Boulder, Boulder, CO 80309, USA}
\affil[e]{Applied Physics Laboratory, Johns Hopkins University, 11001 Johns Hopkins Rd, Laurel, MD 20723, USA}
\affil[f]{Space Telescope Science Institute, 3700 San Martin Drive, Baltimore, MD 21218, USA}
\affil[g]{Department of Physics and Astronomy, Vanderbilt University, Nashville, TN 37235, USA}
\affil[h]{Southwest Research Institute, San Antonio, TX 78238, USA}
\affil[i]{Center for Astrophysics $\vert$ Harvard \& Smithsonian, 60 Garden Street, Cambridge, MA 02138, USA}
\affil[j]{Lunar and Planetary Laboratory, University of Arizona, Tucson, AZ 85721, USA}
\affil[k]{National Solar Observatory, 3665 Discovery Drive, Boulder, CO 80303, USA}
\affil[l]{Astronomy Department and Van Vleck Observatory, Wesleyan University, Middletown, CT 06459, USA}
\affil[m]{Austrian Space Weather Office, GeoSphere Austria, Graz, Austria}
\affil[n]{University of Texas at San Antonio, 1 UTSA Circle, San Antonio, TX 78249, USA}
\affil[o]{INAF -- Osservatorio Astronomico di Brera, Via E. Bianchi 46, 23807 Merate (LC), Italy}
\affil[p]{Space Research Institute, Austrian Academy of Sciences, Schmiedlstrasse 6, 8042 Graz, Austria}
\affil[q]{Lockheed Martin Solar and Astrophysics Laboratory (LMSAL), 3251 Hanover Street, Palo Alto, CA 94306, USA}
\affil[r]{SUPA, School of Physics and Astronomy, University of Glasgow, Glasgow G12 8QQ, UK}
\affil[s]{Center for Astrophysical Sciences, Johns Hopkins University, Baltimore, MD 21218, USA}
\affil[t]{INAF -- Osservatorio Astronomico di Palermo, Piazza del Parlamento 1, 90134 Palermo, Italy}
\affil[u]{Institute of Physics, University of Graz, 8010 Graz, Austria}
\affil[v]{Kanzelh\"ohe Observatory for Solar and Environmental Research, University of Graz, Kanzelh\"ohe 19, 9521 Treffen, Austria}
\affil[w]{Leiden Observatory, Leiden University, PO Box 9513, 2300 RA Leiden, The Netherlands}
\affil[x]{Bridgewater State University, 131 Summer St., Bridgewater, MA 02325, USA}
\affil[y]{Department of Physics and Astronomy, Michigan State University, East Lansing, MI 48824, USA}
\affil[z]{NASA Langley Research Center, Hampton, VA, USA}
\affil[aa]{Science Systems and Applications, Inc., Hampton, VA, USA}
\affil[ab]{JILA, University of Colorado Boulder, Boulder, CO 80309-0440, USA}
\affil[ac]{2452 Delmer Street Suite 100, Oakland, CA 94602-3017, USA}
\affil[ad]{Centro de Astrobiología, CSIC-INTA, Camino bajo del Castillo s/n, 28692 Villanueva de la Cañada, Madrid, Spain}
\affil[ae]{Centre for Exoplanets and Habitability, University of Warwick, Gibbet Hill Road, Coventry CV4 7AL, UK}
\affil[af]{Department of Physics, University of Warwick, Gibbet Hill Road, Coventry CV4 7AL, UK}
\affil[ag]{Mary W. Jackson NASA Headquarters, Washington, DC, USA}
\affil[ah]{NASA Goddard Space Flight Center, Greenbelt, MD 20771, USA}

\cftpagenumbersoff{figure}
\cftpagenumbersoff{table} 
\begin{document} 
\maketitle
\begin{abstract}

The long-term stability of exoplanetary atmospheres depends critically on the extreme-ultraviolet (EUV) photon and high-energy particle fluxes from the host star, which are poorly constrained. To address this key gap in our understanding of atmospheric retention, we present the Extreme-ultraviolet Stellar Characterization for Atmospheric Physics and Evolution (\escape) mission, a NASA Small Explorer concept proposed in 2026. \escape\ employs extreme- and far-ultraviolet spectroscopy (80 - 1650 \AA) to provide the first comprehensive study of the stellar EUV history and stellar coronal mass ejection (CME) environments that control atmospheric mass-loss and determine the habitability of rocky exoplanets. This paper outlines both the primary science goals of the mission, the breadth of future general observer investigations, and a detailed design study of the mission's instrumentation. 
The \escape\ instrument comprises a grazing incidence telescope that feeds multiple diffraction gratings and a photon-counting detector. We describe a demonstration of the Hettrick-Bowyer telescope, etched silicon diffraction gratings, the microchannel plate detector and housing, and gold and zirconium coatings. We present a STOP analysis that verifies \escape's ability to meet its structural integrity, thermal stability, and optical performance requirements throughout the mission environment.

\end{abstract}


{\noindent \footnotesize\textbf{*} Address all correspondence to Allison Youngblood,  \linkable{allison.a.youngblood@nasa.gov} }

\begin{spacing}{1}

\section{Introduction}
\label{sect:intro}  

With over 6,000 confirmed extrasolar planets known, exoplanetary science has progressed from simple detections to population occurrence rates to atmosphere characterization of our closest planetary neighbors, including those that reside within the nominal habitable zone (HZ) of their host stars. One of the highest priorities in the 2020s, 2030s, and 2040s is the search for life beyond Earth, as articulated in the Decadal Survey on Astronomy and Astrophysics report (hereafter Astro2020 \cite{Astro2020}). A foundational aspect to the search for life is establishing the context for habitability on promising targets and determining their ability to sustain surface liquid water over geological timescales. Thus a critical knowledge gap to fill over the coming decade is determining what fraction of temperate, rocky planets around stars of different stellar types can develop and retain atmospheres capable of supporting long-term habitability.

A working definition of a habitable exoplanet is one that resides within the habitable zone of its host star, where the presence of an atmosphere can allow for surface liquid water (Ref\cite{ExoplanetScienceStrategy2018} and references within). Widely used metrics such as planetary effective surface temperature, which depends only on stellar bolometric irradiance and albedo, have little predictive power for determining the presence of an atmosphere capable of sustaining surface liquid water. The stellar energetic radiation environment is a fundamental factor controlling whether a potentially habitable planet can retain its atmosphere, a necessary condition for surface habitability. 

Extreme ultraviolet (EUV; 100-911 \AA) photons dominate the energy input, photoelectron production, total heating rate, and temperature structure in the upper atmospheres of terrestrial planets orbiting both active and inactive stars, driving atmospheric escape \cite{Looveren2024, Solomon2005}. In addition to the present day EUV flux, the integrated EUV flux deposited into a planet's atmosphere over its lifetime is needed to understand the total atmosphere mass lost to space \cite{Zahnle2017,Fulton2018}. The most direct way to determine this is with EUV observations of a large sample of young, intermediate-aged, and old stars. Despite the EUV's importance for driving atmospheric escape, measurements of EUV spectra from exoplanet host stars are scarce. The only previous EUV astronomy mission, \euve\ \cite{Bowyer1991}, obtained spectra of approximately a dozen cool main sequence stars (FGKM dwarfs), including 5 young, active M dwarfs \cite{Craig1997}. Previous EUV observations lacked the sensitivity to survey exoplanet host stars in this spectral band, and as a result, widely discrepant techniques have been employed to approximate the critical stellar EUV inputs into exoplanet atmosphere models (Figure~\ref{fig:EUV_spectral_sim_mass_loss_small}).  

High-energy particles influence the evolution of exoplanetary atmospheres through the actions of stellar winds and impulsive events (e.g., CMEs). While stellar winds play a part in atmospheric escape, CMEs are thought to be a dominant source of long-term erosion in planetary atmospheres \cite{Cherenkov2017,Khodachenko2007,luhmann_space_2007,Lammer2018}. The influence of CMEs depends strongly on their frequency, which is unknown for all stars except the Sun. While stellar flare frequencies can be measured \cite{Pye2015}, observations indicate that stellar flares and CMEs may not be correlated or may have a different relationship compared to the Sun \cite{Drake2013,Moschou2019,Leitzinger2020}. Solar CME characterization techniques with `Sun-as-a-star' (i.e., disk-integrated) EUV spectroscopy with the Solar Dynamics Observatory (SDO) EVE instrument \cite{Mason2016,Veronig2021} demonstrate the promise of EUV spectroscopy in measuring the occurrence rates and kinetic energies of stellar CMEs. 

In this paper, we present the Extreme-ultraviolet Stellar Characterization for Atmospheric Physics and Evolution (\escape) mission, a NASA Small Explorer concept proposed in 2026. \escape\ obtains high-fidelity EUV measurements of the spectral and temporal behavior of a large sample stars for the first time, providing exoplanetary science the analogous stellar inputs long recognized as essential by planetary science, heliophysics, and Earth science.

\escape\ provides a bridge between NASA’s existing exoplanet detection and characterization missions and the future Habitable Worlds Observatory (HWO) by establishing which combination of stellar (mass, age, activity, CME frequency) and planetary (composition, orbital distance, magnetic fields) parameters provide the most favorable conditions for the long-term stability of temperate, terrestrial atmospheres. Importantly, far-UV (FUV; 912-1800 \AA) and near-UV (NUV; 1800-3200 \AA) emissions, measurable by HST (today) and UVEX \cite{Kulkarni2021} (in $\sim$2031), do not drive atmospheric loss on planets around FGKM stars, nor do they trace CMEs \cite{Feinstein2022,Loyd2022,Shkolnik2025,Park2025}. \escape\ gives us access, for the first time, to a broad sample FGKM stars in the EUV, the wavelength range primarily responsible for heating the upper atmosphere and powering atmospheric escape.

This paper is organized as follows: Section~\ref{sec:sci_obj} describes \escape's primary science questions and survey strategy.  Section~\ref{sec:implementation} describes the \escape\ instrument and expected performance. Section~\ref{sec:extendedmission} outlines additional science areas that \escape\ could address in an extended, general observer focused mission, demonstrating scientific impact that exceeds the primary science questions. Section~\ref{sec:conclusions} concludes.

\section{Primary Science Questions and Implementation} \label{sec:sci_obj}

ESCAPE’s primary science goal is to identify and understand the star-planet systems that are conducive to the formation and long-term retention of habitable environments. The \escape\ mission addresses this goal by answering three key science questions:

\begin{enumerate}

    \item What is the EUV irradiance in the habitable zone?
    \item How does stellar EUV irradiance evolve in time?
    \item What are the properties of stellar coronal mass ejections?

\end{enumerate}

\escape\ answers these science questions through a comprehensive observing program supported by state-of-the-art planetary atmosphere models that take ESCAPE’s observables as inputs and uses them to estimate atmospheric mass loss rates from representative terrestrial planets. The \escape\ instrument: (1) achieves $\approx$ 25–100$\times$ the EUV efficiency of previous missions, providing the sensitivity to conduct the first statistical survey of EUV irradiance and resolving current order-of-magnitude uncertainties on exoplanet radiation environments, (2) explores EUV variability from flares, stellar rotation, and stellar evolution with its  monitoring survey of F, G, K, and M stars with a range of ages, and (3) executes the first survey of stellar CMEs using techniques validated in Sun-as-a-star measurements by SDO/EVE \cite{Woods2011, Mason2014, Mason2016,Veronig2021, Veronig2025}, providing direct constraints on exoplanet particle environments.

During its two-year prime mission, \escape\ will conduct two science surveys, SEEN and DEEP (Table~\ref{tab:SEENDEEP}). The SEEN survey measures EUV and FUV (80-825 \AA, 1250–1650 \AA) luminosity as a function of mass and age (using rotation period as an age proxy) of a statistical sample of FGKM stars. The DEEP survey conducts time-intensive monitoring of two dozen stars of interest to measure CME and flare rates. Section~\ref{subsec:surveys} describes the surveys in more detail.

\begin{table}[]
    \centering
    \begin{tabular}{|c|c|c|c|}
    \hline
         {\bf Survey} & {\bf Sample} & {\bf Exp Time} & {\bf Data} \\
    \hline
    Stellar Euv & 276 F, G, K, and 
M dwarfs.   & Median:  &  EUV and FUV  (80–1650 \AA)    \\
    ENvironments  & Rotation Periods (age proxy):  &  $\sim$12 ksec  & luminosity as a   \\
    (SEEN) & 0.5 – 100 days. & per star & function of mass and age \\
    \hline
    Dedicated Euv  & 24 F, G, K, and M dwarfs. & $\sim$1 Msec & CME rate via coronal \\ 
    Eruption Program & Rotation Periods (age proxy): & per star &  dimming, EUV flare \\
    (DEEP) & 0.5 – 20 days. &  &   frequency distributions \\
    \hline
    \end{tabular}
    \caption{\escape\ conducts two science surveys during a two-year prime mission. The snapshot survey (SEEN) addresses the first and second science questions, and the monitoring survey (DEEP) addresses the second and third science questions.}
    \label{tab:SEENDEEP}
\end{table}

\subsection{Stellar EUV Irradiance in the Habitable Zone}

The long-term habitability of a terrestrial atmosphere is intimately linked to the flux of radiation from its parent star.  Stellar photons and particles play a fundamental role in the life cycle of volatiles in the planetary atmosphere. Optical and near infrared photons, combined with the greenhouse effect, heat the surface and troposphere of Earth-like planets. Near-ultraviolet (NUV; 1800–3200 \AA), FUV (912–1800 \AA), and X-ray (5–100 \AA) photons are absorbed in the middle and upper atmosphere where they photodissociate molecules and drive photochemistry.  EUV photons (100–911 \AA) are absorbed high in the atmosphere (i.e., in the thermosphere and exosphere) where they ionize atoms and molecules. Liberated photoelectrons and exothermic photochemistry heat the surrounding gas and increase the scale height of the atmosphere, potentially  leading to extreme thermal (e.g., hydrodynamic) escape of the atmosphere\cite{Tian2008}, depending on the incident stellar EUV flux.

While stellar X-rays are often readily observable due to lower levels of ISM attenuation and wide availability of X-ray observatories, EUV photons are the key drivers for atmospheric mass-loss. First, EUV photons are absorbed in the highest (lowest density) layers of the atmosphere where radiative losses are minor and the heating efficiency is highest. Second, while the X-ray and EUV luminosities (and subsequent atmospheric ionization) are comparable for young stars, stars of all ages produce a much larger total number (3-90$\times$) of EUV photons than X-ray photons  \cite{Fontenla2016,Garcia-Sage2017, King2021, Woods2009}. At longer wavelengths, FUV and NUV emissions, measurable with HST, SPARCS \cite{Shkolnik2025}, or MAUVE \cite{MAUVE2026} (today) and UVEX \cite{Kulkarni2021} (in $\sim$2031), do not drive atmospheric loss on planets around FGKM stars. 

\begin{figure}
    \centering
    \includegraphics[width=\linewidth]{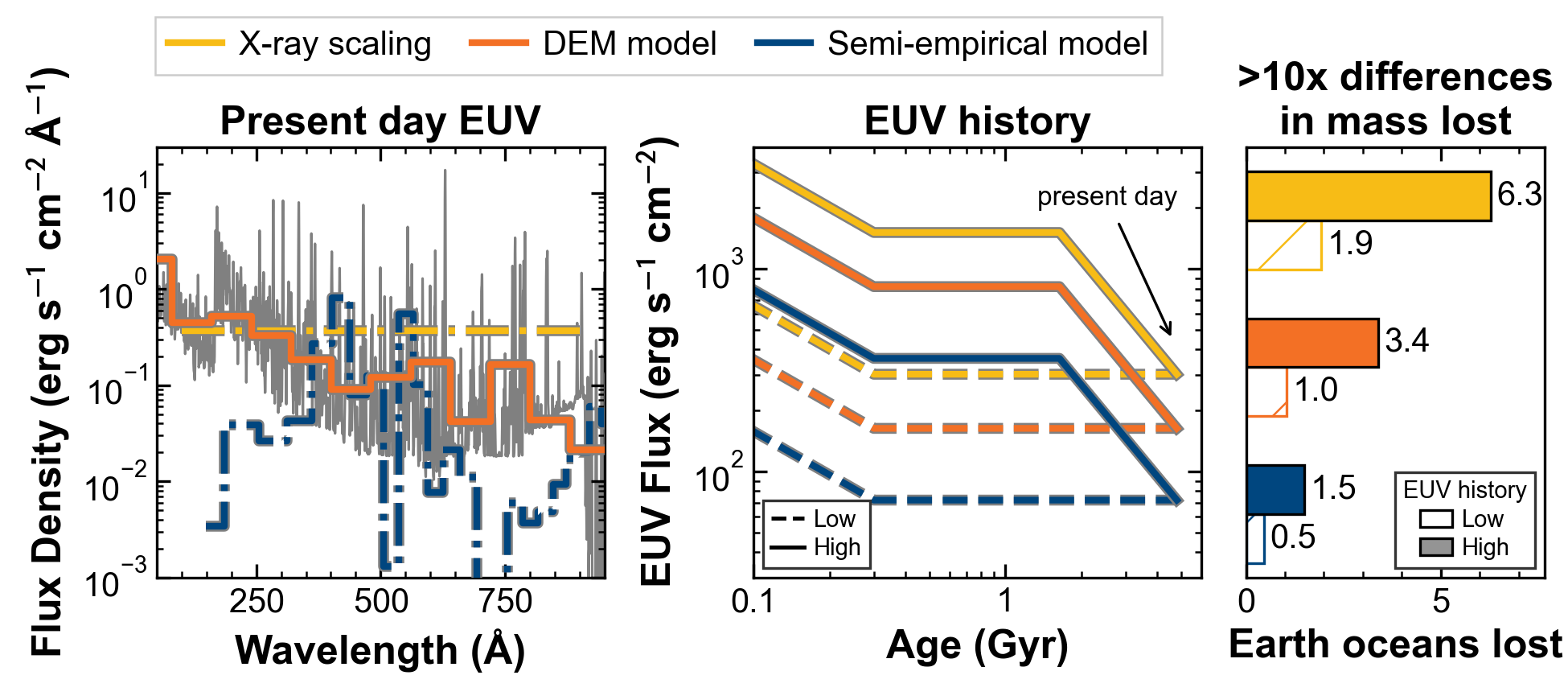}
    \caption{
    \escape\ addresses the factor of $>$10 differences in predicted atmospheric mass loss rates by directly measuring both the present-day EUV irradiance from specific FGKM stars and the stellar EUV evolution as a function of mass. (left) Present day estimated EUV spectra of the M dwarf Proxima Cen used as inputs for the mass loss calculations. Different reconstructions show orders-of-magnitude flux discrepancies: two differential emission measure models in gray \cite{Drake2020} and orange \cite{Duvvuri2021}, X-rays in yellow \cite{SanzForcada2025}, and a semi-empirical model in blue \cite{Peacock2020}. (middle) Possible EUV histories of Proxima Cen \cite{Ribas2017}. (right) Hydrogen mass lost from an Earth-like planet (in units of Earth oceans up to 4.8 Gyr) for high (shaded) and low (hatched) EUV histories from the middle panel. Archival \euve\ spectra of Proxima Cen yielded a handful of emission line detections and thus cannot fully resolve these discrepancies.
    }
    \label{fig:EUV_spectral_sim_mass_loss_small}
\end{figure}

\begin{figure}
    \centering
    \includegraphics[width=0.75\textwidth]{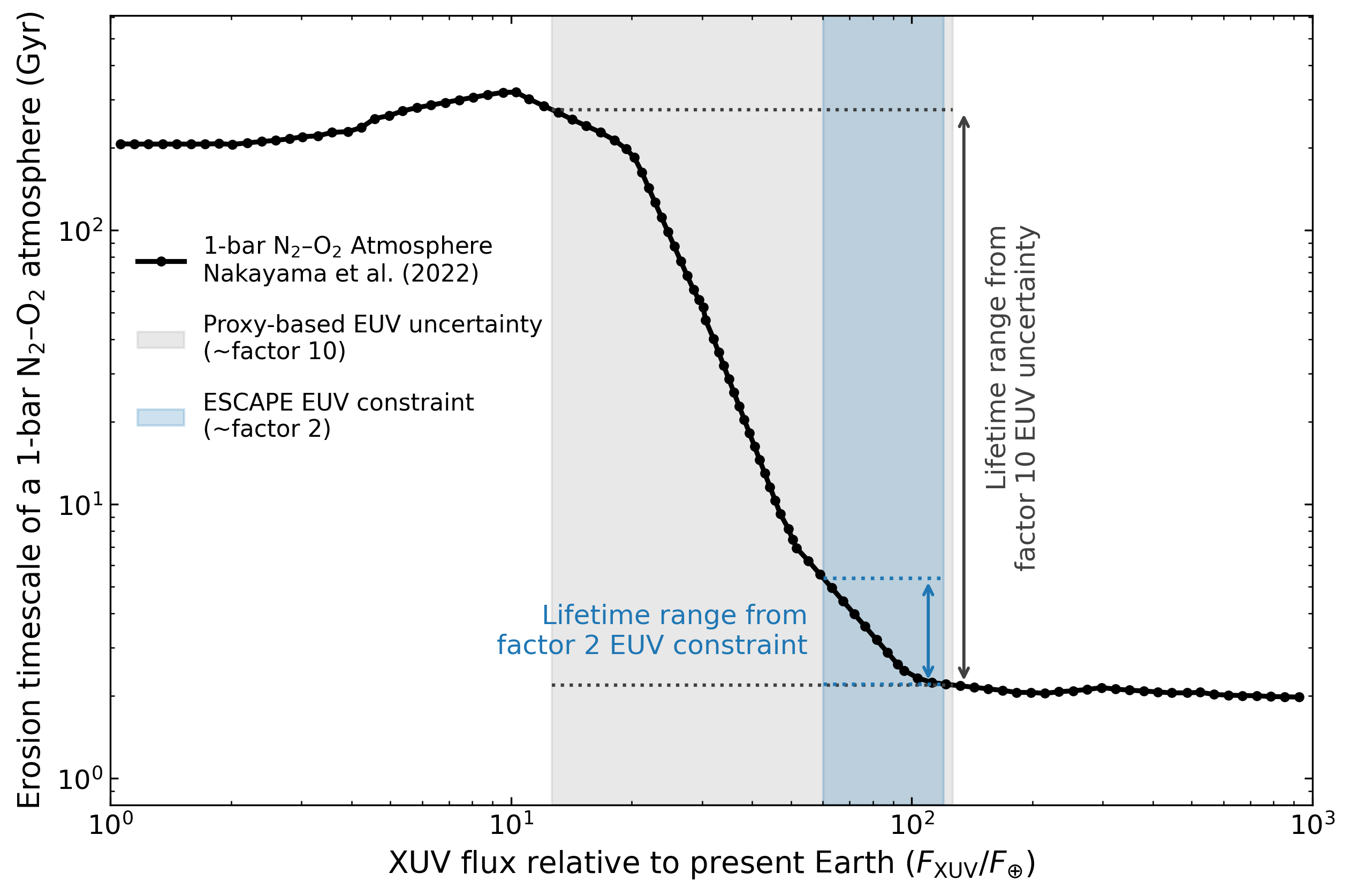}
    \caption{Propagation of stellar EUV uncertainty into atmospheric lifetime predictions for a 1-bar N$_2$–O$_2$ atmosphere (adapted from Ref.~\cite{2022Nakayama}). The black curve shows the erosion timescale of an Earth-like atmosphere as a function of incident XUV flux. The gray and blue shaded regions represent factor of 10 and factor of 2 uncertainties in stellar EUV irradiation, respectively. The factor of 10 uncertainty is representative of present-day proxy-based EUV estimates and encompasses the combined effects of scatter in Lyman $\alpha$–EUV scaling relations \cite{2014Linsky}, differences between reconstruction methodologies \cite{2011Sanz,2016France}, and uncertainties in stellar activity evolution histories \cite{2021Johnstone}. The vertical arrows indicate the corresponding ranges of atmospheric erosion timescales, illustrating how improved stellar EUV measurements can significantly reduce uncertainties in atmospheric escape and retention timescales for rocky exoplanets.}
    \label{Fig:Atm survival lifetime}
\end{figure}

However, stellar EUV fluxes are not well characterized, and uncertainties in stellar EUV flux propagate directly into uncertainties in atmospheric loss rates and atmospheric retention timescales. Figure~\ref{fig:EUV_spectral_sim_mass_loss_small} shows possible EUV spectra for our nearest stellar neighbor, Proxima Cen, which has an Earth-sized planet in the liquid water habitable zone \cite{AngladaEscude2016}. The order-of-magnitude disagreements between the predicted spectra drive major changes in predicted hydrogen loss from Proxima Cen b (Figure~\ref{Fig:Atm survival lifetime}). Figure~\ref{Fig:Atm survival lifetime} additionally demonstrates how current uncertainties in stellar EUV irradiation translate into essentially unconstrained predictions of erosion timescales of Earth-like atmospheres.  By reducing the uncertainty in EUV flux measurements from the current uncertainty levels of a factor of $\sim$10 to a factor of 2 or better, \escape\ significantly narrows the range of permissible atmospheric evolution scenarios, enabling more reliable predictions of the long-term evolution and habitability of rocky exoplanets. This will enable us to interpret rocky exoplanet observations from JWST today \cite{Glidden2025,Lustig-Yaeger2023} to identify planetary systems with the greatest potential for sustaining habitable environments.

\subsection{Stellar EUV Evolution and Variability in Time}

Planetary atmospheres respond to changing stellar conditions~\cite{Schunk2009,Solomon2011}. \escape\ quantifies how EUV emission varies over short and long timescales to assess how flares, rotation, and rotational evolution drive planetary atmospheric escape and evolution.

\subsubsection{Minute-timescale variability}

Solar and stellar FUV/EUV emission varies by factors up to $\sim$100 on minute timescales due to flares \cite{Woods2012,Loyd2018a,Loyd2018b,MacGregor2021,Veronig2021}. Models using estimates of the EUV irradiation of the Great AD Leo Flare of 1985 \cite{Hawley_Pettersen_1991_ApJ} indicate that the atmospheres of planets orbiting active stars with frequent flares never achieve a steady state because the timescales for atmospheric recovery \cite{Tilley2019,Venot2016} ($\sim$years) are much longer than the time between successive flares \cite{Feinstein2022,Hawley2003} (hours). \euve\ photometry provided constraints on EUV flare frequencies for a small sample of young, active stars \cite{Audard2000}; however, the stellar hosts most suitable for habitable planet searches are older and less active. The lack of direct constraints on EUV spectra and flare activity on older, less active stars hinders our ability to model the atmospheric response to stellar flares for the most promising star-planet systems to host habitable worlds. Empirical knowledge of the EUV variability of stars older than the Sun is essentially non-existent; \escape\ will provide this data for the first time.

\begin{figure}
    \centering
    \includegraphics[width=0.7\linewidth]{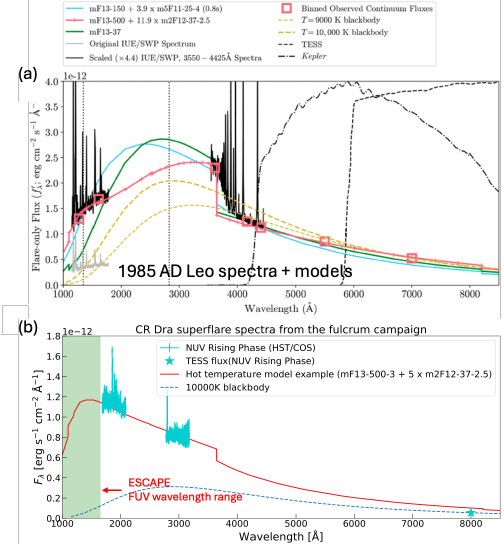}
    \caption{
(a) Observed spectra and models of the early-impulsive phase of the Great AD Leo Flare of 1985 show the broadband continuum flux distribution from the FUV to the optical (adapted from Ref \cite{Kowalski_2022_FrASS}). The wavelength-binned, flare-only fluxes that are used to fit the models are indicated by square symbols with a best-fit, two-component RHD model continuum spectrum shown as the solid red line. Other model predictions are scaled to the observations as follows: the blackbody functions are scaled to the R band flux observation, and the other two RHD models are scaled to the average continuum flux at $\lambda$=4155 -- 4185\AA.
(b) The example spectra of ``rising NUV spectra" \cite{Kowalski+2025_ApJ}, along with an example continuum RHD model spectra with ``hot temperature" component \cite{Kowalski+2024_ApJ} (see also Notsu et al. submitted). 
Importantly, models based on this data predict the peak flux is located in the FUV, as opposed to the conventional expectation from the Great AD Leo 1985 flare in (a) where the modeled flux peaks in the NUV.
}
    \label{fig:flare_FUVspec_example}
\end{figure}

While early constraints on planetary atmospheric impacts relied heavily on historical benchmark events like the Great AD Leo Flare of 1985 mentioned above \cite{Hawley_Pettersen_1991_ApJ,Segura+2010,Tilley+2019,Miranda-Rosete+2025_ApJ}, recent advancements in multi-wavelength spectroscopic observations have broadened our understanding of flare emissions \cite{Tristan+2023_ApJ,Notsu+2025_ApJ,Kowalski+2025_ApJ}. For example, recent HST observations from the Fulcrum Flare Campaign have revealed unique spectral signatures, most notably the ``rising NUV spectra" during M-dwarf superflares (radiation energy $\gtrsim10^{33}$ erg) \cite{Kowalski+2025_ApJ}. Crucially, these rising NUV spectra suggest that the actual NUV and FUV radiative output can be a factor of six or more larger than what conventional models and the traditional 1985 AD Leo flare paradigm predict (cf. Figure \ref{fig:flare_FUVspec_example}). This massive enhancement strongly implies that historical estimates of EUV emission during extreme flares may also be severely underestimated and require substantial revision (see also FUV flares with hot temperature values \cite{Loyd2018b,Froning+2019_ApJ}). To accurately assess exoplanetary irradiation, it is no longer sufficient to simply scale historical spectra; we must investigate the underlying physical mechanisms driving lower atmospheric heating and the resulting EUV emission.

To explore these driving mechanisms, modern stellar flare research utilizes 1D Radiative-Hydrodynamic (RHD) modeling with community-standard codes like RADYN \cite{Carlsson_Stein_1992_ApJL,Allred+2015_ApJ,Kowalski+2024_ApJ}. Within this framework, magnetic reconnection in the stellar corona accelerates non-thermal particles. RHD simulations investigate how these non-thermal electron and proton beams propagate along magnetic field lines and deposit energy into the denser stellar chromosphere \cite{Kowalski+2024_ApJ,Kowalski+2025_ApJ,Kerr+2023_ApJ}. The recent discovery of the rising NUV spectra highlights the importance of a massive electron or proton flux reaching the lower atmosphere. Consequently, the properties of these large particle fluxes are a critical factor in estimating the overall high-energy output, including the revised EUV and X-ray emissions (Notsu et al., submitted).

As these particle beams penetrate the lower atmosphere, they heat the local plasma via Coulomb collisions, driving an atmospheric expansion known as chromospheric evaporation. This evaporation process is a primary source of the enhancements in EUV and soft X-ray emission observed during flares. However, while recent multi-wavelength observations have significantly improved our understanding of the NUV, optical, and X-ray bands, they currently lack simultaneous EUV and FUV coverage over extended baselines. Given the indication that lower-atmospheric radiative outputs (like NUV and FUV) are drastically larger than previously assumed (cf. Figure \ref{fig:flare_FUVspec_example}), relying solely on theoretical RHD models to extrapolate the unobserved EUV flare emission has inherent limitations; direct empirical data are necessary. 

\escape\ will fill this observational gap by providing simultaneous EUV (80-825 \AA) and FUV (1250–1650 \AA) spectroscopy of stellar superflares.  
\escape's FUV channel captures both the FUV continuum—-allowing us to infer flare blackbody temperatures—-and the relative enhancements of intermediate-temperature spectral tracers like C II, Si IV, and C IV during massive events. Furthermore, based on multi-wavelength flare occurrence rates from the Fulcrum Flare Campaign with HST and AU Mic XMM-Newton programs \cite{Kowalski+2025_ApJ,Tristan+2023_ApJ,Notsu+2025_ApJ}, the 15-day stares per target planned for \escape's DEEP survey will capture multiple instances of these large, rare superflares on active stars. 
Ultimately, these simultaneous, long-baseline observations will provide the critical constraints needed to validate RHD beam models and accurately define the newly enhanced EUV radiation environments shaping planetary atmospheres.

\subsubsection{Stellar rotation and cycle variability}

Rotation, active region evolution, and activity cycles introduce wavelength-dependent variations in solar EUV emission on timescales of days to years \cite{woods_solar-cycle_2022}.
Variability amplitudes are highest at the shortest wavelengths, exceeding 100\% at 100~\AA.
A single stellar observation, if unknowingly timed near a minimum or maximum of these periodic variations, will not represent the average irradiance a planet experiences.
This can lead to inaccuracies when interpreting observations of atmospheric escape collected at a different epoch.
\escape's DEEP survey largely overcomes the limitation imposed by rotational variability by constraining the EUV variability of cool stars due to spots and active regions rotating in and out of view on timescales from $\sim$25\% to several times the rotation period for active stars. 

Longer term activity cycle trends, typically in the range 4–12 years \cite{Isaacson2024}, could be readily studied should \escape\ include an extended mission phase \cite{Ayres2020}.
Activity cycle variations can explain nearly all scatter in soft X-ray emission observed from similar, Sun-like stars \cite{Johnstone2021}, yet, recent analysis of multi-epoch observations show they do not account for the soft X-ray scatter of M~dwarf stars (P. Wheatley, private comm.).
An extended \escape\ mission could clarify the timescales associated with the greatest amplitude variations in M~dwarf EUV emission and help resolve the mystery behind star-to-star scatter noted in soft X-ray observations. 

\subsubsection{Stellar evolutionary timescale variability}

\begin{figure}
    \centering
    \includegraphics[width=\linewidth]{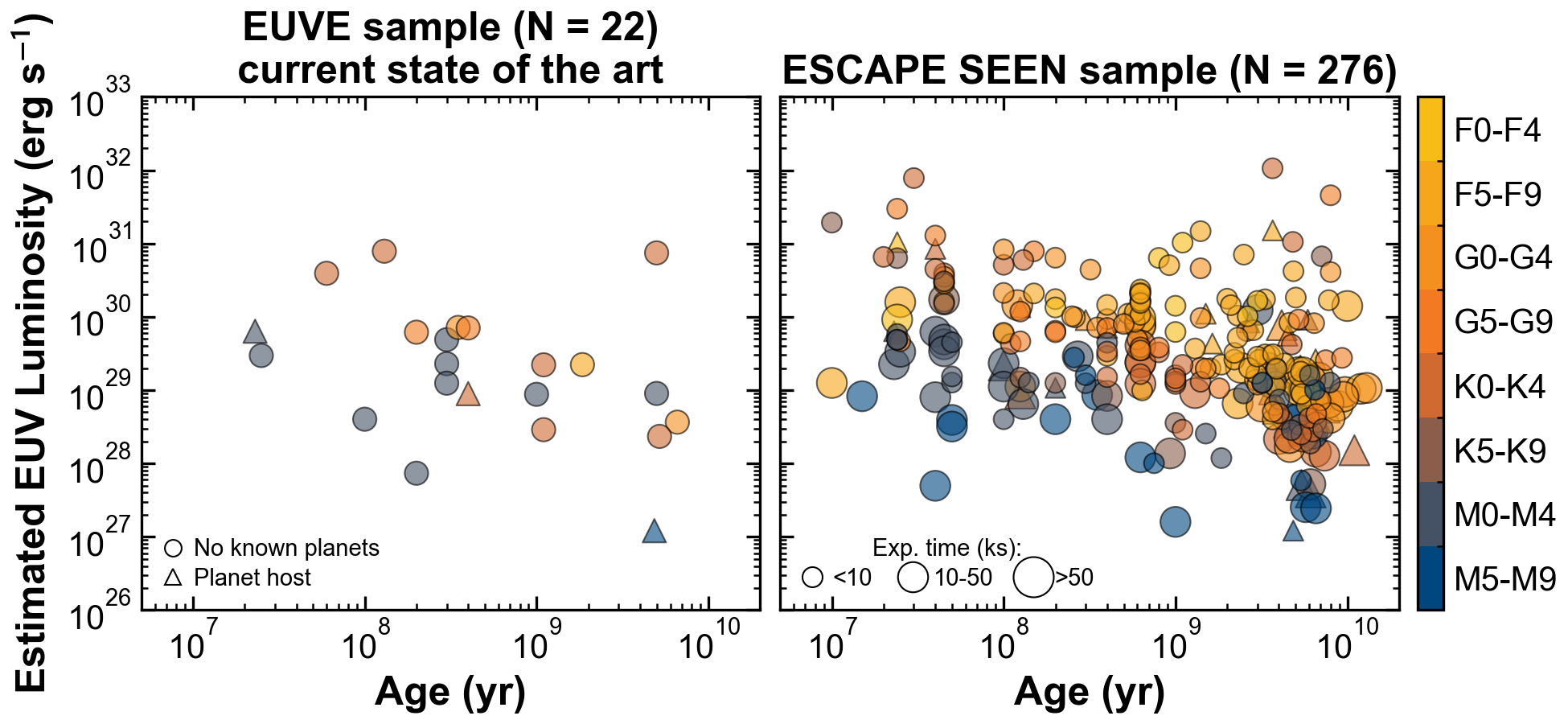}
    \caption{\escape\ characterizes the time-variability of the EUV radiation field on stellar evolutionary timescales with the SEEN survey. This data is essential to creating the first empirical maps of the cosmic shoreline for planetary systems. In both panels, the EUV luminosity is the estimated 169–179 \AA\ intrinsic luminosity and the symbol sizes represent the exposure time for a SEEN observation (applicable to right panel only). Stars with confirmed exoplanets are shown as triangles. The right panel shows the current SEEN target list (276 stars), with resolution in both age and mass (color-coded). For comparison, the left panel shows the 22 FGKM stars of the \euve\ spectroscopic sample \cite{Craig1997}, where the majority of stars were only weakly detected in 1-2 emission lines.}
    \label{fig:EUV_age}
\end{figure}

The integrated EUV flux deposited into the atmosphere over the planet’s lifetime is potentially even more important to the cumulative escape rate than the short-term effect of flares and rotational variability, for both terrestrial \cite{Zahnle2017} and gaseous planets \cite{Fulton2018}. The `Sun-in-Time' program \cite{Ribas2005} observed five solar-type stars of different ages, including the Sun, in the EUV (with \euve\ and solar telescopes); however, the study was limited to one star per age bin because of the lack of available EUV spectra. There is no comparable study of EUV time evolution for any other spectral type.

Moreover, studies of young open clusters reveal both fast and slow rotators, demonstrating that a wide range of initial rotation rates is possible under nearly identical star forming conditions (e.g., Ref.\cite{Johnstone2021} and references within).
To fully characterize the impact of the uncertainty of initial rotation rate on the EUV irradiance history of planets, EUV observations are needed for a large sample of stars spanning a wide range of rotation periods and spectral types.
These can then be combined with models of rotational evolution that factor in scatter in initial rotation rates to quantify the expected range of plausible exposure a planet has experienced over its life.

\escape\ measures both the critical pre-main sequence evolution of M dwarf EUV luminosities \cite{Ribas2017} and the main sequence evolution of F, G, K, and M star EUV luminosities (Fig. \ref{fig:EUV_age}) via the SEEN survey, which targets stars of a wide range of spectral type and age (Section~\ref{sec:seen}).

\subsection{Stellar Coronal Mass Ejections}

CMEs deliver heat to the upper atmospheres of orbiting planets through spikes in dynamic pressure, Ohmic heating from rapidly changing magnetic fields, and charge exchange reactions \cite{Chassefiere1996,Cherenkov2017,strugarek_ohmic_2025}. The space weather generated by CMEs has led several authors to propose the exoplanet ``space weather habitable zone" \cite{Airapetian2020} as more meaningful than the traditional liquid water HZ. Approximately 90\% of large (X-class) solar flares are associated with CME-like particle eruptions \cite{Yashiro2009}; however, despite predictions that analogous eruptions could be observed on other stars, this connection has not been borne out with recent observations \cite{Crosley2018,Odert2017,Osten2015}.

For the Sun, we can observe CMEs directly with coronagraphs and in-situ measurements. These traditional methods are not feasible for CMEs from other stars. However, other techniques for detecting CMEs have been proposed for very active or young stars, including blue-shifted H$\alpha$ \cite{Inoue2023,Namekata2025}, Type-II radio bursts \cite{Callingham2025}, and variations in the FUV or X-ray light curves \cite{Haisch1983,Loyd2022,Moschou2017}. The connection between H$\alpha$, FUV Doppler shifts, X-ray/FUV variability and CMEs remains inconclusive, and these techniques are only sensitive to CME-like events larger than those observed on the Sun, i.e., on the youngest and most active stars, not the more evolved star-planet systems expected to be the optimal habitable planetary system candidates. 

\begin{figure}
    \centering
    \includegraphics[width=\linewidth]{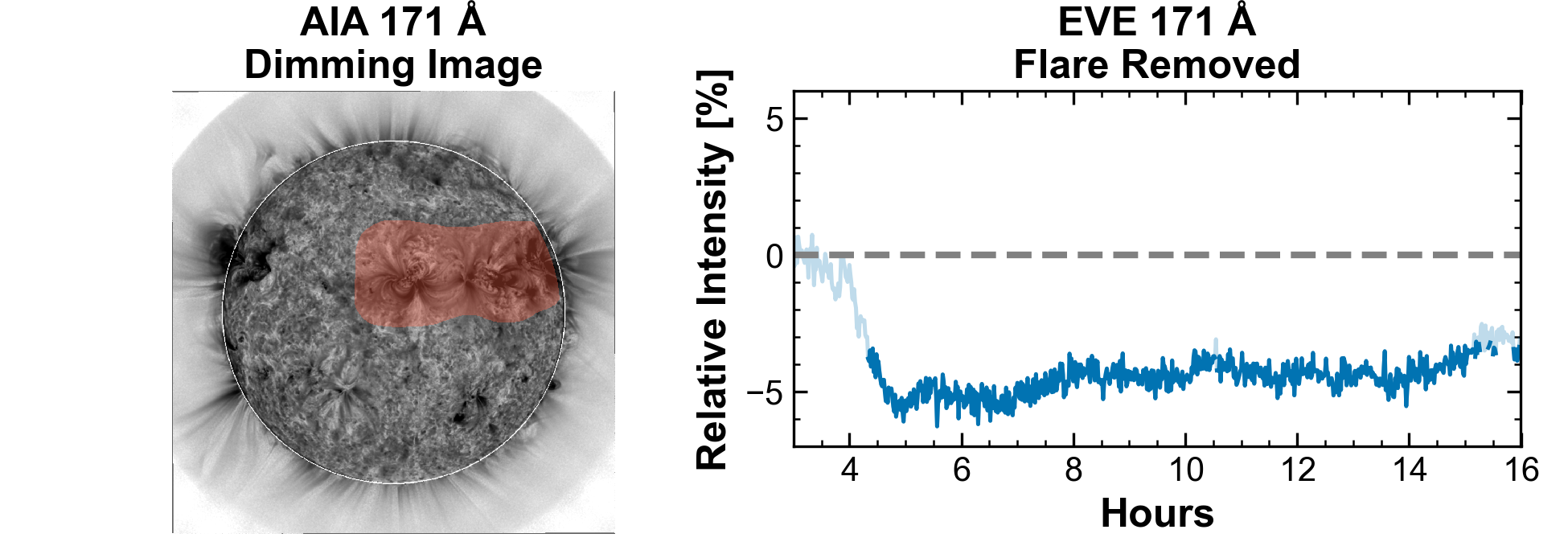}
    \caption{Observation of coronal dimming (location highlighted in red where it occurred in this particular event) in angularly resolved filter imaging (left) and in full-disk solar spectra (right). The SDO/AIA image at left shows dark `holes' where the quiescent corona has been evacuated by a CME, leading to dimming in disk-integrated spectra. The spectrally resolved lightcurve (right) from SDO/EVE shows an example flare-subtracted Fe IX 171 \AA\ using the methods in \cite{Mason2014}. The highlighted portion of the lightcurve indicates the trough, where the intensity is more than halfway down to the max depth, indicating the dimmed portion. \escape\ is sensitive to similar events on F-M dwarfs in the solar neighborhood. More than 200 dimming events are expected in the DEEP survey.}
    \label{fig:CMEfig}
\end{figure}

Conversely, a coronal dimming signal occurs when emission lines associated with the quiescent corona dim as the CME evacuates plasma from the stellar atmosphere \cite{Woods2011} (Fig. \ref{fig:CMEfig}). The Solar Dynamics Observatory EVE instrument \cite{Woods2012} (SDO/EVE) demonstrated that dimming can be characterized in disk-integrated solar EUV spectra; these data also contain information about the kinematics of the CME that produced it \cite{Mason2014,Mason2016,Mason2019}. 

Intriguingly, Ref\cite{Veronig2021} published initial evidence from archival \euve\ photometric observations of young stars that coronal dimming in broad EUV bands ($\sim$100 Å wide) with depths up to 50\% may be detectable from `extreme' dimming events on highly active stars. These extreme dimming events are significantly different than the individual coronal line dimmings of $\leq$ 5\% observed in `solar-like' CMEs \cite{Mason2016}. Hints of extreme CMEs at X-ray wavelengths have been observed only on the most active stars, but high levels of stochastic stellar variability hinder current searches in existing, relatively short observing campaigns \cite{Tristan+2023_ApJ}.

Community reviews of stellar CME detection in 2016 \cite{Harra2016} and 2025 \cite{Loyd2025} concluded that EUV coronal dimming is a primary feature consistently associated with solar CMEs that can be employed to detect and quantify both solar-like and extreme CMEs on other stars \cite{Loyd2025}. 

\escape\ employs the same `Sun-as-a-star' stellar CME characterization techniques validated on SDO/EVE to measure the frequency and properties of CMEs on nearby stars for the first time. Specifically, \escape\ is designed to distinguish between solar-like ($\sim$5\%) and extreme ($\sim$50\%) dimming. \escape\ will vastly expand the empirical basis for understanding stellar CMEs with its $>50\times$ spectrocopic sensitivity compared to \euve\ and ultra-long monitoring campaigns as part of the DEEP survey. These data will quantify the frequency of CMEs with $\geq 3 \sigma$ detections of coronal dimming on individual stars\cite{Mason2025}, demonstrate the connection (or lack thereof) to EUV and FUV flares, and enable us to establish how CME activity changes with stellar mass and age for the first time.

\subsection{\escape\ Surveys and Expected Performance} \label{subsec:surveys}

\subsubsection{The SEEN Survey: EUV Snapshots of 276 FGKM stars} \label{sec:seen}
The Stellar Euv ENvironments (SEEN) survey measures the 80–1650 \AA\ irradiance and time variability for 276 stars (with $\geq$60 stars per spectral type bin) in a nominal two-year prime mission (Table~\ref{tab:SEENDEEP}). These snapshot observations will assess how
similar the EUV outputs of two stars of comparable mass and age are, and importantly, allow us to develop a range of EUV flux estimates for a given set of basic stellar parameters. The SEEN survey also enables the first comprehensive empirical mapping of the
cosmic shoreline (Fig. \ref{fig:EUV_age} and see Section~\ref{sec:nonHZexoplanets}), where current estimates for the cumulative EUV flux experienced by planets are only certain to within a factor of $\sim$10. While it is not the SEEN survey's primary purpose to measure time variability due to flares and CMEs, \escape's photon-counting detector ensures serendipitous flare and CME detections in the SEEN survey (12 ks median exposure time per star).  

The nominal SEEN target list of 276 stars was selected to: 1) cover the majority of the HWO target list \cite{Mamajek2024} and key exoplanetary systems observed by JWST, 2) maximize the number of targets with prior ISM knowledge (Section~\ref{sec:ISM}), 3) sample a broad distribution of spectral type and ages (Figure~\ref{fig:EUV_age}), and 4) minimize the number of unresolved sources in \escape’s aperture. Additionally, 60 days of science exposure time is available to allocate to the highest priority exoplanet host stars at the time of launch. These additional targets will be determined with community input prior to launch. 

\begin{figure}
    \centering
    \includegraphics[width=0.8\linewidth]{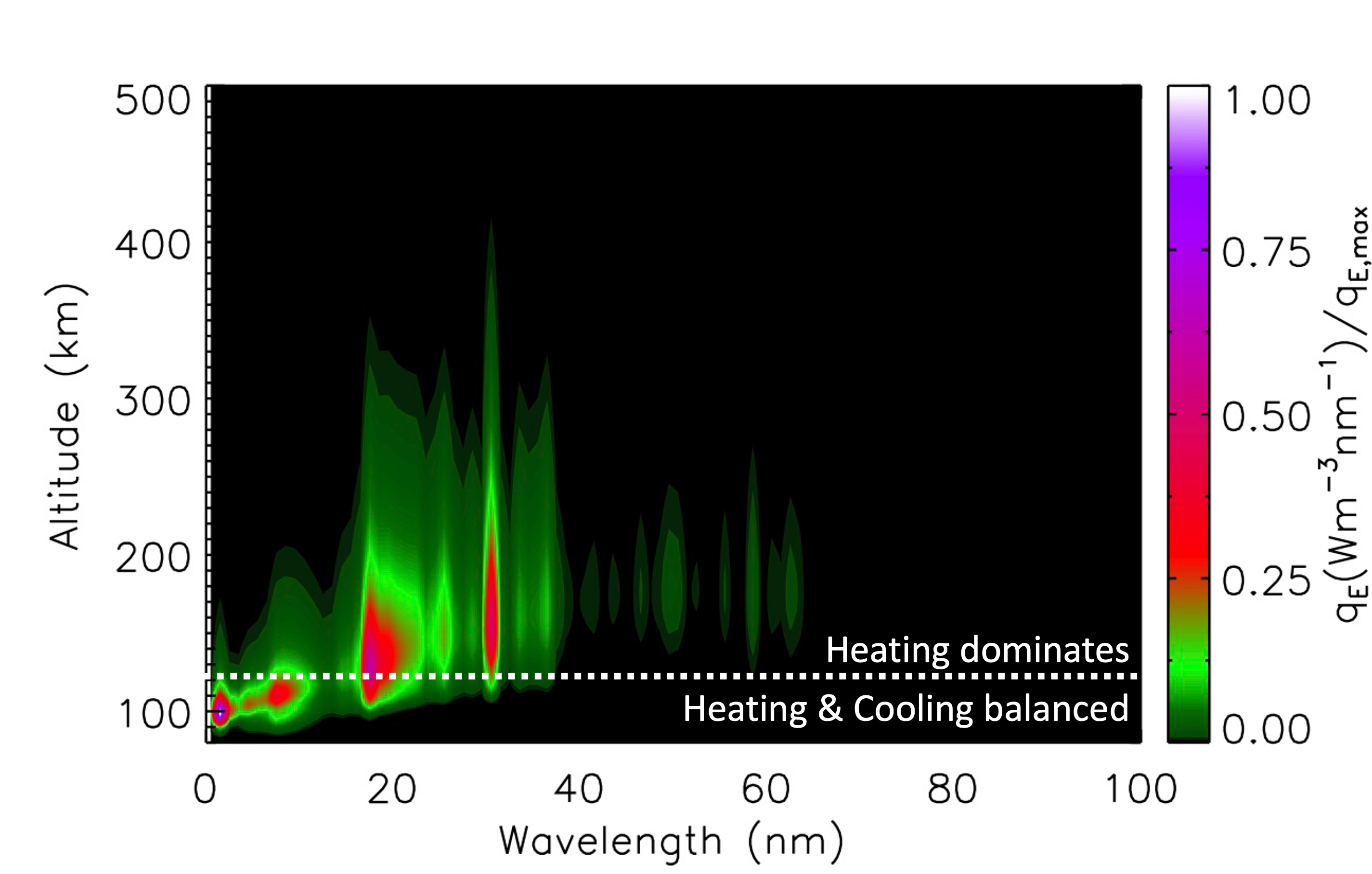}
    \caption{Figure reproduced from Ref.\cite{Youngblood2025}. The contour plot shows, for the Earth’s atmosphere, the specific energy input (qE) of primary photoelectrons (a proxy for heating) normalized by the peak of this energy input (qE,max) as a function of wavelength and altitude. This would represent the photoelectron heating rate for 100\% heating efficiency. In reality, the heating efficiency depends on atmospheric properties. An NRLMSIS \cite{Emmert2021} atmosphere at 60$^{\circ}$ solar zenith angle and noon local time and a TIMED/SEE \cite{Woods2000} solar spectrum at moderate activity were used. The horizontal dotted line at $\sim$120 km demarcates the altitude above which heating dominates and escape occurs. This figure illustrates the relative importance of different wavelength regions to the EUV energy input. Despite the greater number of stellar photons from 400 to 911 \AA, the 100–400 \AA\ photons produce more energetic photoelectrons and dominate the energy input at high altitudes where heating dominates over radiative cooling processes on the Earth ($>$120 km).}
    \label{fig:Tommi}
\end{figure}

\escape\ is optimized for the 90–500 \AA\ bandpass that dominates energy deposition for rocky planet upper atmospheres (Figure~\ref{fig:Tommi}). Long wavelength EUV spectra (600–825 \AA) are acquired directly for bright stars with low interstellar attenuation (Section~\ref{sec:ISM}), and the FUV spectral coverage (1250-1650 \AA) supports long-wavelength EUV reconstruction for fainter targets via differential emission measure (DEM) models. 

The DEM distribution method \cite{Craig1976,SanzForcada2025} relies on the observation of emission lines originating throughout the upper stellar atmosphere. The DEM method is more accurate than scaling relations based on individual UV or X-ray tracers \cite{Duvvuri2025} and has been successfully applied to both EUV and X-ray stellar observations \cite{Bowyer2000,SanzForcada2003,SanzForcada2002,Wood2018}. DEM calculations provide an accurate means to ``fill in” the long wavelength EUV spectrum of stars where the opacity of the interstellar medium prevents a direct measurement (Section~\ref{sec:ISM}). ESCAPE's spectral coverage and sensitivity enable robust DEM-based synthetic spectra to be produced for all stellar targets.

\subsubsection{The DEEP Survey: EUV Monitoring of 24 FGKM stars for CMEs and Flares}
\label{sec:deep}
The Dedicated Euv Eruption Program (DEEP) survey monitors 24 select stars each for 15 calendar days to measure EUV and FUV flare frequency distributions and CME rates. The notional DEEP target list comprises bright stars spanning a range of masses of ages (Table~\ref{tab:SEENDEEP}) with known activity levels in order to maximize the predictability of flare and CME detection. \escape's photon-counting detector records all observations in a `time-tagging' mode that provides higher temporal resolution ($<$1 s) than the characteristic UV flare duration on low-mass stars.  The 15-day temporal baseline is $\sim$20-40$\times$ longer than the best available dataset from HST and enables comprehensive flare frequency and CME statistics.

The stellar flare-frequency distribution (FFD) is a power law with more low energy flares than high energy flares \cite{Hilton2010}. The integral of the FFD gives the total EUV flux enhancement delivered to orbiting planets. \escape\ will determine the power law slope of the FFD, $\alpha_{\rm FFD}$, to an accuracy of 15\%, reducing its uncertainty by a factor of $\sim$3 compared to existing constraints \cite{Audard2000}, allowing clear differentiation on the frequency of ``big" vs ``small" flares \cite{Chen2021}. Atmospheric responses to large, individual events and the cumulative effects of a steady delivery of continued low-level flaring can differ significantly \cite{Chen2021,France2020,Tilley2019}; \escape\ will quantify the frequency and type of flares that deliver excess EUV radiation to exoplanetary atmospheres. 
Forty-five (45) flares per star are required to develop a flare-frequency distribution with $\sim$15\% accuracy on the power-law slope \cite{Kowalski2024}. Scaling the average EUV FFD of Ref\cite{Audard2000} to the distance and quiescent X-ray luminosity of each target, the DEEP survey is projected to detect $>$65 flares per star.

Additionally, \escape\ plans to include a Student Contribution instrument called the Aligned \escape\ Student-led Optical Photometer (AESOP), which will enable direct connection between the optical and EUV responses to superflares observed with Kepler and TESS \cite{Namizaki2023,Notsu2018,Paudel2024,Jackman2024} for the first time. \escape's EUV and FUV channels will similarly yield correlations between the EUV and FUV responses of flares that can be employed by FUV observatories long past \escape\ operations.

\escape's predicted CME detection yield is based on direct observations of CME-driven dimming events on the Sun. The number of solar-like CMEs detected per DEEP target $N_{det}$ is given by

\begin{equation}
    N_{det} = \dot{N}_{CME}~t_{exp}~f_{dim}~\eta 
    \label{eq:CME}
\end{equation}

\noindent where $\dot{N}_{CME}$ is the rate of CMEs for DEEP targets, $t_{exp}$ = 15 days is the exposure time per target, $f_{dim}$ is the fraction of CMEs with an associated solar-like coronal dimming depth $\geq$5\%, and $\eta$ = 78\% (based on `day in the life' orbital analyses that incorporate spacecraft charging, `keep out' zones of avoidance, and communication periods) is the DEEP survey observing efficiency. We baseline a conservative estimate of $\dot{N}_{CME}$ = 5 CMEs/day for the DEEP targets. This is based on the average solar rate of 3.5 CMEs/day (time-averaged) \cite{Barlyaeva2009,Gopalswamy2009} or 10 CMEs/day at solar maximum and the fact that Ref\cite{Reinhold2020} demonstrated the Sun to be a factor of $\sim$5 less active than similar solar-type stars. Ref.\cite{Reinard2009} find that all solar CMEs with velocity $>$800 km s$^{-1}$ and 54\% of CMEs with velocity $<$800 km s$^{-1}$ are associated with dimming events. 94\% of all solar CMEs have velocity $<$800 km s$^{-1}$ \cite{Gopalswamy2009,Barlyaeva2009}. Additionally, solar coronal dimming with depths $>$ 5\% in Fe IX 171 \AA\ make up approximately 33\% of all dimming events \cite{Mason2019}. We combine these numbers to find $f_{dim}$ = ((0.06 $\times$ 1 + 0.94 $\times$ 0.54) $\times$ 0.33) = 19\%. 

Combining these values with Equation~\ref{eq:CME}, we find $N_{det}$ = 11 CMEs per target, demonstrating that \escape\ is expected to detect $>$200 `solar-like' dimming events for the 24 target stars in the DEEP survey. This value is conservative for the mission because it does not include other dimming events caught by the SEEN survey and excludes any $<5$\% dimmings that may be detectable for bright sources.

The high magnetic fluxes on active M stars (factors of 10–1000 times higher than global solar magnetic fluxes \cite{Donati2006,Shulyak2017}) may trap the charged particle eruptions associated with CMEs \cite{AlvaradoGomez2018}. Simulations indicate that CMEs with energy $>$ 10$^{34}$ erg will escape magnetic confinement from active M dwarfs and be detectable with \escape\ \cite{Jin2019}. On the other hand, a non-detection of `solar-like' dimming events would indicate a physical mode for eruptive events on M dwarfs that differs from the Sun, at $\approx$5$\sigma$ confidence. The EUV light curves recorded in the DEEP survey will provide the first systematic study of the characteristics of stellar CMEs and their ability to break out of the stellar magnetosphere. 

\subsubsection{Expected Performance and Comparison to Other Missions} \label{sec:comparison}

\begin{figure}
    \centering
    \includegraphics[width=\linewidth]{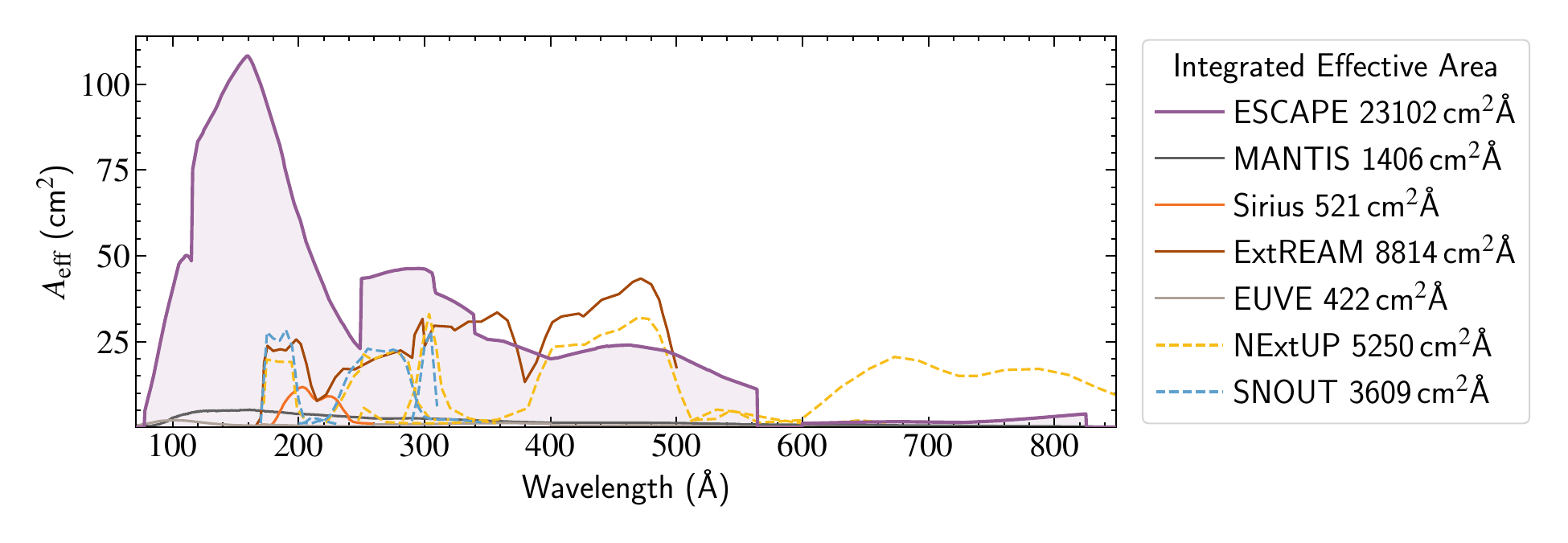}
    \caption{The effective areas of each instrument listed in Table~\ref{tab:missions_properties}. Spectrographs are in solid lines and photometers are in dashed lines. \escape\ is shown as the purple shaded area. The legend also lists the total effective area integrated over each instrument's total bandpass. }
    \label{fig:Aeffs}
\end{figure}

\begin{table}[]
    \centering
    \small
    \begin{tabular}{|c|c|c|c|c|c|c|c|}
    \hline
         & & & {\bf $A_{\rm eff}$} & {\bf Integrated} & & {\bf Total} &   \\
         {\bf Mission} & {\bf Resolving} & {\bf EUV} &  {\bf at 175 \AA} & {\bf $A_{\rm eff}$} & {\bf PSF area} & {\bf background} & {\bf Ref.} \\
         {\bf name} & {\bf power} & {\bf bandpass} & {\bf (cm$^{-2}$)} &  {\bf (cm$^{-2}$) $^a$} & {\bf (cm$^{2}$)} & {\bf rate (ct s$^{-1}$) $^b$} & \\
    \hline
    ESCAPE & $\sim$200 & 80-825 \AA & 94 & 23102 & 1.6$\times$10$^{-3}$ & 1.3$\times$10$^{-3}$ & This work \\
    \hline
    MANTIS & $\sim$20 & 100-500 \AA & 5 & 1406 & 1.6$\times$10$^{-3}$ & 4$\times$10$^{-4}$ & Ref.\cite{indahl2022} \\
    \hline
    SIRIUS & $\sim$5000 & 180-260 \AA\ & 1 & 521 & 1.3$\times$10$^{-4}$ $^\dagger$ & 1$\times$10$^{-4}$ $^\dagger$ & Ref.\cite{Barstow2015} \\
    \hline
    ExtREAM$^c$ & $\sim$5000 & 170-500 \AA\ & 24 & 8814 & 1.3$\times$10$^{-4}$ & 1$\times$10$^{-4}$ $^\dagger$ & Ref.\cite{Drake2025} \\
    \hline
    EUVE & $\sim$200 & 70-760 \AA\ & 1 & 422 & 8$\times$10$^{-5}$ $^\dagger$ & $\sim$3$\times$10$^{-4}$ & Ref.\cite{Bowyer1991} \\
    (DS/S) & & & & & & & \\
    \hline
    EUVE  & photometry & 65-360 \AA\ & 3 & 1440 & 8$\times$10$^{-5}$ & 1.2$\times$10$^{-4}$ & Ref.\cite{Bowyer1991} \\
    (DS/P) & & in 2 bands & & & & & \\
    \hline 
    NExtUP & photometry & 170-921 \AA & 20 & 5250 & 4.2$\times$10$^{-5}$ $^\dagger$ & 1.6$\times$10$^{-5}$ $^\dagger$ & Ref.\cite{Drake2021} \\
    & & in 5 bands  & & & & & \\
    \hline
    SNOUT & photometry & 160-310 \AA\ & 28 & 3609 & 4.2$\times$10$^{-5}$ & 2.5 & priv. comm. \\
    & & in 2 bands  & & & & & \\
    \hline
    \end{tabular}
    \caption{Properties of each mission or mission concept assumed in S/N calculations. $^a$Integrated effective area is the total $A_{\rm eff}$ integrated over the instrument's total bandpass. $^b$Total background rate includes intrinsic detector background (applicable to MCPs), scattered light (applicable to spectrographs, we assume no scattered light for photometers), and dark current (applicable to CCDs). SNOUT (A. Youngblood, private communication) is the only mission concept using a CCD; all others use a MCP. For SNOUT, we assumed a read noise of 10 counts per pixel for a 5$\times$5 pixel PSF. We assume an encircled energy of 90\% for all missions except MANTIS, which uses 82\%. $^c$The ExtREAM mission concept \cite{Drake2025} also includes a photometric channel not listed here. $^\dagger$ denotes that the value was not available in the literature, and we assumed a value that was consistent with other similar missions or concepts. }
    \label{tab:missions_properties}
\end{table}

As evidence of the science gaps that require EUV observations to answer major open questions in astrophysics, a variety of EUV mission concepts have been developed. In this section, we demonstrate that \escape's unique combination of spectral range, spectral resolution, and sensitivity is optimal for detecting CMEs and measuring EUV spectral energy distributions. As stated elsewhere in this work, the FUV and NUV capabilities of current or upcoming missions like HST, SPARCS, and UVEX cannot directly address \escape's science questions.

Figure~\ref{fig:Aeffs} and Table~\ref{tab:missions_properties} show the landscape of EUV astrophysics mission concepts. NExtUP \cite{Drake2021} and SNOUT (A. Youngblood, private communication) are photometry mission concepts, maximizing throughput at the expense of spectral resolution. ExtREAM \cite{Drake2025} and Sirius \cite{Barstow2015} are spectroscopy mission concepts that trade higher spectral resolution (R $\approx5000$) for narrower bandpasses. MANTIS \cite{indahl2022} is a NASA-funded smallsat that is being developed by the \escape\ science and technical team in part as a scientific and technical demonstration mission for \escape, including flying a prototype of \escape's Hettrick-Bowyer telescope. MANTIS covers a similar wavelength range as \escape\ at much lower spectral resolution and sensitivity. \escape\ achieves high sensitivity over a broad bandpass at moderate spectral resolution.

\begin{figure}
    \centering
    \includegraphics[width=0.85\linewidth]{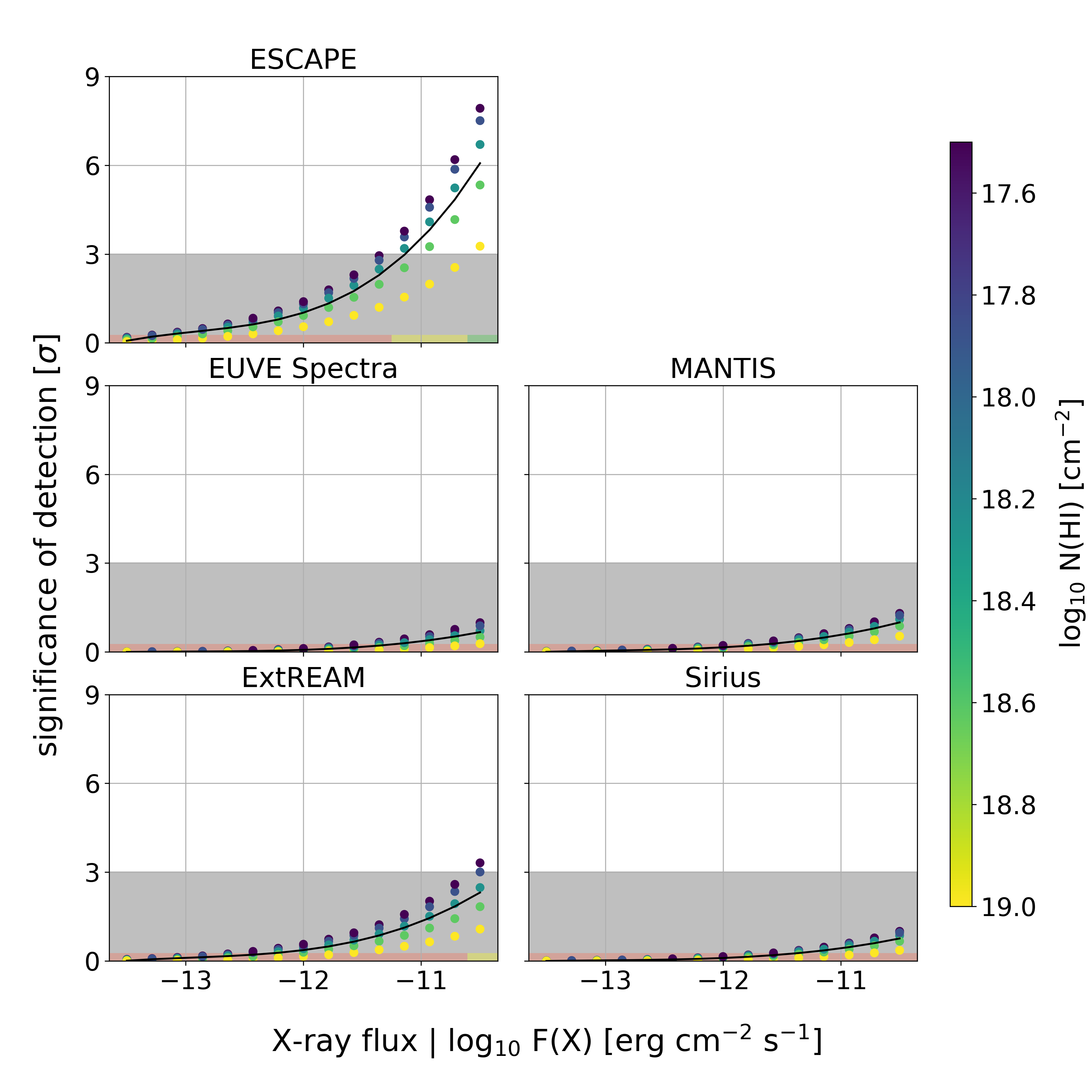}
    \caption{Significance of detecting a stellar dimming in co-added emission line light curves as a function of apparent flux and ISM attenuation for \escape, \euve's spectrometer, the upcoming MANTIS mission, and two additional mission concepts. The five selected emission lines for co-adding in every case are those that result in the most significant detection for that instrument, flux, and attenuation. The different points are colored by the amount of ISM attenuation assumed (see color bar) and the black line is a polynomial fit through the points. The gray shaded region indicates detections below 3$\sigma$. The bottom of each plot has a colored shaded bar: it is red when significance at a particular flux is below 3$\sigma$ regardless of ISM attenuation, yellow when straddling 3$\sigma$, and green when all points are above 3$\sigma$ regardless of ISM attenuation. The \euve\ and \escape\ plots were adapted from \cite{Mason2025}. \euve\ Deep Survey, NExtUP, and SNOUT are not shown because those instruments are filtered photometers that do not observe individual spectral lines.}
    \label{fig:dimming_detection_significance_lines}
\end{figure}

\begin{figure}
    \centering
    \includegraphics[width=0.85\linewidth]{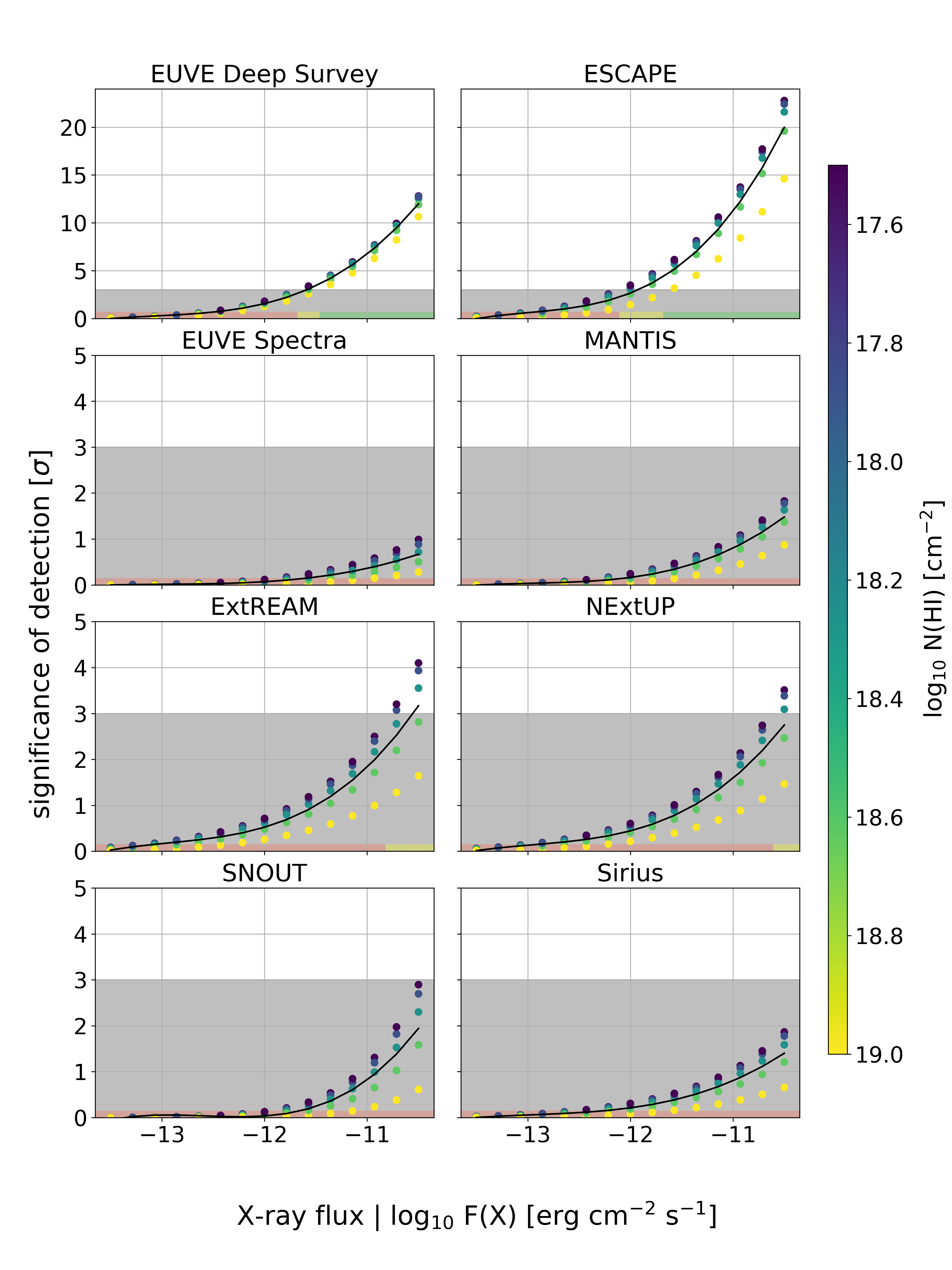}
    \caption{Same as Figure \ref{fig:dimming_detection_significance_lines} but for integrated spectral bands rather than co-added emission lines, including the instruments that are filtered photometers. Each point represents the best detection in the available bands for filtered photometers, and for spectrographs the best detection in the integrated spectral bands defined in \cite{Mason2025}: 80-180 \AA, 150-250 \AA, and 100-300 \AA.}
    \label{fig:dimming_detection_significance}
\end{figure}

We simulated observations of CME-induced coronal dimming signatures for each mission following the framework of Ref\cite{Mason2025}. We scale a solar dimming event observed by SDO/EVE to match a range of fluxes and ISM column densities corresponding to nearby stars. We factor in instrument noise and effective area as a function of wavelength to simulate data (Table~\ref{tab:missions_properties}), and we characterize the resultant dimming light curves. We used publicly available instrument parameters wherever possible for each mission (Table~\ref{tab:missions_properties}). Effective area as a function of wavelength was the only parameter uniformly available for all missions (Figure \ref{fig:Aeffs}). When parameters such as PSF size and detector noise were unavailable in the literature, we assumed typical values based on mission architecture or values similar to \escape, whichever was most favorable to the comparison mission. In order to enable direct comparisons between concepts, we adopt a fixed integration time of 600 sec.

The spectrographs (\escape, Sirius, ExtREAM, and MANTIS) can perform spectral integrations post-facto, so we analyzed both individual emission lines (Figure \ref{fig:dimming_detection_significance_lines}) and used the same spectrally-integrated bands for all of them (80-180 \AA, 150-250 \AA, and 100-300 \AA; Figure \ref{fig:dimming_detection_significance}). The imagers (\euve\ deep survey, NExtUP, and SNOUT), however, have fixed bandpasses so dimming detection significance was assessed within their corresponding bandpasses. Figures \ref{fig:dimming_detection_significance_lines} and \ref{fig:dimming_detection_significance} show the best performing dimming detections (measured depth of dimming over the measurement error) for each instrument, assessed over every combination of co-added emission lines or every bandpass, as applicable. 

Our results show that solar-like dimming in individual emission lines is only detectable with \escape, with the marginal exception of ExtREAM for the brightest stars (log$_{10}$F(X) $>$10.5) with the minimum amount of ISM attenuation (log$_{10}$N(HI) $<$17.87); there are only $\sim$15 stars that bright \cite{Freund2024}, only 2 of which are within 10 pc where the median column density is $<$17.87 \cite{Youngblood2025}. The number of possible targets for \escape, on the other hand, is 126. 

Solar-like dimming in broad bands would have been possible with \euve\ (Figure \ref{fig:dimming_detection_significance} top-left), though because it was primarily targeting other science objectives and the typical stellar exposure times were $<$10\% of the \escape\ DEEP monitoring duration, it appears never to have detected a solar-like CME dimming signal (although see Ref. \cite{Veronig2021} for an extreme, 50\% dimming detection). Solar dimming would also be detectable for the brightest possible targets with the lowest ISM columns with ExtREAM and NExtUP, though again there are only a small number of targets that match these conditions. 

Because it is designed and optimized specifically to make stellar CME-induced dimming measurements, \escape\ far outperforms the comparison instruments. \escape\ can detect the relatively shallow, few-percent dimmings in individual coronal iron lines typical of the Sun for stars as faint as $F$(X) = 10$^{-12}$ erg cm$^{-2}$ s$^{-1}$, corresponding to 920 viable stellar targets.  Larger dimming events, of course, would be detected with correspondingly higher S/N, though these events are expected to be rarer, akin to the flare frequency distribution's relationship between event size and rate of occurrence.

\subsection{Interstellar Attenuation and Flux Reconstruction and Calibration} \label{sec:ISM}

Dust and molecular hydrogen extinction is negligible within the local bubble \cite{Lehner2003}, meaning that the only significant sources of line-of-sight ``reddening" to stars $\leq$100pc from the Sun are neutral and low-ionization atomic gases \cite{Redfield2008}. Neutral hydrogen (H I), neutral helium (He I) and ionized helium (He II) contribute, with the H I column density being the dominant parameter for the total line-of-sight EUV/FUV extinction.

\begin{figure}
    \centering
    \includegraphics[width=\linewidth]{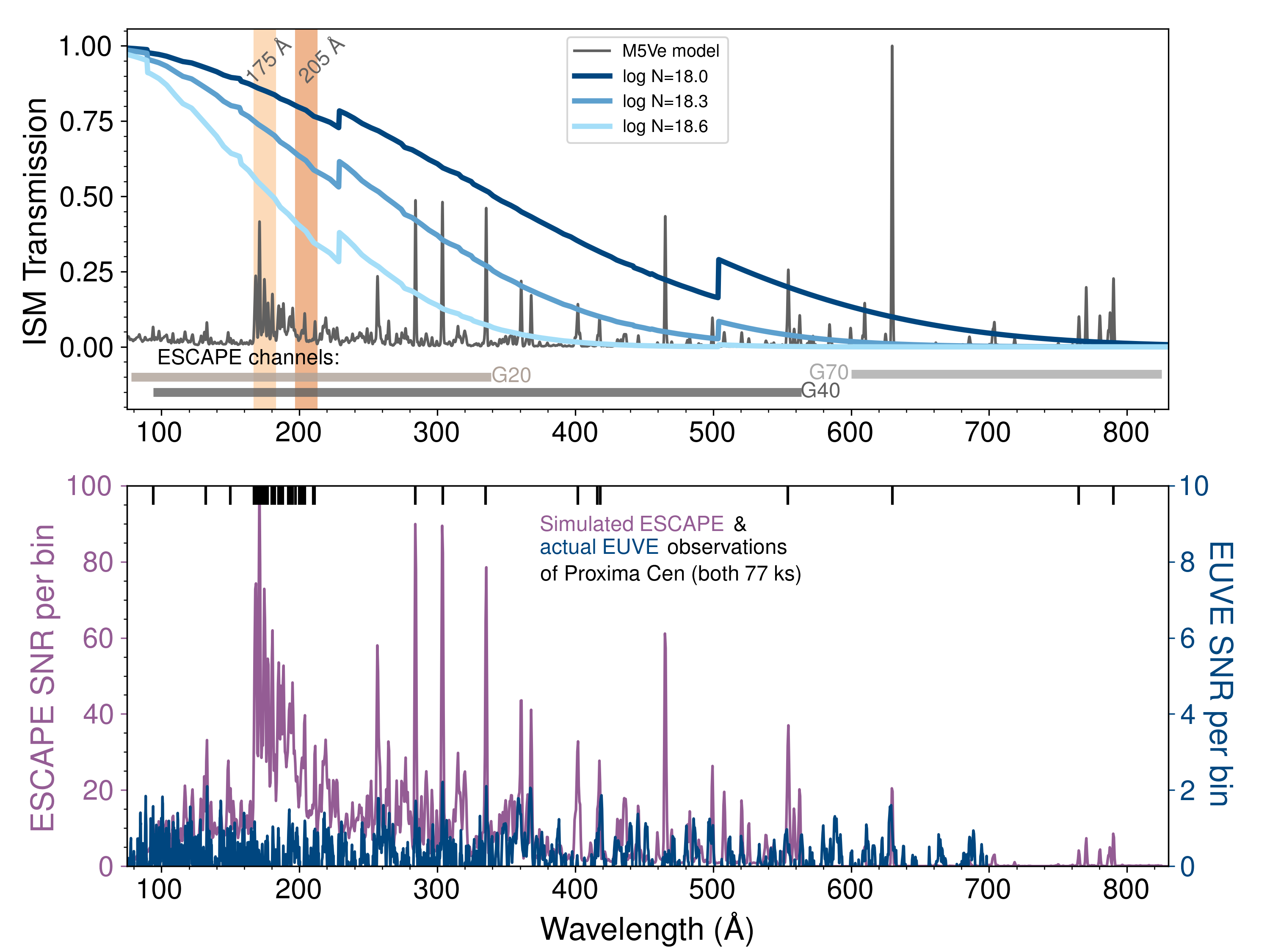}
    \caption{Top: Transmission of the local interstellar medium for H I column densities typical of stars inside the Local Bubble \cite{Wood2005} is shown by the colored curves (10$^{18.0-18.6}$ cm$^{-2}$). The ISM is more than 20\% transparent to EUV photons $\lesssim$300 \AA\ for N(HI)$\leq$10$^{18.4}$, which applies to most stars inside 20 pc \cite{Youngblood2025}. Bottom: The signal-to-noise ratio (SNR) predicted for a 77 ks \escape\ observation of Proxima Cen, including the effects of ISM attenuation (N(HI)=10$^{17.6}$ cm$^{-2}$ \cite{Wood2005}), shows that \escape\ can directly measure high-SNR spectra of nearby stars across the full 80-825 \AA\ spectral region (purple). By comparison, the SNR of the 77 ks \euve\ spectrum of Proxima Cen (blue) was sufficient to detect only a handful of emission lines at signal-to-noise $<$ 3 (corresponding to the right axis in blue).}
    \label{fig:ISM}
\end{figure}

Fig. \ref{fig:ISM} illustrates the ISM transmission for various H I column densities. The nearest stars--the $\alpha$ Centauri system--have H I absorbing columns of 10$^{17.6}$ cm$^{-2}$ and integrated H I column densities rarely exceed 10$^{18.4}$ cm$^{-2}$ within 20 pc of the Sun \cite{Peacock2025,Wood2005}. Stars within 20 pc encompass almost all of the Tier A and B HWO targets for biosignature detection, as well as a majority of the most favorable M dwarf planet hosts studied by JWST \cite{Hord2024}. The density structure of the ISM reveals that column densities rise slowly beyond $\sim$20 pc \cite{Youngblood2025}, and interstellar attenuation does not prevent transmission of EUV photons from cool stars until H I columns exceed 10$^{19}$ cm$^{-2}$ (occurring at distances $\geq$100 pc). Thus, the interstellar medium is not the limiting factor in our ability to study EUV emission from the most important exoplanet hosts (Figure~\ref{fig:ISM}).

Nevertheless, corrections for interstellar H and He attenuation are critical to computing the intrinsic flux of the emergent stellar spectrum. 35\% of the nominal \escape\ target list have direct N(H I) observations from HST-STIS (Ly$\alpha$) or indirect constraints on H I via Mg II or Fe II observations from HST-STIS \cite{Linsky2019}. \escape\ also employs all-sky maps of the local cloud structure and gas density developed by Redfield and collaborators for estimation of H I column densities \cite{Redfield2008,Youngblood2025} with an uncertainty of $<$0.4 dex, translating to $\leq$60\% a priori uncertainty on the ISM attenuation at 200 \AA\ \cite{Youngblood2025}. Combined with the $\leq$10\% photometric uncertainty goal, \escape\ characterizes the absolute EUV luminosity in the key 175~\AA\ and 205~\AA\ line complexes to an accuracy better than a factor of two in all cases (and to better than 10\% towards stars with well-characterized interstellar sightlines).

Additionally, \escape\ spectra constrain the interstellar H I column density directly. Resonance line doublets such as Fe XVI 335, 361 \AA\ have intrinsic flux ratios set by fundamental atomic physics and have been used to derive independent ISM constraints \cite{SanzForcada2003}. H I column density can also be used as a free parameter in the DEM models, improving the corrected flux uncertainty to $<$50\% at 200 \AA\ for stars without any {\it a priori} ISM constraints. The underlying principle is that the observed EUV spectrum encodes information about N(H I) that can be recovered during a DEM fit that assumes a typical DEM morphology dictated by atomic physics and collisional equilibrium. We perform two tests to demonstrate this. 

\begin{figure}
    \centering
    \includegraphics[width=0.95\linewidth]{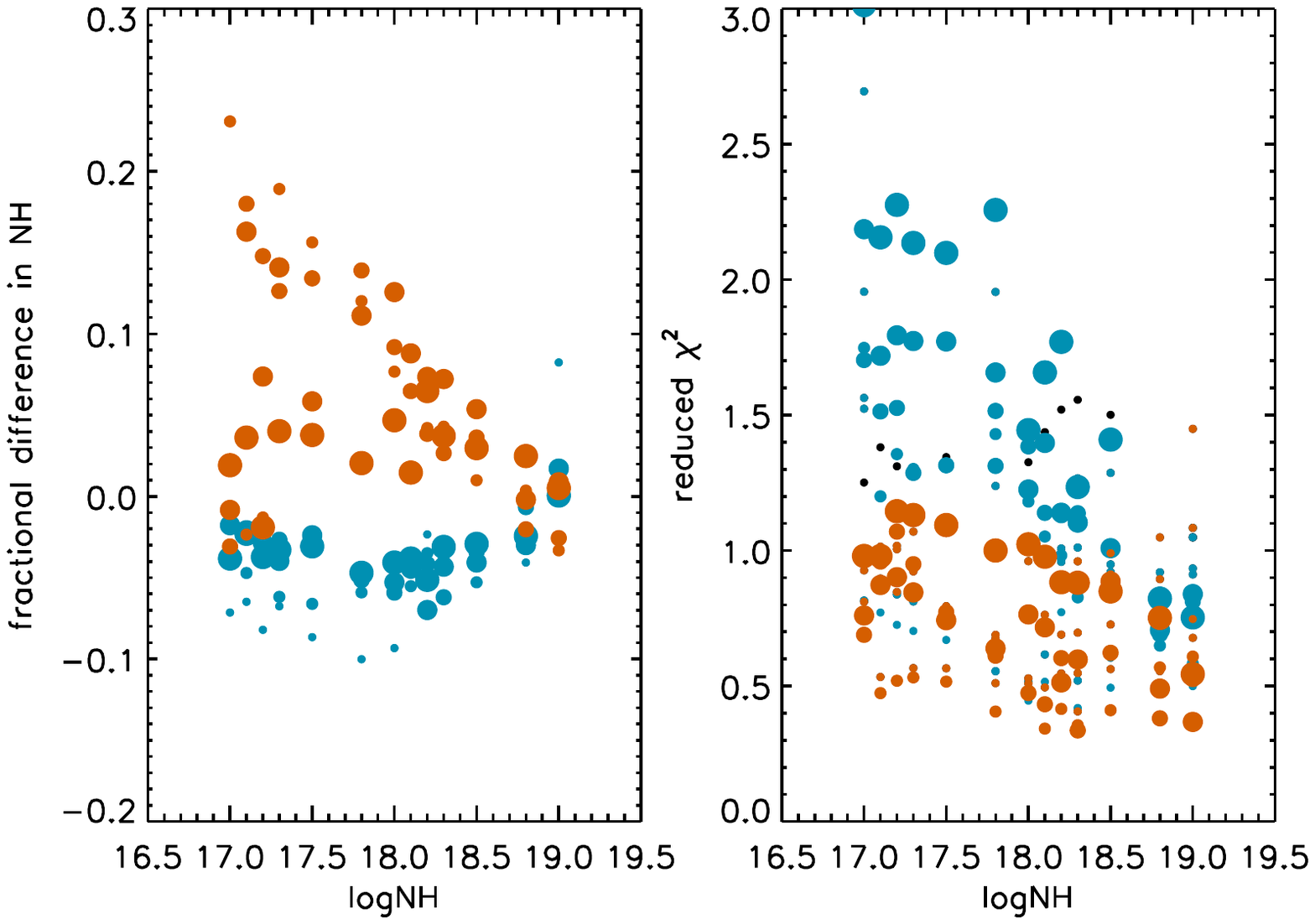}
    \caption{Post hoc assessment of N(HI) recovery simulations, for the fractional difference in estimated N(HI) relative to input ({\sl left}), and the reduced $\chi^2$ of the detectable lines ({\sl right}) as a function of the input N(HI).  Each point represents a separate simulation setup, with larger point sizes denoting inputs with more counts.  The cases with solar active region style DEMs are shown as red points and the Proxima~Cen style DEMs as blue. The systematic offsets seen for the different DEMs is a consequence of fitting a generic piecewise DEM\cite{Drake2020}.}
    \label{fig:assess_mcmc}
\end{figure}

In the first test, we generate simulated counts for a grid of ISM columns ($\log$N(HI)$=17.0:19.0$) using realistic shapes for the DEM (taken from solar active regions\cite{Brosius1996} and Proxima~Cen\cite{Drake2020}).  We simulate \escape\ spectra, extract 20 strong emission lines in the wavelength range 93-500~\AA, and use them to reconstruct parametric DEMs and obtain best fit estimates of N(HI).  We consider a variety of brightness values, ranging from $\approx$50 to $\approx$10\,000 counts in these spectral lines, covering the gamut of observational scenarios.
We adopt a flexible DEM model\cite{Drake2020} which usefully captures the gross shape of DEMs in a parametric form.  The rationale for using a different DEM shape from the input (true) DEM shape \cite{Brosius1996} is that the true DEM shape will not be known for real \escape\ targets, and the parametric DEM will allow us to characterize and compare large sets of stars.  Despite the mis-specification of the fitted DEM model, we find that both the absorption column and the intrinsic fluxes are recovered with high accuracy.  The absorption columns N(HI) are typically estimated to be within 10\% of the input values (see Figure~\ref{fig:assess_mcmc}), and the fluxes of detectable lines (those with $S/N>2$) are recovered with an average reduced $\chi^2$ of $\lesssim$1.5 for a given N(HI).

In the second test, we repeat the polynomial-based DEM calculations of Ref\cite{Duvvuri2025} for $\epsilon$ Eri, a young K2 dwarf at 3.22 pc that has been observed by \xmm, \euve, and \hst. When allowing N(HI) to float as a free parameter during the calculation, we find that the recovered N(HI) value (log N(HI) = 18.2) is 0.3 dex higher than the value assumed from \hst\ Ly$\alpha$ data (log N(HI) = 17.9), but enables absolute flux reconstructions to an accuracy considerably better than a factor of two.

Both tests demonstrate the ability of DEM calculations to provide strong constraints on the ISM corrections for \escape\ targets. Additionally, the \escape\ flux calibration program monitors bright white dwarfs of different temperatures, providing additional empirical constraints on the hydrogen columns, H/He ratios, and helium ionization fractions of the local ISM.

\section{Implementation: \escape's Science Instrument} \label{sec:implementation}

\begin{figure}[!t]
    \centering
    \includegraphics[width=0.99\linewidth]{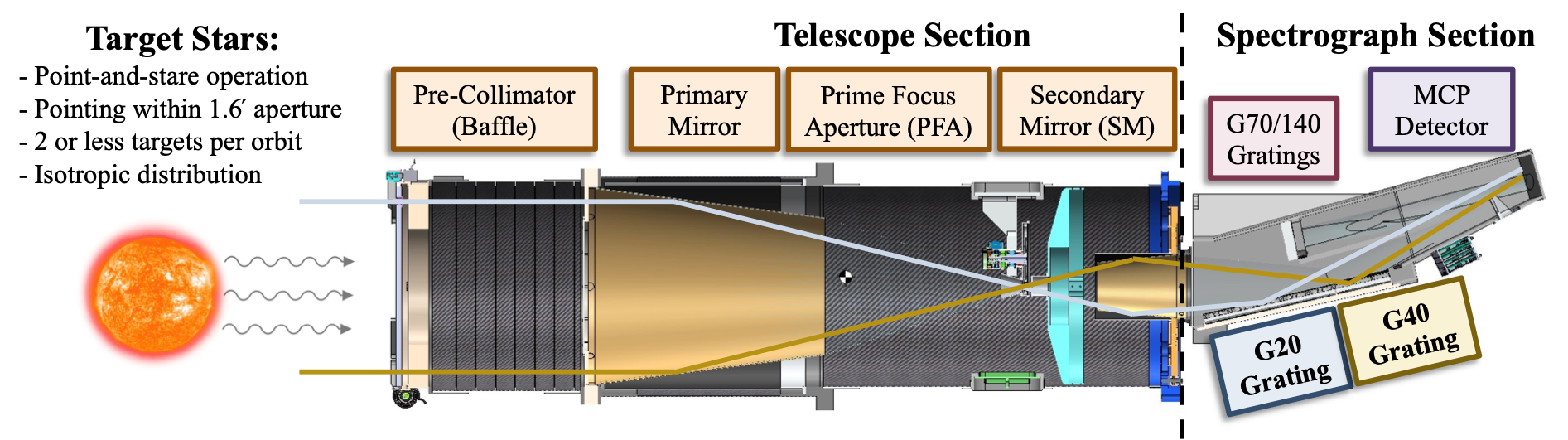}
    \caption{Side view of the \escape\ science instrument depicting the major components. The ray-tracing lines show how light from approximately one half of the telescope collecting area is dispersed by the G20 grating (shown in light blue, and then G70) while the other half is by the G40 (shown in gold, and then G140) \cite{Fleming2021,France2025_spie}. Section~\ref{sec:implementation} largely describes the technology development and flight qualification of this design and several of the optical elements that enable the \escape\ performance.}
    \label{fig-ESCAPE}
\end{figure}

\begin{figure}
    \centering
    \includegraphics[width=0.69\linewidth]{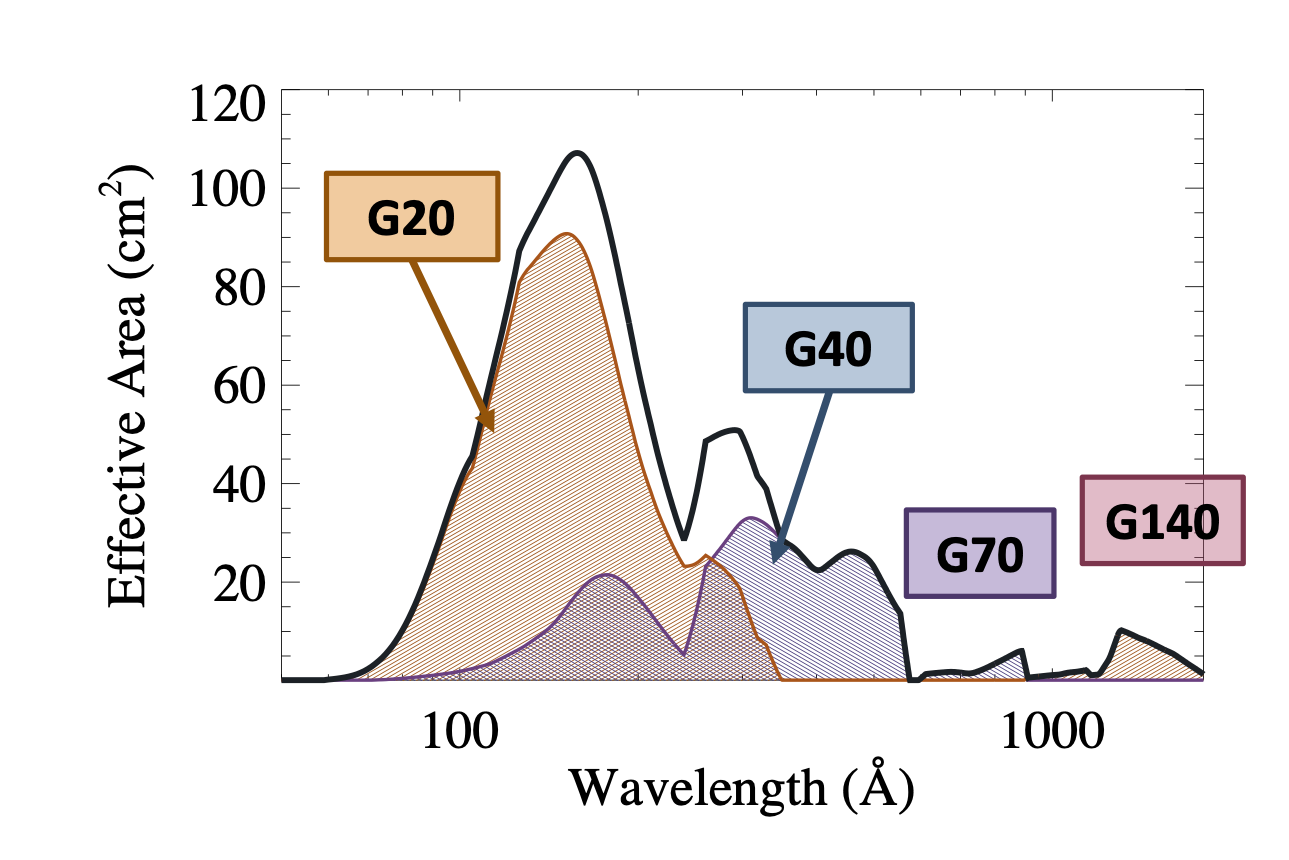}
    \caption{The effective area of the \escape\ instrument for each of the four channels. This curve is calculated using a model of the grazing incidence reflection efficiency across the shaped optics (as the incidence angle changes), and measured and modeled reflectances of the mirror coatings, gratings, and detector efficiencies. The black line represents the sum of the four channels. The last major EUV instrument, \euve, had an effective area close to 1 cm$^{2}$ over the 100 - 600 \AA\ bandpass, with no sensitivity coincident with the \escape\ G70/G140 channels.}
    \label{fig-aeff}
\end{figure}

The \escape\ science payload is optimized for the observations described in the preceding sections, with the design balanced for sensitivity and spectrally-resolved broad bandpass coverage. Light is collected by a large 0.5 meter clear aperture outer diameter Hettrick-Bowyer (HB) telescope, the grazing incidence equivalent of a Gregorian. At the focus of the primary mirror is a field stop that restricts the \escape\ field-of-view to a diameter of 1.6\amin. A grating array fabricated from lithographically etched grating segments (the G20 and G40 gratings) with near-zero groove roughness disperses the light \cite{Grise2021, Kruczek2022}, which is imaged onto a 95 $\times$ 40 mm microchannel plate (MCP) detector with a potassium iodide (KI) photocathode \cite{Ertley2025}. The zero-order reflections off the grating array are directed by a flat fold mirror to normal incidence gratings that provide the longer wavelength FUV and EUV (G70 and G140) coverage (Figure~\ref{fig-ESCAPE}). Together, the \escape\ instrument provides a spectral range simultaneously spanning 78 $\lesssim$ $\lambda$ $\lesssim$ 1650 \AA\ at a projected sensitivity $>$ 50$\times$ that of the last EUV-sensitive astrophysics mission, \euve\ \cite{Bowyer1991} (Figure~\ref{fig-aeff}). 

The technology and design innovations that enable \escape\ have undergone a development and testing program, both in the laboratory and on other funded flight programs. Most notably, the MANTIS SmallSat is effectively \escape\ in miniature. MANTIS covers the same EUV bandpass with a Hettrick-Bowyer telescope, etched silicon grating, and MCP detector, just as \escape, but features less sensitivity and spectral resolution (Table~\ref{tab:missions_properties} and Figure~\ref{fig:Aeffs}). In the rest of this section, we describe the main optical components of \escape, including some of the technology development activities and a preliminary STOP and stray light analysis of the opto-mechanical packaging of the instrument. 

\subsection{The Hettrick-Bowyer Telescope: High throughput for stellar spectroscopy}\label{section-telescope}

The \escape\ parabolic primary mirror (PM) is a 50 cm clear-aperture diameter, 1 mm thick $\times$ 585 mm long electro-formed nickel alloy shell coated in gold. The prime focus aperture (PFA) restricts the field-of-view of \escape\ at the PM focus to a 1.6\amin\ diameter, with the inner 50\asec\ diameter having $>$ 99\% transmission. This restricted FOV is essential for limiting geocoronal airglow contamination without requiring bandpass limiting filters in normal operations; the bandpass filters employed for calibration are described later in this section. The diverging beam is re-focused by an elliptical secondary mirror (SM) that is 1 mm thick $\times$ 188 mm long with a 14 cm outer diameter. The PM and SM are both oversized by $>$10 mm of unilluminated length at each end for bonding and to mitigate the impact of edge distortion. 

The shells are fabricated using electro-formed nickel replication processes that are standard for many grazing incidence optical systems (e.g., Swift/XRT, XMM-Newton, IXPE \cite{IXPE}). The \escape\ telescope partner, Media Lario (hereafter ML), machines and polishes an aluminum mandrel with the negative of the optical figure and $\sim$ 2 \AA\ RMS surface roughness. Many shells can be replicated off each mandrel, enabling flight spares and engineering models (EMs) for minimal cost. The telescope optics are coated in gold, which has sufficient reflectance in the \escape\ bandpass. The gold is applied to the mandrels prior to replication as a separation layer that prevents the nickel shell from sticking to the mandrel. As the shell is removed, the gold transfers to the inner surface of the shell and maintains a surface roughness only slightly degraded ($\sim$5-7 \AA\ RMS) from that of the mandrel. 

The overhanging ends of the PM and SM are inserted in wire-EDM cut slots in the spider assemblies for bonding. The bonding process at ML has been employed for many successful missions, including the nested shells of XMM-Newton \cite{XMM}, several of which were larger than the \escape\ PM shell. The PM and SM are metered by a graphite epoxy tube (Figure~\ref{fig-ESCAPE}), with fine XY thread adjustment bolts on the SM mount for aiding in alignment. The alignment tolerances of the telescope are all within the family of prior programs and have margins $>$ 30\% against the worst-case hot/cold limits of the preliminary STOP analysis (Section \ref{sec:STOP}). The aligned telescope produces a PSF of $\lesssim$ 20\asec\ half-energy width (HEW), as estimated by ML. 

\subsubsection{Geocoronal light reduction enabled by the HB telescope design}\label{section-geocoronal}

Reducing the geocoronal background into the spectrograph is of prime importance for the \escape\ science mission. The prime focus of the HB telescope design enables an aperture stop to restrict the field-of-view (FOV) without requiring a collimator (as on \euve) or re-focusing optics after the telescope, both of which would reduce throughput. Geocoronal He I and He II (584 \AA\ and 304 \AA , respectively) emissions are each comparable to the expected stellar signal from our targets, while H I Lyman alpha 1216 \AA\ will account for the majority of detectable stray-light photons scattered in the \escape\ optical system (Section \ref{section-straylight}). While neither He I or H I Lyman alpha directly land on the detector, controlling the field-of-view of \escape\ with an aperture minimizes the total number of potential scatter photons that could contribute to the background, while also limiting the footprint of He II 304 \AA\ feature on the detector. 

A wire-grid collimator was used on \euve\ to control the FOV to roughly 2\amin , which is $\gtrsim$2$\times$ larger than the effective \escape\ FOV. The majority of the geocoronal emission in \euve's DS/S mode was removed by bandpass limiting thin film foil filters that were used to sort spectral orders into discrete channels. These foil filters are opaque to H I Lyman alpha and nearly the entirety of the G70 and G140 \escape\ bandpasses \cite{Bowyer1991}. Although these were effective in reducing the out-of-band signal for \euve, they severely limited the effective bandpass that can be observed at any time. To overcome this, \euve\ effectively split the telescope beam into six channels, resulting in a major loss of throughput. This also served as a means of spectral order separation, which \escape\ does not require in the G20/G40 channels; individual orders are cleanly separated in two-dimensional retrievals because of the known wavelengths of the coronal emission lines that make up the vast majority of the signal from magnetically active, cool stars\cite{France2022}. 

A preliminary stray light analysis has been carried out on the \escape\ model that shows that the scattered light levels, and all other detector backgrounds, are well within the requirement for science observations (see Section~\ref{section-straylight}). The HB telescope design effectively reduces the FOV, and therefore the geocoronal light into the spectrograph, eliminating the need for the order sorting and geocoronal mitigation measures employed by \euve.

\subsubsection{Hettrick-Bowyer Telescope - Early Demonstrations and the MANTIS SmallSat}\label{section-telescopeMANTIS}

The first Hettrick-Bowyer telescope ever demonstrated was a laboratory model aligned at the Smithsonian Astrophysical Observatory (SAO) in 2021 as part of the prior \escape\ Phase A study \cite{Fleming2021}. This test telescope consisted of two electro-formed nickel optics separated by roughly 1.3 meters with a primary mirror outer diameter of $\sim$ 15 cm \cite{Champey22}. This first demonstration of the HB architecture met expectations, but the two optics were metered only by ground test equipment, which motivated the ESCAPE team to develop an integrated and flight-compatible prototype telescope. 

The HB design and other critical \escape\ technologies will be demonstrated in flight by the Monitoring Activity of Nearby sTars with uv Imaging and Spectroscopy (MANTIS) SmallSat, a NASA-funded ESPA-class satellite project currently in development (see Section~\ref{sec:comparison}). MANTIS carries two telescopes and three spectral channels, covering the majority of the 100-6400 \AA\ bandpass simultaneously, albeit at far lower spectral resolution ($\sim$22 \AA\ with a requirement to be $\leq$70 \AA ) and EUV sensitivity than \escape. MANTIS is led by members of the \escape\ team at CU-LASP and is providing essential experience in the alignment, calibration and use of the EUV telescope and gratings, both at LASP and the partner institutions. 

\begin{figure}
    \centering
    \includegraphics[width=0.99\linewidth]{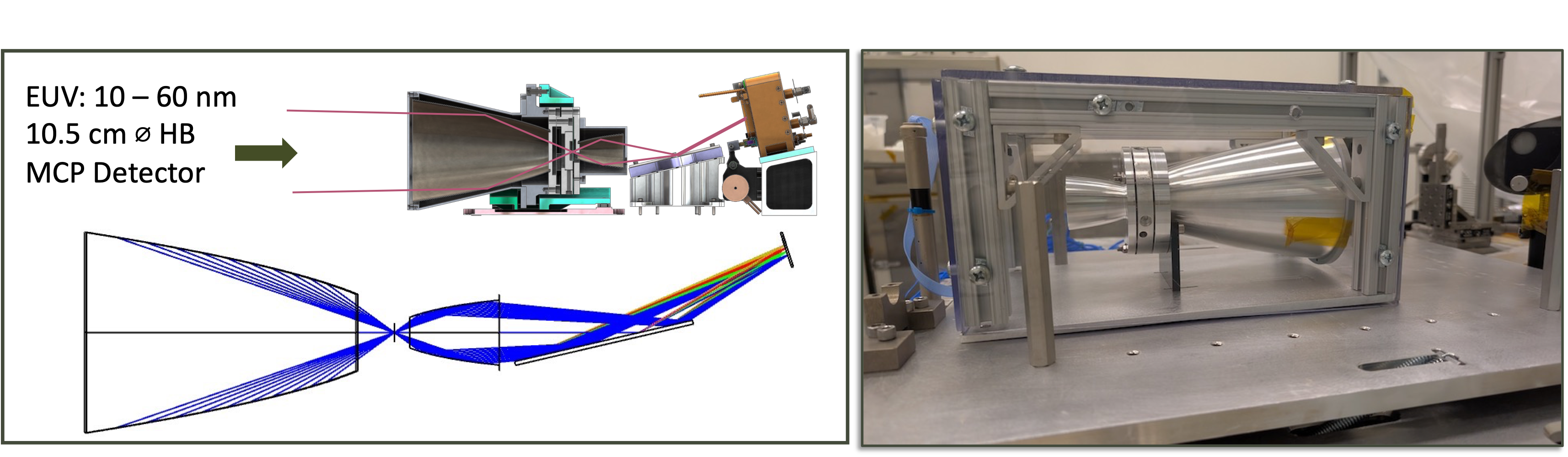}
    \caption{(Left) CAD drawing (top) and raytrace (bottom) of the MANTIS EUV channel - which is a scaled-down prototype of \escape. Light is collected and focused by the 10.5 cm diameter HB telescope, diffracted off of a variable line-spaced grating, and the spectrum is imaged onto an MCP detector. (Right) An all-aluminum test model of the MANTIS telescope after delivery to CU-LASP. This telescope was fabricated primarily to test and refine the optical alignment plan for the replicated nickel flight optics.}
    \label{fig-mantisscope}
\end{figure}

The MANTIS HB telescope is 10.5 cm in diameter with a 380 mm focal length. The optics are provided by a team affiliated with the Italian National Institute for Astrophysics (INAF) at the Osservatorio Astronomico di Brera in Merate, Italy. Several of the fabrication steps occur at ML and follow procedures similar to those used on \escape. The mandrels for the MANTIS HB are in their final polishing stage at the time of this publication, with the delivery of the engineering model (EM) and flight units by early 2027. A prototype EM produced with diamond-turned and polished aluminum optics has already been fabricated, providing a demonstration system to test the alignment procedure for the flight telescope. This EM has been measured to have a half-energy width of $<$ 60\asec\ already, far exceeding expectations given the manufacturing methods employed for this test telescope (Figure~\ref{fig-mantisscope}). 

\subsection{Advanced Etched Silicon Diffraction Gratings}
\begin{figure}
    \centering
    \includegraphics[width=0.99\linewidth]{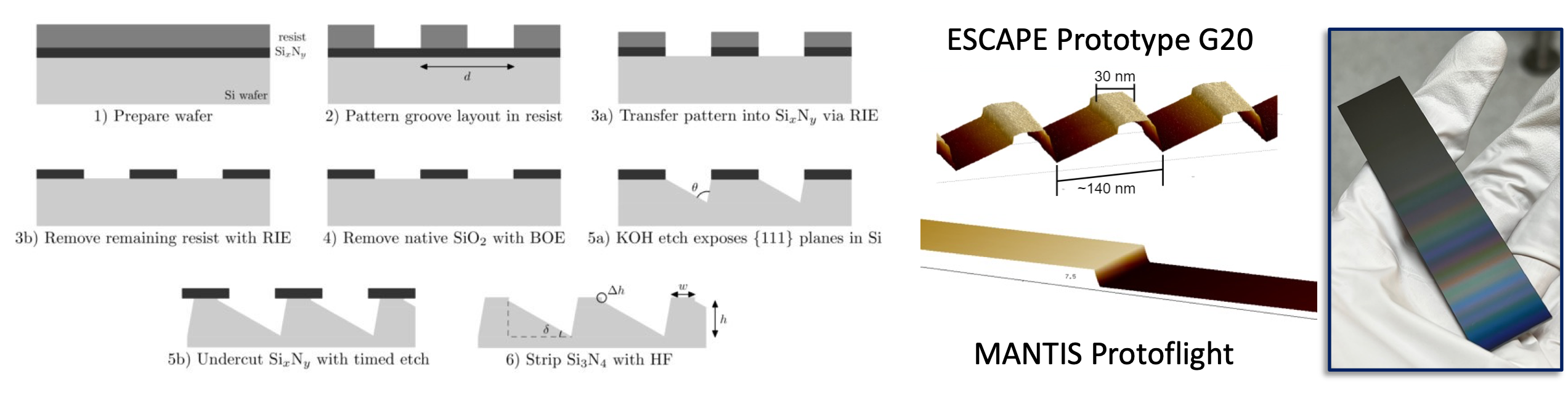}
    \caption{Left: The flow of the grating fabrication process for an etched silicon grating (courtesy PSU, see also \cite{Miles2025}). (Right) AFM measurements of the G20 grating prototype segment ruled at PSU, above the same measurements for the MANTIS protoflight grating (far right photograph), which has a far smaller blaze angle of only $\sim$ 1\adeg. The residual nitride tabs create a small flat at the top of the blaze which is reduced through process optimization steps for each grating. The tabs are nearly undetectable on MANTIS, while on the \escape\ test grating they are $\sim$ 20\%\ of the prototype groove width. Test gratings are planned for Phase A that will provide the process development optimization to reduce the tabs significantly.}
    \label{fig-gratingprocess}
\end{figure}

The \escape\ gratings are fabricated from single crystal silicon wafers etched to include the ruling pattern \cite{Kruczek2022, Grise21}. This technology produces extremely smooth grating facets and enables shaped grooves for a radial pattern to match the reciprocal linear dispersion (d$\lambda$/dx) along the converging beam of the telescope. The basic processing steps for the grating fabrication are shown in Figure~\ref{fig-gratingprocess}, alongside atomic force microscopy (AFM) images of an \escape\ prototype and one of the early MANTIS protoflight gratings fabricated at Penn State University (PSU) using this process. 

\begin{figure}
    \centering
    \includegraphics[width=0.99\linewidth]{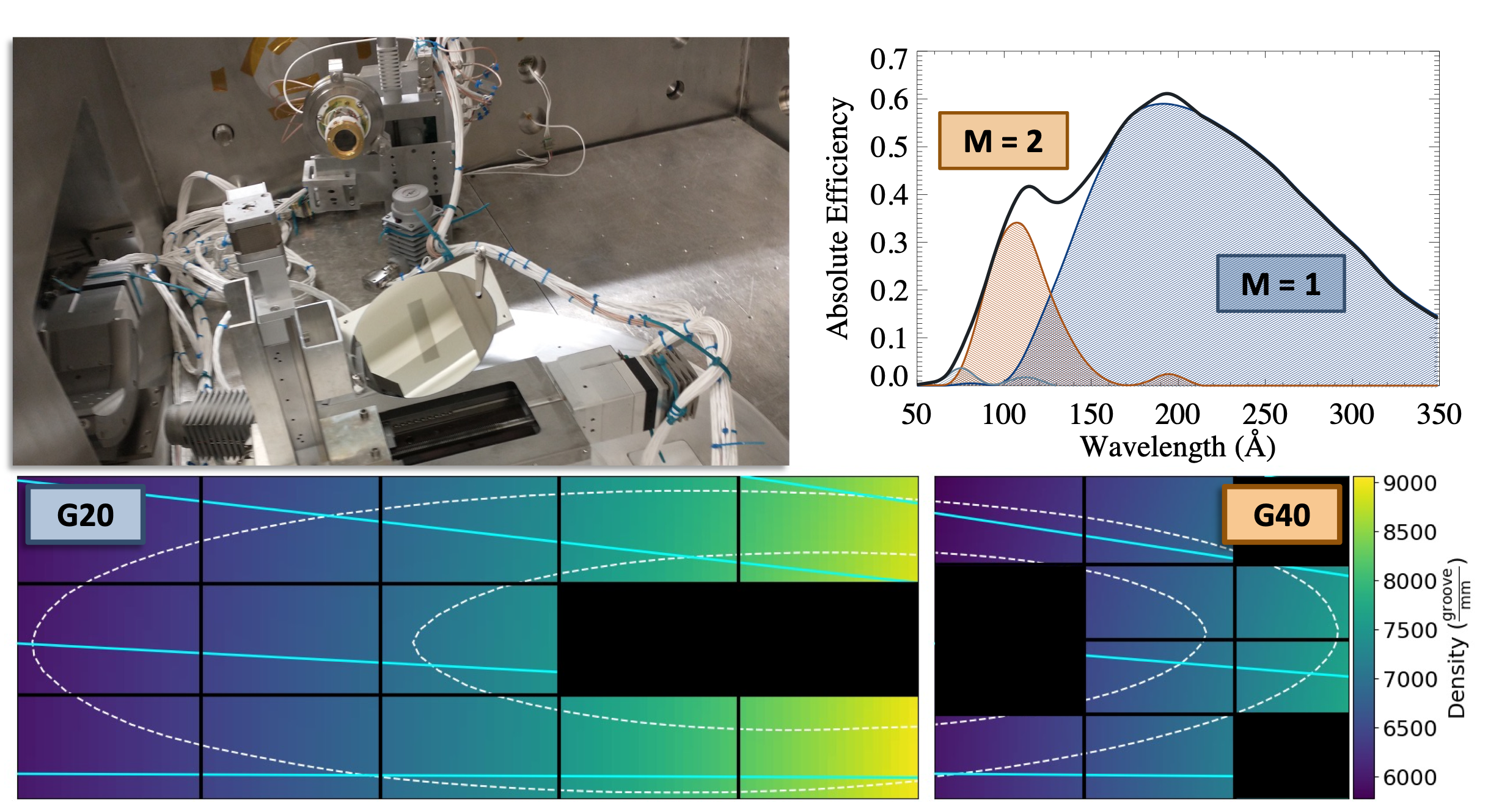}
    \caption{Top left: An \escape\ G20 grating segment in the CU-LASP Square Tank calibration facility \cite{Kruczek2022}. This grating was measured at several discrete wavelengths in-band and found to be consistent with the modeled grating efficiency predictions using grating analysis software. The grating was also measured at the Advanced Beamline Source (ALS) at the short wavelength end of \escape's spectral range. Top Right: First and second order (M = 1 and 2, respectively) efficiency of the prototype G20 grating segment. Bottom: The \escape\ grating has a radial-like groove pattern that increases in line density as the focusing telescope beam approaches the detector in order to maintain a constant dispersion. The grating arrays have a preliminary segmentation as-shown to meet the size requirements for \escape. This scheme will be optimized as the \escape\ design matures. }
    \label{fig-gratings}
\end{figure}

The G20 and G40 channels of \escape\ each have a grating array made up of segments co-aligned to direct the spectra onto the science detector (Figure~\ref{fig-gratings}). The G20 is made up of 13 segments with a ruling density increasing from approximately 6000 to 9000 grooves mm$^{-1}$ and subtends $\approx$ 50\%\ of the telescope beam. The G40 is made up of 8 smaller segments and subtends the other half of the telescope beam. The G40 segments are smaller because the groove density changes more rapidly close to the detector, and the smaller segments better optimize the writing quality. The G40 is optimized for longer wavelengths than the G20, with the groove density spanning only $\sim$5800-8300 grooves mm$^{-1}$ despite being at the extreme end of the grating. 

A G20 grating segment spanning the $\sim$ 7500 grooves mm$^{-1}$ portion of the array was fabricated and coated in zirconium in 2021. The grating was illuminated at CU-LASP in several discrete wavelengths of interest in-band using a Manson soft X-ray lamp coupled to a grazing-incidence EUV monochromator. Light was recorded in both incident mode (no grating) as well as after diffracting off the grating segment. The ratio of these two measurements minus any dark current is a measure of the product of the absolution diffraction efficiency and the reflectance of the zirconium coating.
The measurements are taken at a high angle of incidence to simulate the incident angle in \escape\ for that segment. The efficiency curve in Figure~\ref{fig-gratings} meets \escape's requirements.

The MANTIS spectrograph design mirrors that of ESCAPE, with a flat grating intercepting the converging beam from the HB telescope to diffract the light before reaching focus at the MCP detector (Figure~\ref{fig-mantisscope}). The MANTIS grating is also fabricated at PSU following the same procedure, however given the compact nature of MANTIS and a far lower spectral resolution requirement ($\lesssim$70 \AA\ instead of $\leq$1.5 \AA\ for ESCAPE), the MANTIS grating is a variable-line-spaced (VLS) design rather than a radial profile. The groove density increases from 136-831 grooves mm$^{-1}$ along the 94 mm length of the grating. While \escape\ has a simpler 20\adeg\ blaze angle to optimize for the centers of the G20 and G40 bandpasses, the low resolution of MANTIS reduces this blaze angle to as little as 1\adeg , a shallow blaze beyond the capabilities of many other grating fabrication techniques. Nevertheless, the PSU process has produced exquisite groove profiles with high uniformity on several process development and engineering model test gratings (Figure~\ref{fig-gratingprocess}). The final flight gratings are to be delivered later in 2026. 

\subsection{Photon-counting MCP detector}
Science spectra from all \escape\ channels are recorded on a 95 x 40 mm microchannel plate detector (MCP) from Sensor Sciences LLC. This MCP is similar in size to a single 85 $\times$ 10 mm HST-COS MCP segment \cite{Green12} and smaller in size than prior CU-LASP detectors sourced from Sensor Sciences, such as the 2 $\times$ 115 $\times$ 40 mm SISTINE detector array \cite{Nell24} and the 200 $\times$ 200 mm DEUCE detector \cite{GreenDeuce}. The detector is read out by a version of the same detector electronics developed for the Juno-UVS spectrograph, and also used on Europa-UVS, JUICE-UVS, as well as CU-LASP SmallSats SPRITE and MANTIS \cite{Gladstone2018, Retherford2024, Bowen2023}. The electronics are a slightly updated version from those used on the CU-LASP Mars instrument EMUS and on New Horizons Alice \cite{EMUS, NewHorizons}. Because the \escape\ optical design yields fairly coarse PSFs by design, the \escape\ electronics readout only needs to be optimized to an average of $\leq$80 $\mu$m full width at half maximum resolution at the detector, compared to $\sim$50 $\mu$m achieved on SPRITE. The data for \escape\ is digitized to 13 bits of X-axis resolution and 12 bits of Y-axis resolution, which over the active area corresponds to a $\sim$15 $\times$ 15 $\mu$m digital pixelation (sampling resolution).

\begin{figure}
    \centering
    \includegraphics[width=0.99\linewidth]{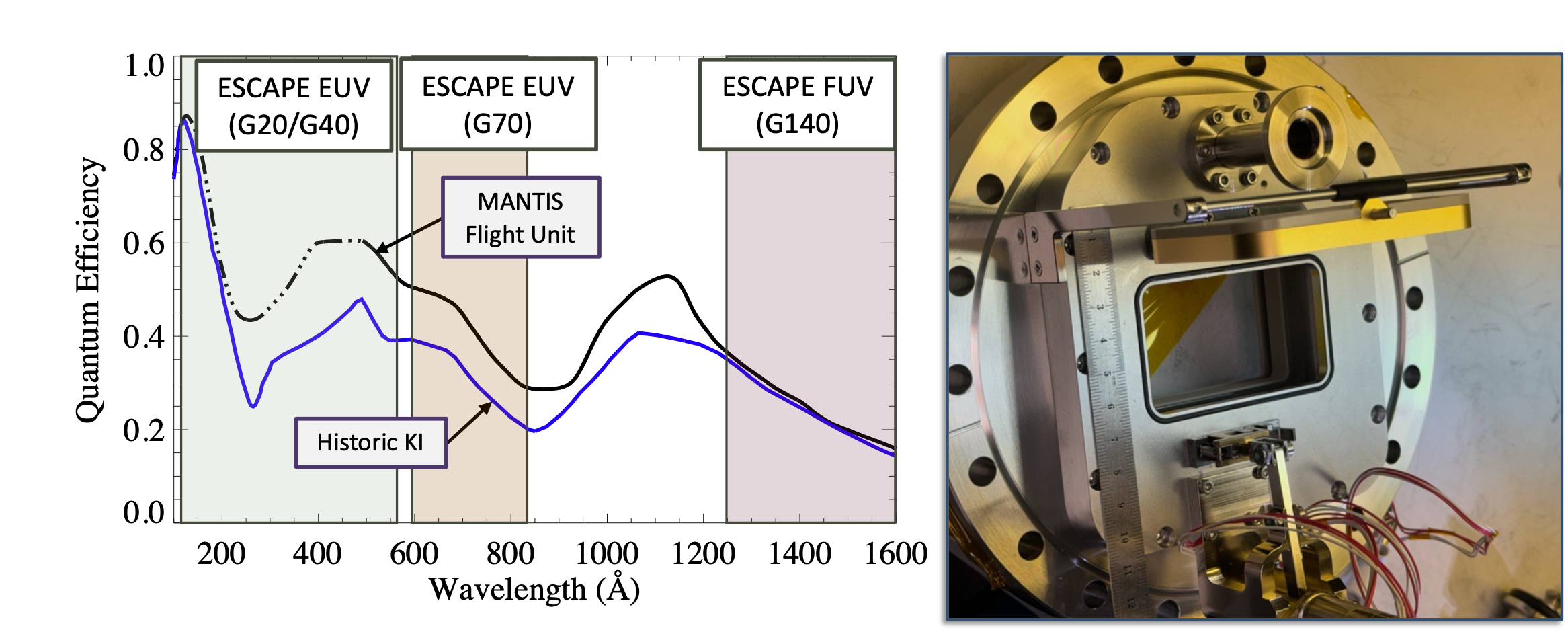}
    \caption{(Left) The \escape\ quantum efficiency estimation for the purposes of our performance calculations is taken from historical KI photocathode performance (blue curve). The MANTIS EM under-performs this estimate, however it did not undergo rigorous optimization, while the MANTIS flight unit uses advanced MCPs with 20 -- 35\% higher QE than historical data. \escape\ baselines the historical results, but will pursue this new plate technology if merited in Phase A, potentially increasing effective area by as much as 35\% . (Right) The UMIS MCP housing with a 70 x 50 mm door area, 86\% the size of the expected \escape\ door.}
    \label{fig-detectors}
\end{figure}

Photons incident on the MCP surface are converted to photoelectrons by an opaque potassium iodide (KI) photocathode on the top plate surface. KI is optimized for the \escape\ and MANTIS science objectives, with a peak quantum efficiency near 100-200 \AA . We baseline a conventional KI photocathode QE curve for effective area estimates (Figure~\ref{fig-aeff}), however recent developments in MCP fabrication have improved the quantum efficiencies by 20-35\% for both cesium iodide (CsI), as used on HST-COS, and KI photocathodes. This development is too new for us to baseline for \escape\ at this stage, however the MANTIS flight detector and the upcoming CU-LASP led FLUID sounding rocket \cite{FLUID} do carry this advancement (Figure~\ref{fig-detectors}), raising the TRL of this technology to TRL 6 in the first half of 2027. These advanced plates will be further studied during Phase A and potentially adopted if ready, increasing the predicted effective area of \escape\ by up to 35\%\ for a significant addition to the sensitivity margin at no cost. 

Like many prior MCP detectors for the ultraviolet, including HST-COS, the photocathode is protected on the ground in a hermetic housing with an aperture door that opens once in space. The \escape\ detector is protected by a scaled version of the JUNO-UVS, SPRITE, and MANTIS housing, as well as the upcoming EUVST and CU-LASP UMIS detectors (Figure~\ref{fig-detectors}). These housings are fabricated from aluminum with a one-time open (manually re-settable) spring loaded door with a MgF$_{2}$ window for transmission of \escape\ G140 bandpass on the ground during testing. The door will be deployed with a custom pin-puller mechanism in-flight and for certain on-ground vacuum testing. The \escape\ door is projected to be 14\%\ larger by area than the UMIS/EUVST doors (the largest made by Sensor Sciences to-date at 70 $\times$ 50 mm). A prototype door is planned to be made in early 2027 and tested to ensure proper vacuum sealing, deployment, and vibration stability. 

\subsection{Mirror Coatings and development for the \escape\ Optical System}\label{section-coatings}

\begin{figure}
    \centering
    \includegraphics[width=0.85\linewidth]{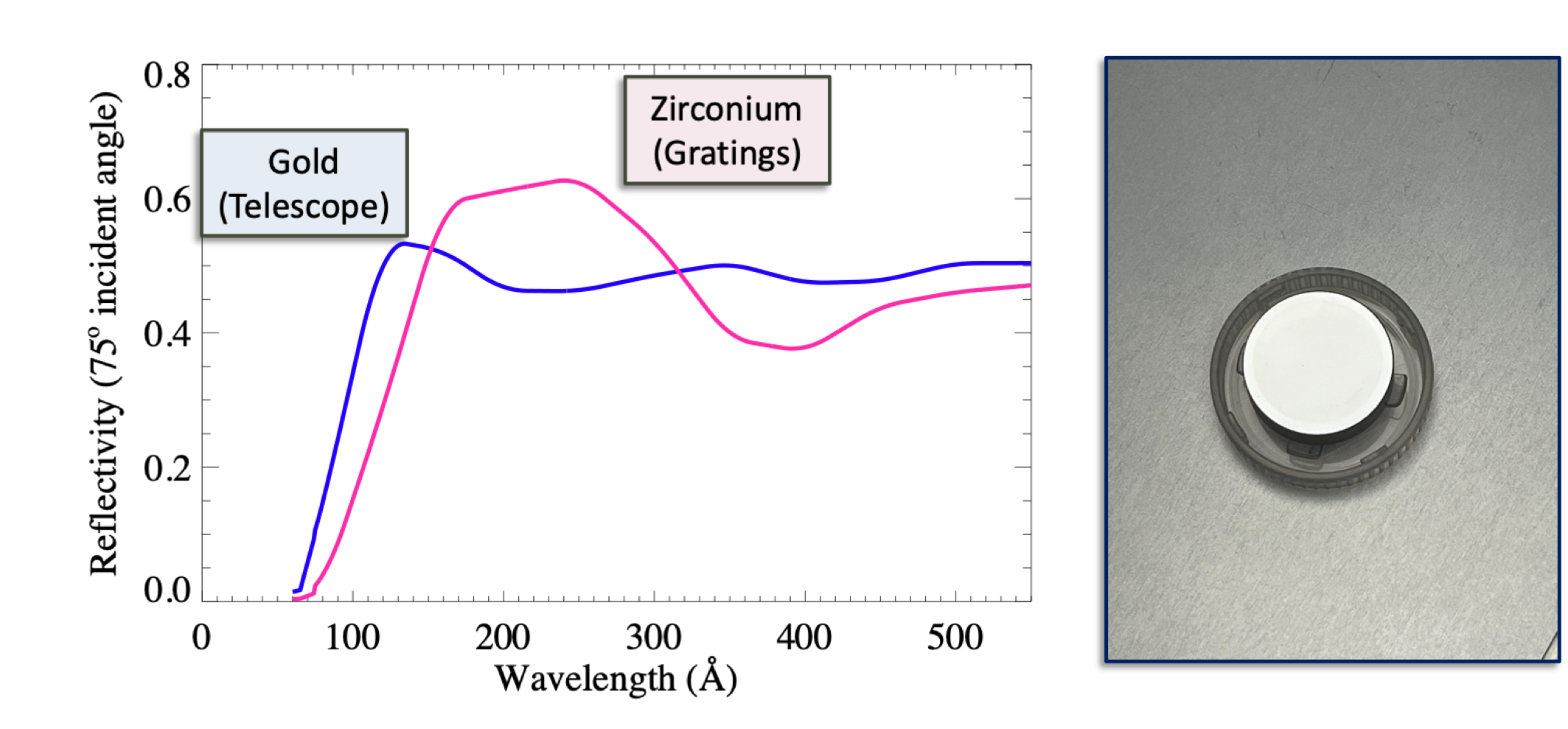}
    \caption{(Left) The \escape\ telescope is coated in gold (Au), while the gratings are baselined to be coated in zirconium (Zr). For the same grazing reflection angle, Zr exceeds the reflectance of Au by $\gtrsim$ 20\% in the essential 180 -- 260 \AA\ bandpass, while Au is superior at both shorter and longer wavelengths. (Right) A photograph of an \escape\ Zr sample after exposure to atomic oxygen on the International Space Station. A slight darkening of the center is evident where the sample was exposed. This would represent a far worse environment than any Zr coating would experience on ESCAPE, as \escape\ carries a ram direction avoidance requirement as well as having the grating encapsulated in the spectrograph.}
    \label{fig-coatcomp}
\end{figure}

As noted in Section \ref{section-telescope}, gold is an effective mirror coating for \escape\ and is baselined for the telescope mirrors due to its high heritage on grazing incidence telescopes. Zirconium (Zr) would be a superior coating relative to gold for the \escape\ bandpass, however, because it peaks in grazing incidence reflectivity between 160 -- 300 \AA\ (Figure~\ref{fig-coatcomp}). With Zr coatings, the throughput in this critical band would be $\gtrsim$ 20\% higher, adding significant margin to the science performance. Zr would be a new coating for use on grazing incidence optics and ML does not have experience depositing it on \escape - like mirrors. The costs of a test deposition and TRL demonstration has been deemed to prohibitive given that gold already meets all the science requirements for \escape , therefore the telescope mirrors are planned to be gold. 

We do baseline Zr on the G20/G40 gratings, however, as PSU has a mature Zr deposition method and has used it to coat the G20 prototype grating and the MANTIS gratings.  Zr is subject to oxidization, first rapidly to a thickness of $\sim$ 2-3 nm, and then slowly until the entire Zr thin film is converted to ZrO$_{2}$. A fully oxidized ZrO$_{2}$ coating only reduces the expected reflectance of the Zr coating in the \escape\ band by $\sim$ 9\%, however the oxide layer may change shape as a result of the oxidization process, adding roughness to the grating. We have measured this process over five years since the production of the G20 prototype. For the first six months on the ground in Colorado with no specific protection (stored in ambient laboratory conditions) we observed no detectable change in diffraction efficiency. This indicates that the initial rapid oxidization had stopped and the slow oxidization had a very long timescale. 

While storage in dry N$_{2}$ and eventual deployment into the vacuum of space in the protection of an enclosed spectrograph will likely result in no additional oxidization, we have elected to carry out additional studies to ensure this is the case prior to the end of a planned Phase A. A PSU-coated sample was sent to the International Space Station as part of the MISSE experiments from August 2022 -- March 2023, where it was exposed to the ram direction fluence of atomic oxygen for nearly 8 months. The reflectance decreased slightly over this time, but maintained over 90\%\ of the pre-flight performance. A slight darkening of the coating was observed and is apparent in Figure~\ref{fig-coatcomp}. In late 2026 we plan to expand this study further by re-coating the G20 prototype with fresh Zr and subjecting it to an accelerated aging test at 70\%\ relative humidity and 60\adeg\ C for up to six months, measuring the efficiency in-band at least once per month over that time.  

An (unlikely) descope of coating the diffraction gratings in gold would result in a $\sim$10\%\ decrease in the \escape\ effective area in the 175 - 260 \AA\ band, with an increase in effective area at $\lambda$ $<$ 160 \AA\ and only a minor change for $\lambda$ $>$ 400 \AA\ (Figure~\ref{fig-coatcomp}). These are relatively minor impacts and this descope could be executed at any point up to the final bonding of the grating segments into arrays. 

\subsection{Opto-mechanical design and preliminary STOP analysis} \label{sec:STOP}
A comprehensive Structural, Thermal, Optical, and Performance (STOP) analysis was performed to verify that \escape\ satisfies its structural integrity, thermal stability, and optical performance requirements throughout the mission environment. The analysis employed an integrated workflow in which structural and thermal predictions were coupled to an optical performance model to quantify the effects of mechanical and thermal distortions on system performance.

\begin{figure}
    \centering
    \includegraphics[width=0.99\linewidth]{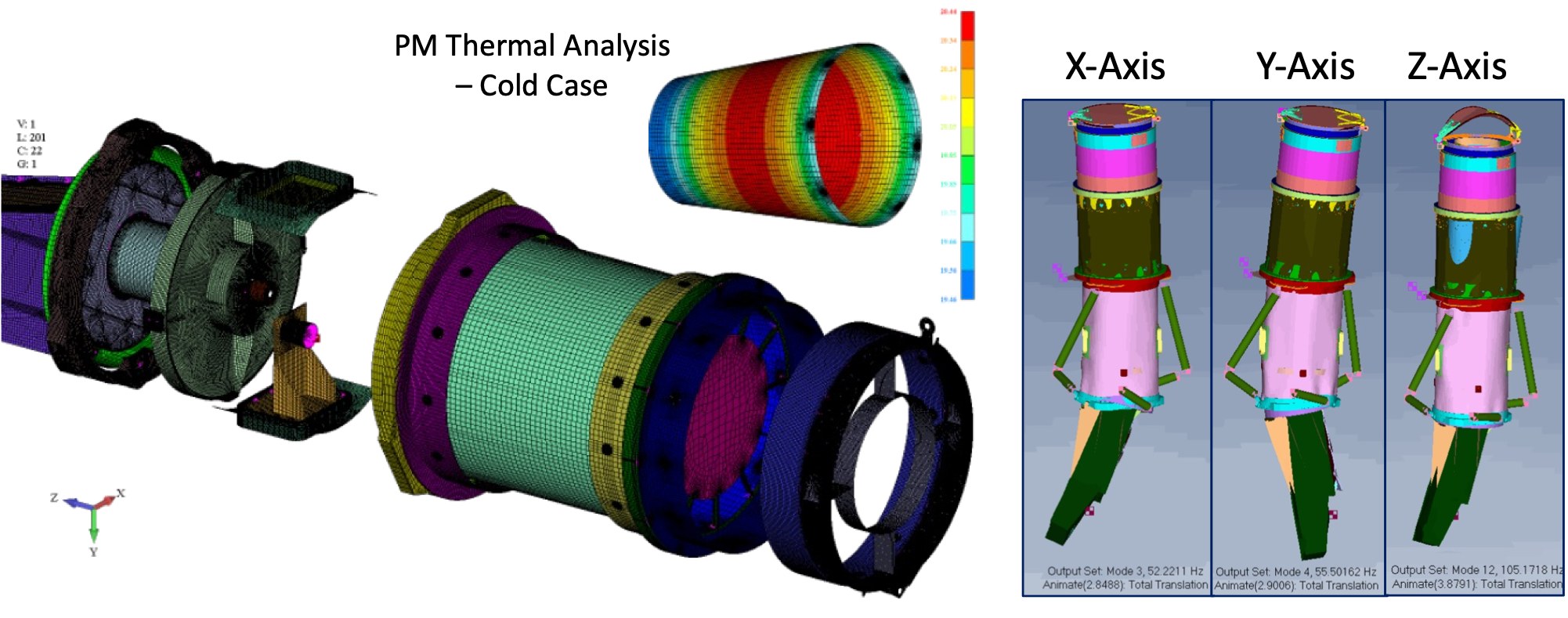}
    \caption{(Left) Structural and thermal model of the \escape\ instrument, including individual optics such as the PM (inset). In the cold case, the PM has a gradient of $<$ 2 \adeg C, which has a minimal impact on the optical performance. (Right) The structural analysis animation screenshots for the X (lateral), Y (lateral) and Z (axial) first modes.}
    \label{fig-structual}
\end{figure}

The structural finite element model (FEM) was developed using Siemens Simcenter NASTRAN with FEMAP. The high-fidelity model consisted of approximately 2.5 million nodes and 1.8 million finite elements, and an approximate mass of 186.4 kg (Figure~\ref{fig-structual}). The FEM provided detailed representation of the instrument structure, optical bench, mounting interfaces, and major assemblies. Static, modal, and thermoelastic analyses were performed to characterize structural response and generate mechanical deformation data for optical performance evaluation. Modal analysis was conducted to verify compliance with the structural dynamic requirements, which specified minimum fundamental natural frequencies of 35 Hz in the lateral (X and Y) directions and 50 Hz in the axial (Z) direction. The predicted modal frequencies exceeded ESCAPE’s requirements demonstrating adequate structural stiffness and dynamic margin for the anticipated launch.

{\bf Thermal Analysis and Optical Impact:} The thermal model was developed using Thermal Desktop® to predict the instrument's on-orbit thermal behavior for a 580 km circular orbit. Bounding thermal environments were evaluated using a 73\adeg\ solar beta angle for the hot operational case and a 0\adeg\ solar beta angle for the cold operational case. The instrument exterior was insulated with 15 layers of multi-layer insulation (MLI) to minimize radiative heat exchange and reduce thermal gradients across the structure. We assumed an internal spacecraft temperature of 25\adeg C$\pm$10\adeg C. Active thermal control was provided through three independently controlled heater zones. The pre-collimator assembly utilized heaters with a maximum power of 100 W, while the primary and secondary mirror assemblies were controlled using heaters with peak powers of 40 W and 30 W, respectively. The heater control architecture was incorporated into the thermal model to verify component temperature regulation and power requirements under both orbital environments. The PM and SM are the most sensitive optical elements and require dedicated thermal control. We baseline a thermal control of 20\adeg C $\pm$ 2\adeg C with minimal gradients as that level of control is readily achievable, though the design is tolerant to larger gradients. 

A non-sequential optical model was developed in Ansys Zemax OpticStudio and interfaced with the MATLAB Application Programming Interface (API) to automate the import of structural and thermal deformation results, execute optical simulations, and compute performance metrics for each operational thermal case. Thermally induced deformation fields predicted by the Thermal Desktop were mapped into the optical model to evaluate the effects of structural distortion on system alignment and performance. \escape\ utilizes M55J carbon fiber to meet performance metrics, including optical throughput, image quality, line-of-sight stability, and focus sensitivity, were evaluated for both the hot- and cold-case orbital environments. The results of this model for each optical element are presented in Table~\ref{tab:optic_motions}. When all factors are included in the optical models for either the cold or hot case, the G20 spectral resolution only degrades from $\sim$0.78 \AA\ to $\sim$1 \AA, well under the required resolution of 1.5 \AA\ with a margin of more than 50\%.

\begin{table}[htbp]
\centering
\caption{Optic translations and rotations under cold and hot operating conditions.}
\label{tab:optic_motions}
\footnotesize 
\setlength{\tabcolsep}{0pt} 
\begin{tabular*}{\textwidth}{@{\extracolsep{\fill}} l cccccc cccccc }
\toprule
 & \multicolumn{6}{c}{\textbf{Cold - Absolute Translation/Rotation}} & \multicolumn{6}{c}{\textbf{Hot - Absolute Translation/Rotation}} \\
\cmidrule(lr){2-7} \cmidrule(lr){8-13}
\textbf{Optic} & \multicolumn{1}{c}{\textbf{DX}} & \multicolumn{1}{c}{\textbf{DY}} & \multicolumn{1}{c}{\textbf{DZ}} & \multicolumn{1}{c}{\textbf{RX}} & \multicolumn{1}{c}{\textbf{RY}} & \multicolumn{1}{c}{\textbf{RZ}} & \multicolumn{1}{c}{\textbf{DX}} & \multicolumn{1}{c}{\textbf{DY}} & \multicolumn{1}{c}{\textbf{DZ}} & \multicolumn{1}{c}{\textbf{RX}} & \multicolumn{1}{c}{\textbf{RY}} & \multicolumn{1}{c}{\textbf{RZ}} \\
 & \multicolumn{1}{c}{($\mu$m)} & \multicolumn{1}{c}{($\mu$m)} & \multicolumn{1}{c}{($\mu$m)} & \multicolumn{1}{c}{(arcsec)} & \multicolumn{1}{c}{(arcsec)} & \multicolumn{1}{c}{(arcsec)} & \multicolumn{1}{c}{($\mu$m)} & \multicolumn{1}{c}{($\mu$m)} & \multicolumn{1}{c}{($\mu$m)} & \multicolumn{1}{c}{(arcsec)} & \multicolumn{1}{c}{(arcsec)} & \multicolumn{1}{c}{(arcsec)} \\
\midrule
Primary Mirror & 0.00 & -5.08 & -11.94 & 1.75 & -0.11 & -5.54 & 0.25 & -6.35 & -5.84 & 2.39 & 0.02 & -2.74 \\
\addlinespace
PFA & 0.51 & -0.25 & -19.56 & 0.21 & -0.20 & -5.54 & 0.25 & 0.00 & -8.89 & -0.16 & -0.11 & -2.74 \\
\addlinespace
Secondary Mirror & 1.27 & 0.76 & 12.95 & 1.15 & -0.75 & -5.51 & 0.76 & 0.25 & 7.62 & 0.43 & -0.43 & -2.71 \\
\addlinespace
G20 Array & 2.03 & -0.51 & -0.25 & -3.13 & -1.60 & -5.58 & 1.52 & -1.78 & -0.51 & -2.41 & -1.04 & -2.74 \\
\addlinespace
G40 Array & 1.52 & -0.25 & -1.52 & 2.29 & -2.04 & -7.31 & 1.02 & -0.51 & -1.02 & 0.21 & -1.26 & -3.59 \\
\addlinespace
G70/G140 & -1.78 & 1.02 & -0.25 & -1.02 & -1.33 & -5.47 & -0.51 & -0.25 & 0.00 & -1.35 & -0.89 & -2.75 \\
\addlinespace
Fold Mirror & -1.02 & -4.32 & 2.29 & 1.25 & -5.31 & -9.38 & 0.25 & -5.33 & 2.54 & 1.42 & -4.84 & -6.76 \\
\addlinespace
Detector & -2.29 & -5.59 & 2.29 & -2.02 & -1.55 & -5.38 & -0.25 & -5.59 & 2.29 & -1.91 & -1.03 & -2.67 \\
\bottomrule
\end{tabular*}
\end{table}

The STOP analysis demonstrates that the structural design provides sufficient stiffness while the passive and active thermal control systems maintain the optical assembly within its required operating temperature range. Thermoelastic deformations remained within allowable limits, resulting in negligible degradation of optical performance over the evaluated mission environments. The coupled structural, thermal, and optical analyses confirm that the instrument satisfies its performance requirements with adequate engineering margin for the anticipated on-orbit conditions. 

\subsection{Stray Light Analysis and Control}\label{section-straylight}
\escape\ is designed to achieve sensitivity $>$50$\times$ greater than \euve, the last EUV astrophysics spectrograph.
The majority of the effective area enhancement is provided by the HB telescope design and lack of bandpass-limiting filters (Section~\ref{section-telescope}). 
Equally important as effective area is limiting the detector backgrounds and other noise sources. 
In \escape's case, the majority of photons entering the spectrograph (and therefore potential scattered/stray light sources) are not from the science targets themselves, but rather the geocoronal H I, He I, and He II emission from the Earth's exosphere (Section~\ref{section-geocoronal}). H I and He I do not fall directly on the detector, however they are bright sources that represent the majority of photons in the spectrograph that the solar-blind MCP detector is sensitive to. The less bright He II is in-band for both the G40 and G20 channels.

\begin{figure}
    \centering
    \includegraphics[width=0.99\linewidth]{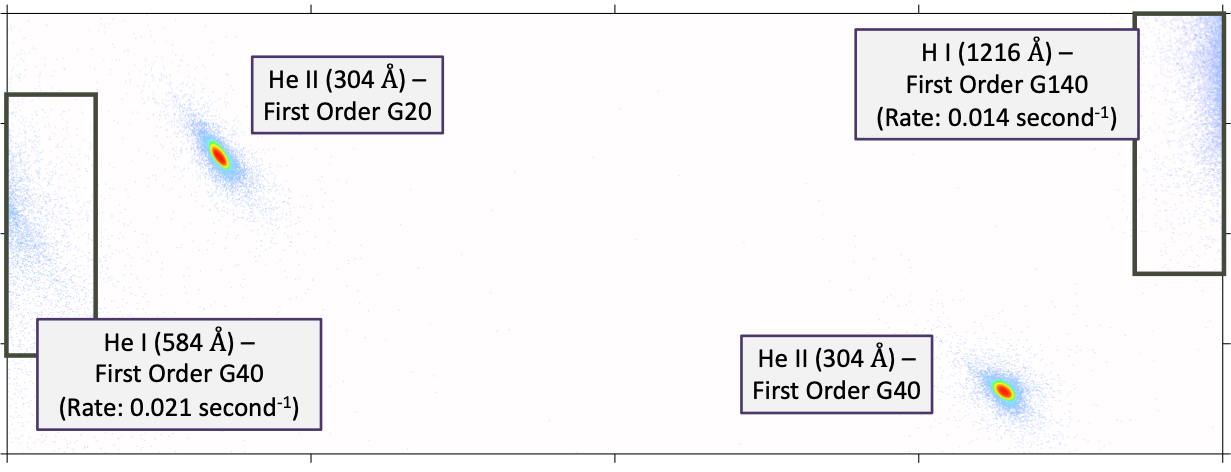}
    \caption{A simulated 10,000 second sky (no target star) exposure of the \escape\ detector showing the in-band He II 304\AA\ features as well as scattered light from the out-of-band He I 584 \AA\ and H I 1216 \AA\ features at the edges. Second order He II 304 \AA\ is also included, but negligible. These edge scatter sources are less than the expected ambient dark rates, therefore the expected intrinsic dark is removed to show the scatter. Total, this scatter plus the intrinsic dark carries 190\%\ margin on the \escape\ background requirement.}
    \label{fig-scatter}
\end{figure}

We modeled these backgrounds by first assuming a fluence of photons into the spectrograph commensurate with \escape 's orbital altitude and assuming a phase position near the limb (5,000 Rayleigh H I Lyman alpha brightness). While this is not a worst-case scenario, the results of this analysis can be scaled for orbital day or night. Each optic in the spectrograph is assigned a total integrated scatter based on the anticipated microroughness, wavelength, and incident angles, and then a Bidirectional Reflectance Distribution Function (BRDF) based on prior measured optics and gratings (e.g., GOLD\cite{GOLD1}). The simulation was run for 10,000 seconds. An image of the projected scattered (He I and H I) and directly measured (He II) geocoronal emission is shown in Figure~\ref{fig-scatter}.

Overall, we find that the total projected scattered light from geocoronal H I and He I emission is less than 0.1 counts s$^{-1}$ and is concentrated near the edges of the detector as scatter from the line cores, which do not directly fall on the MCP. These scatter features do not overlap with any essential spectral features in the \escape\ science bandpass. Even if they did overlap, the rate is low enough to be less than the expected ambient on-orbit global background of 0.38 counts cm$^{-2}$ s$^{-1}$, or $\sim$ 0.07 counts per pixel per day (assuming 15 $\times$ 15 $\mu$m digital pixels). This on-orbit background is from a combination of the intrinsic MCP dark rate and the ambient radiation-induced background on the MCP detector\cite{Siegmund2020}. Geocoronal He II 304 \AA\ does land directly on the detector with a combined G20 and G40 rate of $\sim$ 21.4 counts s$^{-1}$ at 10 Rayleighs \cite{Airglow71}. Of this, a small halo of scatter does extend outwards from the imaged spots, but falling to less than the ambient background level within $\sim$ 750 $\mu$m ($\lesssim$ 1.5 \AA ) of the line core. 

 When \escape\ is fully between the Sun and Earth (where the geocorona is brightest), the scatter rate will increase by a factor of 4-6 from the limb brightness. Given that the mechanical design is still in the early stages, and that we are not considering source-scattering at this time, we adopt this extreme case as our baseline scatter rate, raising the total intrinsic + radiation + scattered light background to 0.8 counts cm$^{-2}$ s$^{-1}$ (0.15 counts per pixel per day). This still carries nearly 400\% margin on the requirement of a total background less than 4 counts cm$^{-2}$ s$^{-1}$.

\subsection{Science closure and margins}

\begin{figure}[htbp]
    \centering
    \includegraphics[width=0.485\textwidth]{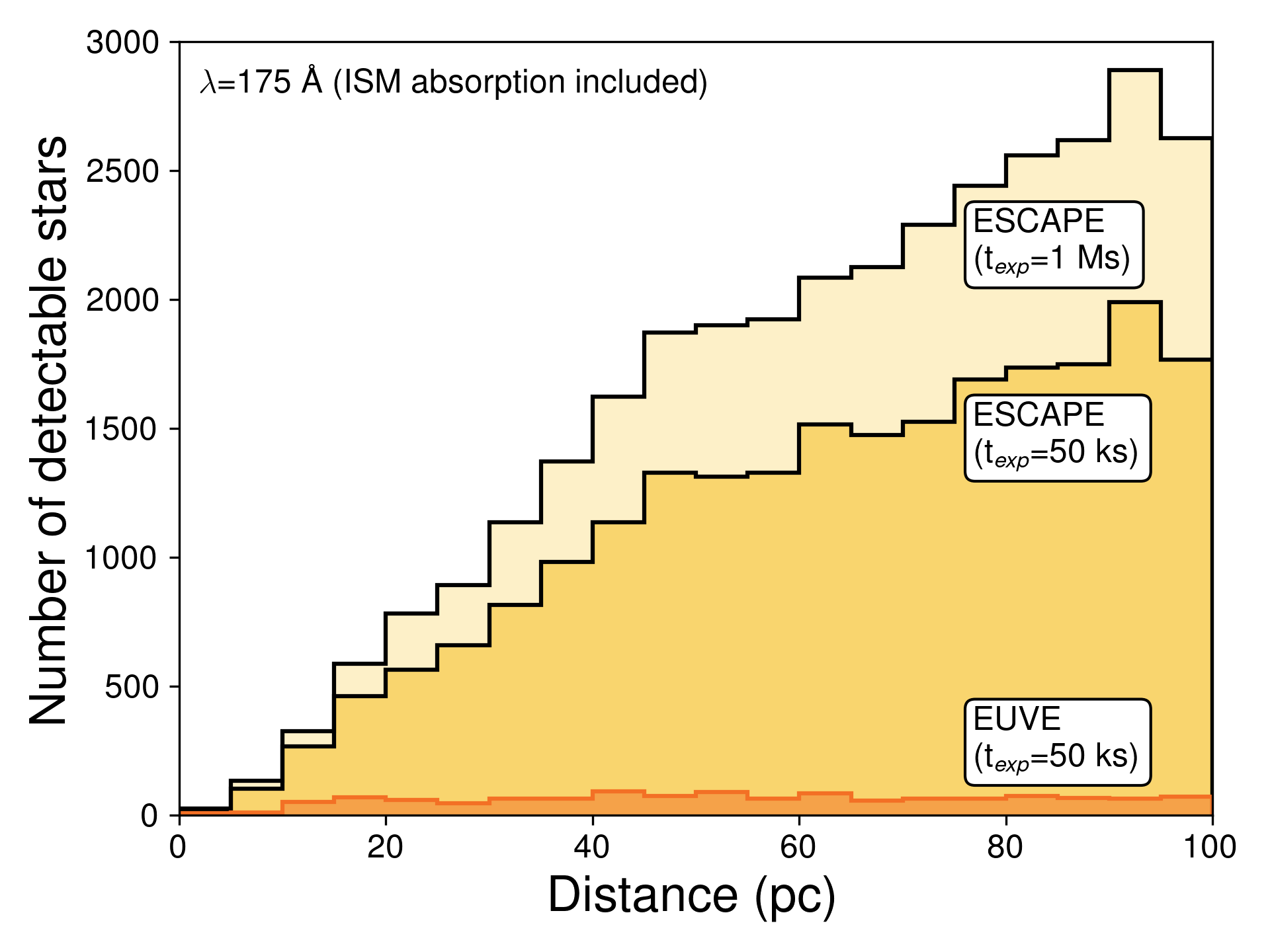}
    \hfill 
    \includegraphics[width=0.485\textwidth]{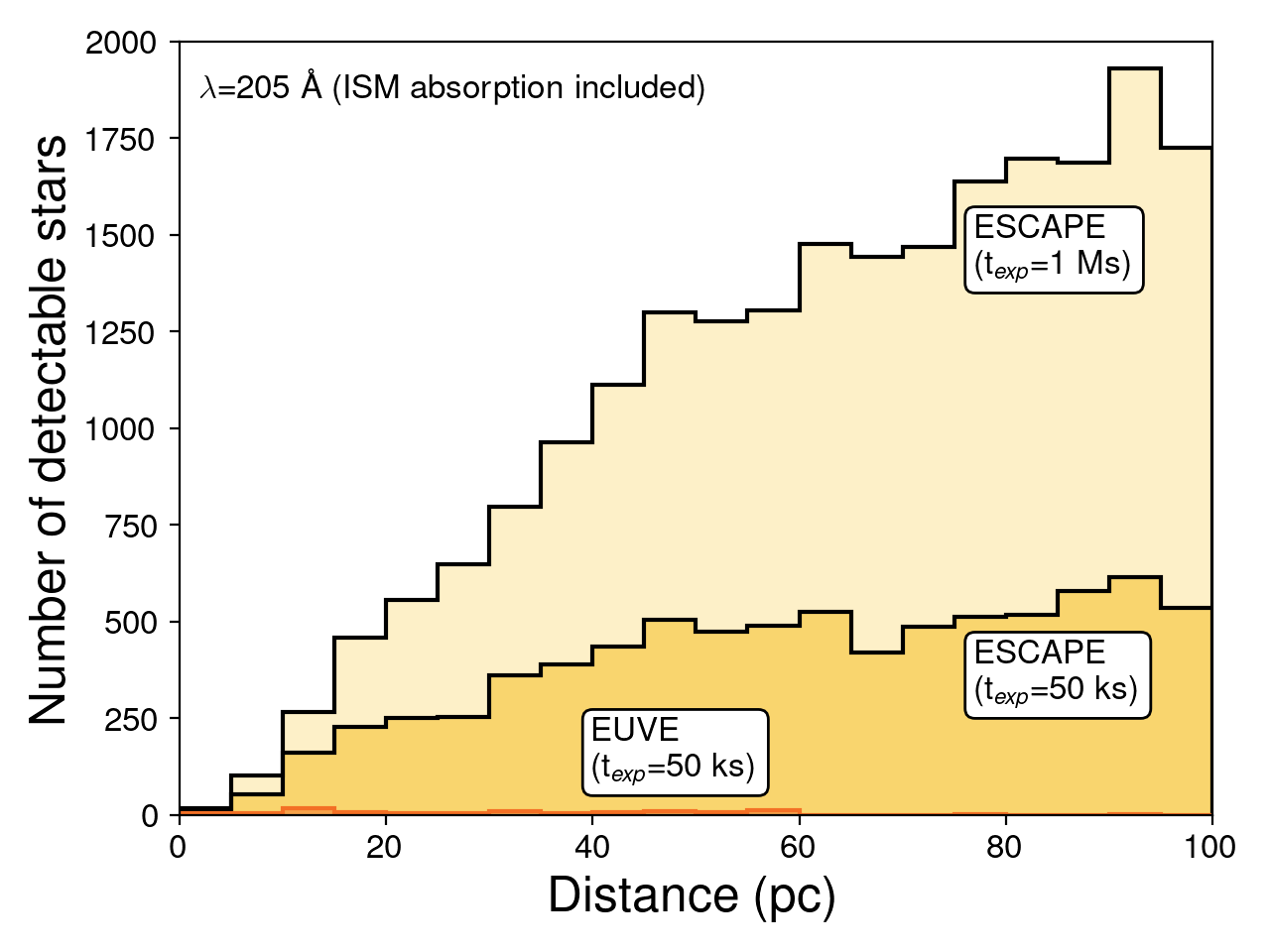}
    \caption{Simulations of the number of stars inside 100 pc detectable by \escape\ demonstrate its broad capability compared to its predecessor \euve\ and the large pool available for target selection. The left panel shows a histogram of the number of stars where S/N$\geq$10 in a 15 \AA\ wide band centered on 175 \AA\ is achievable in the listed exposure times. \euve\ observations with a simulated 50 ks exposure time (a typical value for that mission) is shown in orange. \escape\ observations with a simulated 50 ks exposure time (similar to the longest exposure time of the SEEN survey and enables direct comparison with \euve) are shown in dark yellow and with a simulated 1 Ms exposure (exposure time for DEEP survey targets) in light yellow. The right panel is similar to the left, except the 15 \AA\ bandpass is centered at 205 \AA. In both panels, ISM absorption has been included.}
    \label{fig:Target_sensitivity}
\end{figure}

\escape\ achieves the mission science goals through the execution of the SEEN and DEEP spectroscopic surveys (Table~\ref{tab:SEENDEEP}). Figure~\ref{fig:EUV_age} plots the nominal \escape\ target list and demonstrates that \escape\ can survey stars with a range of ages, activity levels and masses.  The nominal \escape\ target list comprises 300 nearby stars, a factor of $>$ 20 increase in the number of EUV spectra currently available for main sequence FGKM stars. The 300 targets were drawn from a list of thousands of nearby stars with archival X-ray fluxes; more than 1000 backup targets meet the SEEN survey S/N requirement (Figure~\ref{fig:Target_sensitivity}, described at the end of this section).

The science closure goals for \escape\ are captured in four key parameters:  1) the $A_{\rm eff}$ (controlling sensitivity to a target flux), 2) the PSF (controlling spectral resolution and subtended background), 3) the number of stars surveyed, and 4) the total observing time. 

We present an example simulated data calculations for the 167 – 182 \AA\ band here for a typical SEEN survey target ($\zeta^{1}$  Ret) that has EUV brightness approximately equal to the median of the SEEN target list. We assume projected performance values presented earlier in this section.  We require a signal-to-noise ratio (S/N) $\geq$ 10 on the integrated band flux.  Source fluxes are estimated based on X-ray scaling relations \cite{SanzForcada2011} and computed or directly measured interstellar attenuation factors \cite{Youngblood2025}.  The calculation flow is as follows: 

\begin{itemize}

\item Estimated integrated 167 – 182 \AA\ flux of $F$(175 \AA)=1.17$\times$10$^{-13}$ erg cm$^{-2}$ s$^{-1}$.  Exposure times ($T_{\rm exp}$) are based on target flux, with $T_{\rm exp}$ $\approx$11.2 ks for $\zeta^{1}$ Ret.  

\item Converting flux units to photons, folding in the $A_{\rm eff}$ (parameter 1) and integrating over $T_{\rm exp}$, we find the total number of source counts Signal(167 – 182 Å) = $F$(175 \AA)$\times$ $\lambda$/hc × $A_{\rm eff}$(175 \AA) × $T_{\rm exp}$ = 7762 counts (assuming a 73\% encircled energy).

\item The background term is a combination of the detector background level, a scattered light contribution, and the PSF size expanded over the 15 \AA\ band. The background counts per $T_{\rm exp}$, accounting for the G20 and G40 channels, are Noise(167 – 182 \AA) = 631 counts. 

 \item The projected S/N at 167 – 182 \AA\ for this star is S/N = Signal/ SQRT(Signal+Noise) = 84.7; combined with the 205 and 1335 \AA\ bands (the calculations are not shown here for brevity but follow the same form as the 175~\AA\ example), we find a $\geq$ 100\% margin over the S/N requirement in key EUV and FUV bands.  

\item Combining this radiometric S/N requirement with the 10 – 40\% uncertainty associated with the interstellar attenuation correction, we arrive at $\sim$40\% EUV luminosity accuracy, approximately 150\% margin on the requirement.  
\escape\ carries large margin on the optical PSF that is held against alignment and fabrication errors, while A$_{eff}$ margin is held against contamination and other degradation.  

\end{itemize}

The SEEN survey requirement is observing 200 stars.  The nominal SEEN survey list has 276 stars to be executed in the prime mission, a 38\% margin on the number of sources.  The total SEEN survey mission observing time duration (factoring in our derived 75.5\% SEEN observing efficiency) is 194.4 days. The DEEP survey requires 15 ‘wall clock’ days of observing, a total duration of 24 stars × 15 days/star = 360 days. Including the 5 days budgeted for collision avoidance, the baseline science mission is completed after $\sim$560 days. The science mission lifetime is 730 days, providing 30.4\% margin on the total observing time. We note that this mission observing time reserve is maintained in addition to the healthy margins on S/N and target numbers. 

Here we provide more details about the reserves on target numbers (Figure~\ref{fig:Target_sensitivity}). We used the eROSITA source catalog of Ref.\cite{Freund2024}, which covers half of the celestial sphere. We applied cuts to the catalog to only include targets with the following properties: inside the Local Bubble (d$<$100 pc), F type or later (B$_P$-R$_P$ $\geq$ 0.327), approximately main sequence (3$\leq$ $\Delta$M$_G$ $\leq$ 1.5, where $\Delta$M$_G$ is the
vertical offset from the main sequence as roughly defined by interpolating the B$_P$-R$_P$ vs. M$_G$ values from Ref.\cite{Pecaut2013}), and high likelihood cross matches with other catalogs (CtpRank = 1). These imperfect cuts are adequate for our rough demonstration of the number of possible targets within \escape's grasp. We estimate 100-360 \AA\ intrinsic fluxes from the eROSITA X-ray fluxes using the scaling relations of Ref.\cite{Johnstone2021} and stellar radii estimated from color \cite{Pecaut2013}. To approximate the other half of the celestial sphere unavailable from Ref.\cite{Freund2024}, we assume the local stellar population is roughly isotropic and reflect every star across galactic longitude, doubling the number of targets in the filtered catalog. 

We assign ISM HI column densities based on each target's galactic coordinates and distance using the maps from Ref.\cite{Youngblood2025}; this step is performed after the catalog reflection because the N(HI) distribution is not isotropic. To estimate each target's apparent flux in narrower bands than the broad EUV bands provided by Ref.\cite{Johnstone2021}, we assign each target a template spectrum: FGK stars use the Sun's spectrum \cite{Woods2009} and M dwarfs (B$_P$-R$_P$ $\geq$ 1.8) use Proxima Centauri's spectrum \cite{Drake2020}. The templates are scaled to match each target's 100-360 \AA\ intrinsic flux \cite{Johnstone2021} and then wavelength-dependent ISM attenuation is added \cite{Youngblood2025}. We determined for two key bandpasses (167-182 \AA, 199-214 \AA) the number of stars where \escape\ or \euve\ could achieve S/N$\geq$10 in a 50 ks or 1 Ms exposure time following the same methods in Section~\ref{sec:comparison}. 

Figure~\ref{fig:Target_sensitivity} shows that even when including the effects of the ISM, there are thousands of suitable FGKM dwarf targets for the SEEN survey and DEEP surveys. The \euve\ mission's sensitivity only allowed access to approximately one thousand (at 175 \AA) or one hundred (at 205 \AA) of targets; however, the mission only actually obtained spectra of approximately two dozen FGKM dwarf stars, most of which had SNR $<$ 5 (Figure~\ref{fig:EUV_age}).

\section{\escape\ Enables Science Beyond the Primary Science Questions} \label{sec:extendedmission}
\escape\ is the first broadly capable EUV spectroscopy instrument ever, achieving $>$50$\times$ the sensitivity of its predecessor \euve\ that flew more than 20 years ago. \escape's two-year prime mission is necessarily centered around addressing its three primary science objectives and working toward the goal of understanding which star-planet systems are the most conducive to supporting habitable conditions. An extended mission consisting of a General Observer (GO) program could address many additional science objectives within and beyond astrophysics. We outline these possible programs here.

\subsection{Solar system science}

\subsubsection{Small bodies}
The EUV is particularly useful for studying comets. Cometary volatiles can carry with them information about primordial materials incorporated in the icy objects formed in the early stages of our Solar System. As the comets’ orbit take them close to the Sun, charge exchange between the solar wind ions and neutral cometary species produces high-energy emissions at X-ray and EUV wavelengths \cite{Cravens1997,Krasnopolsky1997}. SDO observations of sun-grazing comets have shown that comet sublimation results in a high abundance of ions produced through photodissociation by solar radiation and subsequent ionization that are readily measurable at EUV and FUV wavelengths \cite{Bryans2012}. Key cometary species to be observed include He, H, O, Fe, C, N, Mg, Si, S, and Ne (e.g., \cite{Sasseen2006}). Comparison of the relatively understudied set of FUV emission features from comets with the well-studied set of oxygen 1304 \AA, 1355 \AA, CO 4th positive, sulfur 1479 \AA, and carbon 1561 \AA\ neutrals provides important constraints on their outgassing rates as a function of perihelion distance. 

The importance of this science is further underscored by the ability to observe interstellar comets as they become more easily identified, like in the case of 3I/ATLAS. Studying the relative ratios of volatile abundances of these interstellar objects cracks open a new and exciting opportunity to understand the differences in the primordial, volatile makeup of Solar Systems. Changes over time for the water production rates for 3I/ATLAS were determined through FUV and visible (1600-8000 \AA) observations with the Neil Gehrels Swift Observatory’s Ultraviolet/Optical Telescope \cite{Xing2025}. Additional UV observations ($\sim500-2000$ \AA) of 3I/ATLAS were made with the ESA JUICE mission, the NASA Europa Clipper mission, and the NASA Lunar Reconnaissance Orbiter mission. These measurements can be further compared with protoplanetary disks observed around distant stars, potentially connecting interloping comets with the types of systems from which they derive. 

Additionally, EUV surface reflectance measurements of the Earth’s Moon and small bodies such as asteroids would improve constraints to the average geometric albedos made using \euve\ photometer data of 0.15\% ($\pm$0.03\%), 3.1\% ($\pm$0.3\%), and 3.5\% ($\pm$0.3\%), over wavelength intervals of 150-240 \AA, 400-580 \AA, and 550-650 \AA, respectively \cite{Gladstone1994}. Changes to the UV reflectance slopes can be used to study top-level surface processing by space weathering. 

\subsubsection{Terrestrial Planet Aeronomy}

He$^+$ 304 \AA\ emissions from Earth’s plasmasphere reveal the dynamics of inner magnetospheric processes, with similar expectations for Mars, Venus, and potentially Jupiter space environments. The He 537 \AA\ and O$^+$ 539 \AA\ ratio is a diagnostic feature for aeronomy discussed in Ref\cite{Krasnopolsky2005}.  Ne II 461.4 \AA\ could resemble He II 304 \AA\ plasma features, with maybe the best opportunities for new discoveries.

Oxygen 1304 \AA, 1355 \AA, sulfur 1479 \AA, and carbon 1561 \AA\ neutrals reveal copious information about the tenuous atmospheres of the Galilean satellites and help us understand the escaping atmospheres and climate history of Mars and Venus. Io’s angular diameter for comparison is 1-1.2$''$ from Earth; however, the extent of its escaping atmosphere reaches out to several tens satellite radii and beyond.

Neon in the Moon’s exosphere is the second most abundant constituent after Helium. Viewing the darker, nightside surface at or near the terminator or above the limb could enable new measurements of the Ne I 736 \AA\ and 744 \AA\ resonance lines to further investigate this species remotely.

\subsubsection{Giant Planet Thermospheres, Aurora, and Plasma Environment}

Numerous UV studies on H$_2$ dayglow and aurora emissions from Jupiter, Saturn, Uranus, and Neptune have been conducted since the Voyager era. These include recent Juno-UVS measurements at Jupiter \cite{Gladstone2018}, Cassini-UVIS studies at Saturn \cite{Gustin2010}, Hisaki measurements of both \cite{Clare2026}, and Hubble observations of all four giant planets \cite{Lamy2020}.  While the Juno-UVS bandpass extends shortward to 70 nm, its sensitivity to the series of H$_2$ molecule Rydberg band emissions has been elusive. New observations with more sensitivity in the 700-825 \AA\ range would enable improved mapping of the deposition of energy from higher energy electrons \cite{Benmahi2024}.

While Hubble observations of giant planet FUV aurora has emphasized imagery of various morphological and time variable phenomenon, several key results have come from high spectral resolution investigations of H$_2$ emission band structure \cite{Gustin2016,Gustin2017,Barthelemy2014} and from ``color ratios" of longwave to shortwave bandpasses \cite{Gerard2014}. The coordination of imaging and spectroscopy together has proven revealing \cite{Morrissey1997}. \escape\ EUV spectroscopy with Europa-UVS, JUICE-UVS, and Juno-UVS spectral imagery would similarly be powerful in combination for time-domain investigations of auroral processes. 

In the event that occultation opportunities with EUV-bright stars be available for \escape, occultations of all four giant planets would leverage the Rydberg band H$_2$ absorption cross-sections as more sensitive to the key altitudes for the thermospheric temperature profiles (cf. New Horizons Alice Jupiter occultations \cite{Greathouse2010} and Cassini-UVIS Saturn occultations \cite{Koskinen2016}). A benefit of this technique for \escape\ is that spatial resolution is not required and the baseline star spectrum prior to ingress or after egress serves as its own calibration reference.

Studies of ionized sulfur emissions from S$^+$, S$^{++}$, and S$^{+++}$ from the Io plasma torus are needed to follow on to the Japanese Hisaki mission orbiting Earth (5-10$''$ spatial resolution, similar 0.5m diameter) and Cassini UVIS measurements during its Jupiter flyby \cite{Steffl2004}. Regular, repeated observations of the torus help constrain how much and how often Io's volcanism mass-loads the magnetic field of Jupiter, which has an impact on all aspects of the Jupiter system including its aurora, radiation environment, and interaction with its satellites.

The amount of O$^+$ versus O$^{++}$ in the plasma torus has been especially difficult to ascertain \cite{Bagenal2015} even following Juno mission Io flybys, although the brightest O(III) feature and singly ionized oxygen features are at 834 \AA\ and 833 \AA, respectively. Therefore the \escape\ spectral coverage and resolving power is well suited to tackle this conundrum.

\subsubsection{Cross-disciplinary Heliophysics and Planetary Science}

\escape's astrophysical surveys (SEEN and DEEP) measure the stellar EUV and CME evolution of solar-type stars from their infancy to today, creating an unprecedented dataset for understanding the evolution of the stellar corona as a function of stellar age and mass. This supports the Heliophysics 2024 Decadal question ``Why does the Sun and its environment differ from other similar stars?”\cite{Helio2024}. This dataset also provides stellar inputs for modeling the evolution of solar system planets. NASA’s Heliophysics, Earth Science, and Planetary Science Divisions have long utilized measurements of solar EUV irradiance and solar eruptive events to interpret atmospheric observations of Earth’s thermospheric composition and densities \cite{Woods2002}, at Mars and Venus \cite{Paxton1992,Jakosky2018}, and space weather \cite{Luhmann2007}. By observing young solar analogs as part of the primary science surveys and general observer program, \escape\ will provide a new empirical basis for investigating the history and evolution of the Sun and its influence on solar system bodies.

\subsection{Non-habitable exoplanets} \label{sec:nonHZexoplanets}

More than 6,000 exoplanets have been discovered\footnote{\href{https://exoplanetarchive.ipac.caltech.edu/}{https://exoplanetarchive.ipac.caltech.edu/}} and the upcoming Nancy Grace Roman Space Telescope promises to deliver an order of magnitude more \cite{Wilson2023}. These worlds cover a vast parameter space of radius, mass, and orbital distance. We have detected ultra-hot Jupiters orbiting their host stars at a tenth of Mercury's orbital distance \cite{Gaudi2017} and Earth-mass planets orbiting stars half way between the Solar System and the galactic center \cite{Yee2021}. Though many of these worlds would not be considered habitable, what they all have in common is exposure to EUV and FUV energy from their host stars that has sculpted, if not stripped away, their atmospheres. The mission of \escape\ to measure the spectroscopic EUV and FUV output of about 300 FGKM stars will have far-reaching impact on our understanding of all planets and their atmospheres, not just the habitable rocky ones we are pushing towards with HWO.

\begin{figure}
    \centering
    \includegraphics[width=\linewidth]{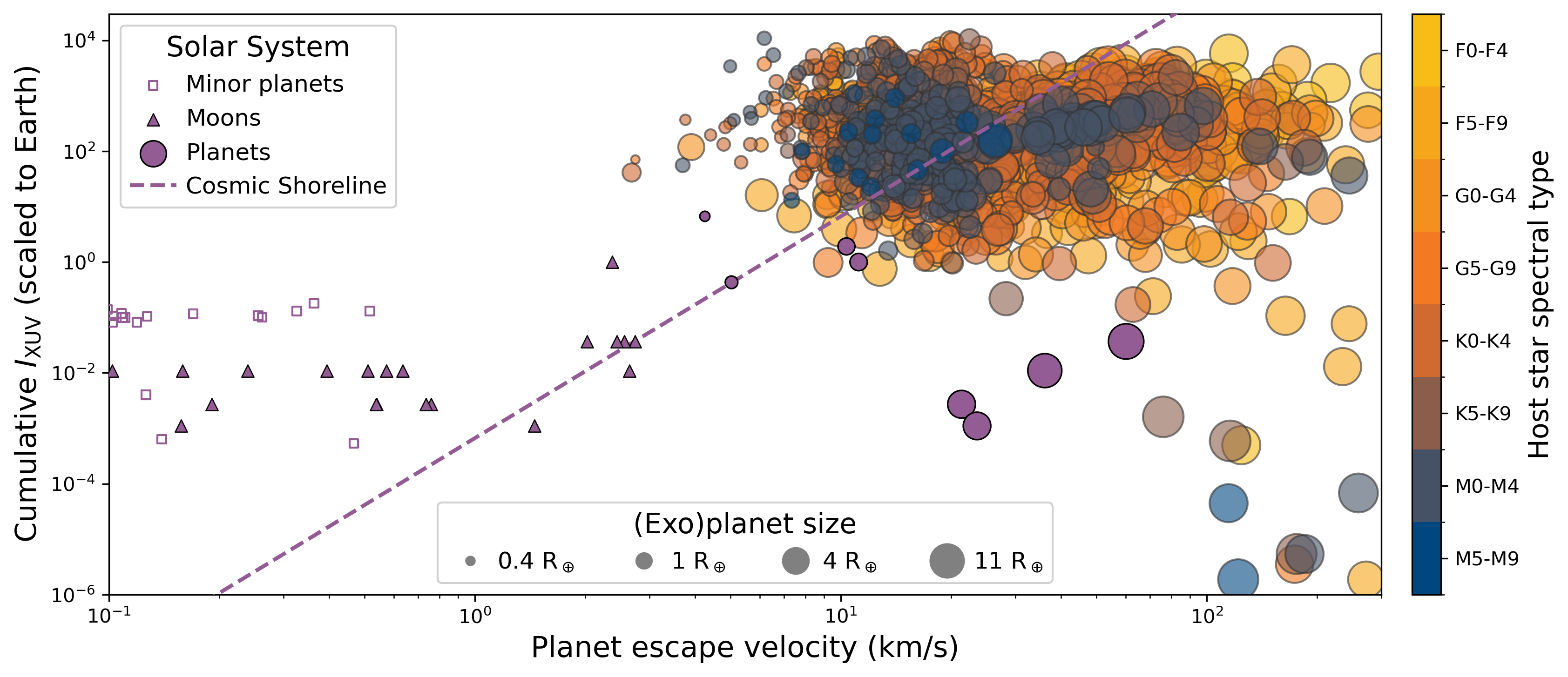}
    \caption{Illustration of the cosmic shoreline framework derived for solar system bodies and plotted against the exoplanet population. Marker size for (exo)planets are log-scaled by planet radius. Exoplanet markers are colored by host star spectral type. The cumulative XUV irradiation relative to Earth's is computed using a simple scaling relation: $I_\mathrm{XUV} = \frac{a_\oplus^2}{a^2}\left(\frac{L_*}{L_\odot}\right)^{0.4}$\cite{Zahnle2017}.
    The uncertainty associated for the cumulative XUV irradiation is not shown for clarity, and because computing all possible sources of y-axis uncertainty is beyond the scope of this work. The present and historical XUV uncertainty prohibits us from determining which star-planet combinations are most likely to result in long-term exoplanet atmospheres. Populations of objects are generated by \texttt{exoatlas}\cite{Berta-Thompson2025}.
    }
    \label{fig-cs-exoplanets}
\end{figure}

Figure~\ref{fig-cs-exoplanets} illustrates the cosmic shoreline concept for Solar System bodies, from which the power law is derived, and extended to cover the parameter space of the much larger exoplanet population. The cumulative XUV irradiation is estimated for exoplanets following a simple scaling law \cite{Zahnle2017}: $I_\mathrm{XUV} = \frac{a_\oplus^2}{a^2}\left(\frac{L_*}{L_\odot}\right)^{0.4}$, where $a$ is the semi-major axis and $L$ is the luminosity. While generally, larger planets are consistent with having atmospheres and smaller planets do not, there is significant blurring of this cosmic line when applied to the thousands of non-Solar System worlds we have detected. The estimated cumulative XUV irradiation is particularly uncertain for exoplanet systems, which prevents us from distinguishing sub-populations and trends in the exoplanet sample. 

\subsubsection{Hot Jupiters} 

Hot Jupiters, i.e. giant planets orbiting their host stars at separations shorter than 0.1 AU, constitute about 1--2\% of the known exoplanet population \cite{Beleznay2022}. Owing to their intense irradiation, these planets are ideal laboratories for studying physical processes such as atmospheric escape, and advancing our understanding of planetary atmospheres and evolution. Ultra-hot Jupiters, that is hot Jupiters with equilibrium temperatures exceeding 2000 K, play a particularly important role in exoplanet science because they currently enable the highest-quality atmospheric observations from both ground- and space-based facilities. These planets typically orbit bright intermediate-mass stars hotter than the Sun (effective temperatures $T_{\rm eff} = $ 6500--10000 K). However, the high-energy emission of these stars, which drives atmospheric escape and photochemistry, remains poorly constrained. This is largely due to the lack of EUV observations of main-sequence stars with temperatures between those of Procyon ($\sim$6500 K) and $\epsilon$\,CMa ($\sim$22500 K)\cite{Craig1997}. Without reliable knowledge of the stellar high-energy spectrum, atmospheric escape and photochemical models remain highly uncertain. 

X-ray, ultraviolet, and near-infrared diagnostics of stellar activity suggest that stars cooler than $\sim$8400 K host chromospheres and/or coronae, while hotter stars appear dominated by purely photospheric emission\cite{Fossati2018}. Furthermore, existing observations and the \euve\ spectrum of Procyon indicate that the high-energy emission of F- and late A-type stars may exceed solar levels even after accounting for stellar radius differences\cite{Fossati2018}. Therefore, the difference in high-energy output between chromospheric/coronal vs photospheric emission may amount to several orders of magnitude. Only EUV spectroscopy of stars in the T$_{\rm eff}$ = 6500--10000 K range can definitively determine the spectral properties and strength of their high-energy emission. Approximately 20 G0--A0 stars lie within 20 pc of the Sun. For these targets, \escape\ will be able to characterize the stellar EUV emission, providing the observational foundation needed to place atmospheric models of ultra-hot Jupiters on significantly firmer ground. These observations may also help us improve our understanding of the origin of stellar high-energy emission and may reveal links with the occurrence and strength of surface convection\cite{Grassitelli2015}.

\subsubsection{Sub-Neptunes}

Close-in sub-Neptune planets constitute a significant fraction of the exoplanet population known to date---about half of all the small planets that dominate the galactic planet population\cite{Kunimoto2020}. These planets are typically low-density objects enveloped in hydrogen/helium-dominated, and possibly water-rich, atmospheres. Based on radius and mass measurements alone, these worlds cover a vast parameter space of possible interior compositions\cite{Lopez2012,Rogers2023}. 

Owing to their low gravity, inflated atmospheres, and short orbital separations, atmospheric escape is thought to play a major role in shaping their long-term atmospheric evolution \cite{Kubyshkina2022}. Based on the \textit{Kepler} dataset with ground-based follow up from the \textit{Keck} observatory, it was found that the sub-Neptune population is statistically distinct from the smaller-sized rocky planet population (called the small planet radius valley\cite{Fulton2017}), leading to questions about how the two populations may be related. The process of atmospheric loss via photoevaporation, driven by stellar flux, is one of the dominant processes in determining whether a small body can retain a substantial low mean molecular weight atmosphere, leaving a modern-day sub-Neptune, or else loses its primordial atmosphere and becomes a rocky planet\cite{Owen2017,Lehmer2017}. This process occurs early in the planet lifetime, soon after formation and dispersal of the gas disc, meaning that the amount of time the host star spends in the highly-active saturation phase where XUV radiation is at a maximum\cite{Wright2016,Wright2018,Pineda2021}, becomes an important term in explaining the observed planet demographics seen today. The span of stellar ages in the \escape\ sample of FGKM stars will therefore fill in a key missing puzzle piece in our understanding of the small planet radius valley. 

\subsubsection{Tracing atmospheric escape with the He {\sc i} line}

Transmission spectroscopy of the metastable He {\sc i} triplet at $\sim$1083 nm from ground-based observatories with high-resolution instruments has become one of the primary techniques for probing exoplanet upper atmospheres and atmospheric escape \cite{Allart2023}, complementing ultraviolet studies performed from space \cite{Spake2018,DosSantos2023}. The metastable He {\sc i} state responsible for the near-infrared triplet is populated primarily through photoionisation of neutral helium by stellar EUV radiation, followed by recombination \cite{Oklopcic2018}. Consequently, the ability to accurately quantify the stellar He {\sc i} photoionising flux, and its spectral distribution, is crucial for reliably modelling the He {\sc i} triplet and interpreting the observations. 

Transmission spectroscopy observations of the He {\sc i} triplet have now been obtained for a growing sample of hot Jupiters and sub-Neptunes, albeit with mixed success, largely because uncertainties in the stellar EUV emission prevent robust predictions\cite{Krishnamurthy2024,SanzForcada2025}. The inflated hot Jupiter WASP-80b provides a representative example. Theoretical models indicate that K-type stars, and particularly active K-type stars hosting inflated gas-rich planets like WASP-80 b, should be favorable targets for detecting He {\sc i} absorption \cite{Oklopcic2019}. However, near-infrared He {\sc i} transmission spectroscopy observations resulted in a non-detection, placing a stringent 95\% confidence upper limit of 0.7\% on the transit absorption (corresponding to 1.11 planetary radii)\cite{Fossati2022}. Several explanations have been proposed for this result, including a low atmospheric helium abundance, a low [Fe/O] coronal abundance dimming the stellar EUV emission \cite{Fossati2023}, diffusive separation of H and He \cite{Taylor2026}, or intrinsically weak EUV emission in the spectral region responsible for He {\sc i} photoionisation \cite{SanzForcada2025}. 

Similar questions around He {\sc i} detection and variability arise for sub-Neptune worlds, where patterns or trends in which sub-Neptunes exhibit escaping He {\sc i} are not identified and defy predictions\cite{Kasper2020,Ahrer2025}. The apparent variability points to a stellar origin in the EUV to which we are currently insensitive, perhaps coupled with other mechanisms of line suppression\cite{GarciaMunoz2025}. The puzzling variability in He {\sc i} may extend further to smaller rocky bodies, as appears to be the case with the habitable-zone M dwarf planet LHS 1140 b\cite{Cherubim2026}. By characterising the strength and spectral shape of the UV emission of late-type stars as a function of effective temperature and activity level, the SEEN survey will provide critical input for substantially improving the reliability of He {\sc i} exoplanet atmosphere models and the interpretation of transmission spectroscopy observations.

\subsubsection{Hot rocky planets}
Because planets orbiting close to their host stars are generally easier to detect, many of our nearby rocky neighbors have equilibrium temperatures closer to that of Venus than the Earth, with some of them having permanent daysides reaching over 1200 K \cite{Kreidberg2019}. A thick atmosphere on a tidally locked rocky planet would be able to advect incoming stellar radiation from its permanent dayside to the cold night side, thereby bringing the planet's measured dayside temperature into alignment with its theoretical equilibrium temperature \cite{Koll2019}. With the launch of JWST in 2021, many of our nearest rocky planet neighbors have become newly accessible in thermal emission measurements \cite{Greene2023,Zieba2023,WeinerMansfield2024,Xue2024}, as well as in more traditional transmission spectroscopy measurements at higher resolution and over a wider wavelength range in the near-infrared \cite{Lustig-Yaeger2023,Moran2023,May2023}. The very hottest known rocky worlds ($T_{\rm eq}>1700\ {\rm K}$) show evidence of atmospheres likely composed of vaporized rock material \cite{Teske2025,ParkCoy2026}, but the majority of more temperate rocky planets ($300\ {\rm K} < T_{\rm eq} < 1500\ {\rm K}$) have returned measurements consistent with bare rock surfaces and allow for only tenuous atmospheres \cite{MeierValdes2025,Fortune2025,Allen2025,Holmberg2026}. The question then becomes why so many terrestrial exoplanets look like barren rocks. 

The answer is inextricably linked to the high-energy radiation these worlds receive from their host stars, particularly the EUV which is most responsible for driving atmospheric escape\cite{Youngblood2025}. We have no reason to think that the geological processes of volatile outgassing are universally suppressed on rocky exoplanets, and it is difficult to concoct planet formation scenarios that would preclude atmospheres at all points in time for M dwarf rocky worlds\cite{Ortenzi2020}. The most likely explanation for the observed dearth of rocky planet atmospheres is stellar EUV irradiation. The rocky planets we have so far been able to investigate almost all orbit M dwarfs due to the favorable signal-to-noise ratios provided by these small, low-mass, cool stars. Though the long-range view of exoplanet science is pushing towards observing rocky planets orbiting Sun-like stars with HWO, these M dwarf rocky planet systems are what are currently accessible, and over hundreds hours of JWST time is dedicated to exploring them \cite{Espinoza2026}, including a 500-hour Rocky Worlds Director's Discretionary Time program \cite{Redfield2024}. The interpretation of the atmospheric status of all of these worlds, both currently in the JWST archive and in the future HWO mission, is incomplete without a comprehensive understanding of the EUV radiation they receive, how it impacts atmospheric loss of high mean molecular weight material, and if stellar flares or CMEs are exacerbating that loss. \escape\ will be able to fill in the EUV picture for all stars hosting rocky planets.

\subsection{Galactic Astrophysics}

\begin{figure}
    \centering
    \includegraphics[width=\linewidth]{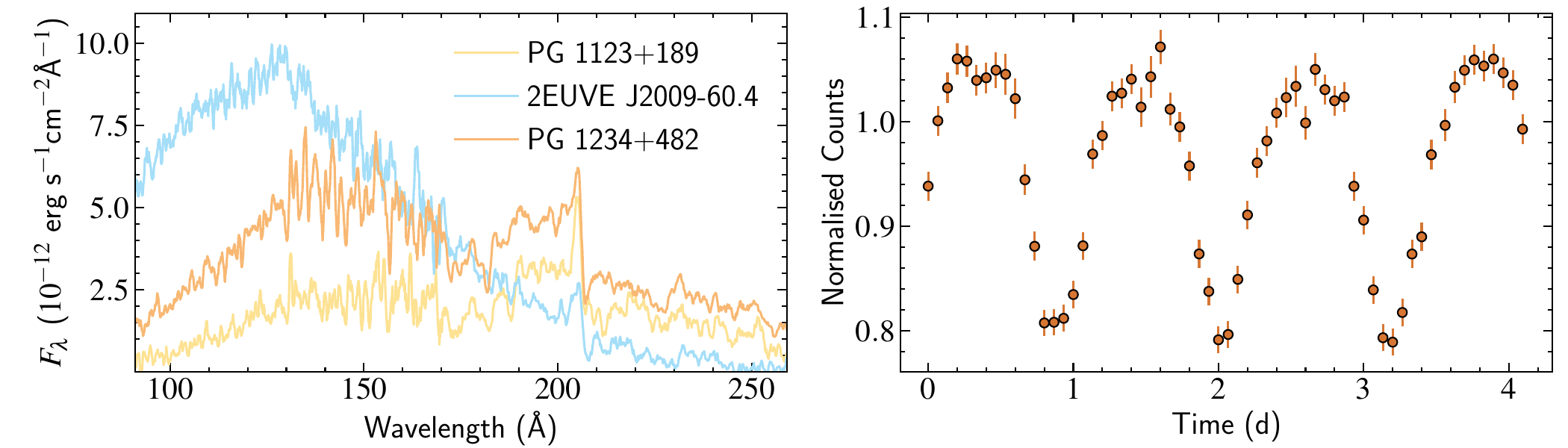}
    \caption{Hot white dwarfs in the EUV. Left: Example EUV spectra of white dwarfs containing a forest of metal lines \cite{schuh02-1}. Right: The EUV light curve of GD\,394 \cite{dupuis00-1}.}
    \label{fig:wds}
\end{figure}

\subsubsection{White Dwarfs}
The spectra of hot ($\gtrsim 40,000$\,K) white dwarfs peak in the EUV (Figure \ref{fig:wds}, left). 26 hot white dwarfs were observed with EUVE\cite{schuh02-1}, suggesting that hundreds of such targets will be available to \escape. At such temperatures, atmospheric metal abundances are governed by the competing forces of gravitational settling removing metals from the visible photosphere and radiative levitation sustaining them, unlike in other temperature regimes where one or the other process dominates\cite{chayer95-1, koester14-1, barstow14-1}. The post-\euve\ confirmation that white dwarfs are accreting rocky objects from their remnant planetary systems has made understanding these processes vitally important, as only after accounting for radiative levitation can the accretion rates and hence abundances of the planetary material be measured for hot white dwarfs. The EUV has the richest collection of metal absorption features in hot white dwarf spectra, making it the key observational input for this effort \cite{barstow97-1}.    

Observations of the hot white dwarf GD\,394 provided one of the most notable mysteries from the \euve\ mission: A roughly 20\,percent variation in EUV flux with a sustained period of 1.15\,days\cite{christian99-1, dupuis00-1} (Figure \ref{fig:wds}, right). This variability mode is thus far unique among white dwarfs and has eluded explanation in the decades since, with suggested explanations including a metal rich spot or a close-orbiting, evaporating planet \cite{wilson19-1, wilson20-1}. \escape\ could not only provide new observations of GD\,394 with time-resolved, high S/N EUV spectra to test these hypotheses, but also search for more examples of EUV variable white dwarfs. Several candidates that share some of GD\,394's oddities in other wavebands have been identified\cite{hallakoun18-1}, and establishing if EUV variation is a common occurrence at hot white dwarfs or a product of extraordinary circumstances would be a notable scientific outcome for \escape.

\subsubsection{Interacting Binaries}

Interacting binaries --- including X-ray binaries (XRBs), cataclysmic variables (CVs), and novae --- are intrinsically UV-bright and observable out to kpc scales for luminous systems along favorable sightlines \cite{hynes20-1,callanan20-1}. \euve\ observed numerous interacting binaries, detecting both thermal and line emission from multiple system components, including neutron star surfaces, the white dwarf, the accretion disk, accreting columns in magnetic systems, and disk winds \cite{long96-1L,christian20-1,mauche20-1,halpern96-1,edelstein20-1}. 

The improvement in effective area and spectral resolution with \escape\ will open up a new phase space for interacting binaries. This population includes the progenitor types of several supernova Type Ia classes, both single degenerate and double degenerate systems. EUV emission from the boundary layer will directly constrain the accretion rate from the disk onto the white dwarf. With the enhanced sensitivity of \escape, short time-series observations of eclipsing systems will allow for spatial mapping of the accretion region, resolving questions of the efficiency of accretion compared to material lost to winds in the system \cite{islam26-1}. In XRBs, the ultimate amount of material actually accreted onto the compact object from this disk, and how this is mediated by disk irradiation and the properties of the outflows, remains inconclusive. Winds can drive off the majority of material, but the outflow signatures change between X-ray and optical wavelengths at different disk accretion states, and it is unclear if these are the same phenomenon or different behaviors. By probing intermediate ionization states between existing observations, EUV spectra will show if the lack of X-ray wind lines in low accretion states is an ionization effect or sign of a transient outflow, with implications for models of wind launching mechanisms and black hole mass growth \cite{buisson20-1,castro22-1,georganti26-1,kosec23-1}.

\subsubsection{Local Interstellar Medium}
The Sun and the nearest stars are moving through a rich and complex interstellar environment. This local interstellar medium (LISM) lies near the center of the Local Bubble that is thought to have been carved out by recent supernovae \cite{Zucker2022}. Indeed, the Sun and the nearest stars appear to be passing through a collection of warm, partially ionized interstellar clouds \cite{Frisch2011}. Due to the low column densities ($\log$ N(HI) = 17.5--18.5) and large subtended angles, it has been difficult to study this environment. Significant broadband absorption by H and He occurs only in the EUV (Figure~\ref{fig:ISM}) and the vast majority of individual absorption lines that are sensitive enough to detect the LISM are located in the FUV and NUV. Indeed, observations in the EUV provide critical observational constraints on the ionization of H ($\sim$20\%) and He ($\sim$40\%) in the LISM via the bound-free ionization edges of He I (504 \AA) and He II (228 \AA) and estimates for H I (912 \AA) \cite{Dupuis1995,Shull2025}. Helium's higher ionization fraction is surprising and argues for a complex morphology of local ISM clouds and for a variety of sources of ionizing radiation.   

\escape\ would provide the first EUV measurements of this unique interstellar environment in over 25 years. The targets in the SEEN and DEEP surveys will likely surpass the number of targets accumulated over decades by HST with high resolution LISM absorption line measurements. \escape\ will provide a dense all-sky map of the HI column density in the LISM. The dramatic variation of HI column density toward even the nearest targets and the strength of the signal in the EUV will make it possible to detect small scale structures and model the global morphology of the material. 

Another development over the last several decades has been the growth in the number of detected exoplanets and an appreciation for the interaction that the LISM can have on planetary systems. The dynamic balance between the LISM and stellar winds leads to variations in the cosmic ray flux and particle deposition on planetary atmospheres \cite{Opher2024,Vannier2025}. The surrounding ISM confines the outward stellar wind, built up in part by stellar CMEs. For our Sun, this produces the heliosphere and for other stars, an astrosphere \cite{Wood2004}. The critical influence of CMEs on planetary atmospheres is discussed in this work and is one of the fundamental scientific motivations of \escape. A dense sample of LISM measurements, as proposed by \escape, is required to develop a three-dimensional morphological map of the ISM surrounding nearby planetary systems. 

\subsection{Stellar Coronae as Laboratories}

The Sun has historically been our best laboratory for understanding how stellar magnetism controls the dynamics of plasma heated to temperatures above $10^6$ K. Our ability to interpret those dynamics hinges upon theoretical calculations of the radiative decay rates and collision strengths for electronic transitions of ion species found at these extreme temperatures. The \texttt{CHIANTI} project is one example of a database collating these transitions' atomic data and cross-checking the results of theoretical calculations against solar observations to assess their accuracy \cite{Young2009a,Young2009b,DelZanna2012}, and one of their findings is that there are still many unidentified transitions in the extreme ultraviolet \cite{Beiersdorfer2018}. By observing the coronae of stars whose magnetic field strengths and topologies differ from the Sun's, \escape\ could provide context that complements existing solar observations and provides both an additional test for our understanding of atomic structure and an opportunity to refine our diagnostics of the solar corona.

One example of this can be found in how \euve\ refined our understanding of the ``first ionization potential'' effect, where Ref. \cite{Pottasch1963} noted that the abundance of Mg, Si, and Fe (all elements with a first ionization potential less than 10 eV) was significantly higher in the low corona compared to the solar photosphere. Later observations revealed that this fractionation was tied to solar magnetism and differed based on the speed of the solar wind and/or the type of active region \cite{Laming2015}. \euve\ made it possible to test for the same effect in other stars and Ref.\cite{Wood2010} found that for most stars (excluding the most active ones), the magnitude of this FIP effect decreased with later spectral type. However, the sample size was fewer than 20 stars, with only a couple of stars of each spectral subtype. \escape\ will have the sensitivity to explore the FIP effect for a more robust sample and its survey programs will have the monitoring baseline to study the evolution of the FIP effect during flares, an experiment that has been done for stellar flares with X-ray spectroscopy \cite{Nordon2007}, but would be much improved by \escape's access to more elements and formation temperatures than \xmm.

\subsection{Extragalactic Astrophysics: The Ionizing Spectrum of Active Galaxies}

The EUV spectra of active galactic nuclei (AGN; Seyfert galaxies, quasars, and QSOs) contribute to the metagalactic ionizing radiation field that has maintained the ionization state of the intergalactic medium for the majority of the history of the universe~\cite{shull12}. The AGN EUV spectrum also drives the physical and chemical evolution of their host galaxies, likely plays a role in the regulation of galactic star-formation~\cite{costa18,Mercedes-feliz23}, and contributes to the ionization state of circumgalactic halos~\cite{shull25}. While it is widely acknowledged that AGN play these critical roles, the detailed amplitude and slope of the EUV spectrum are remarkably unconstrained at the hardest EUV energies (100~--~300~\AA). These shortest wavelengths not only contribute to the overall ionization balance of hydrogen, but also regulate the ionization of helium, owing to their high energies ($>$ 25 eV). Previous observations at intermediate redshifts (z~$\approx$~1) have shown that there is a significant change in the spectral slope from the FUV spectrum to the 600-900 \AA\ EUV spectrum \cite{shull12}. We have very little empirical knowledge of how this spectral slope evolves to shorter EUV wavelengths.

\begin{figure}
    \centering
    \includegraphics[width=0.6\linewidth]{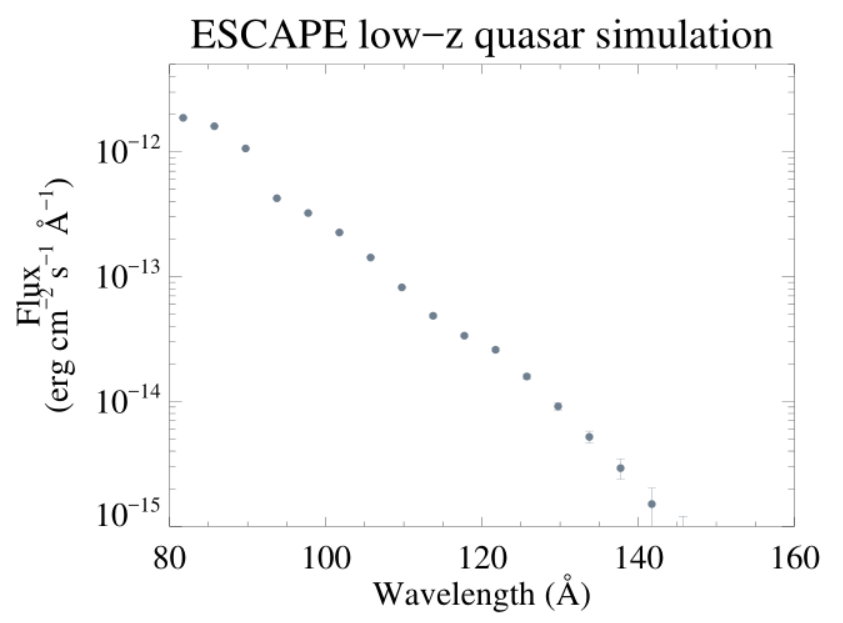}
    \caption{Predicted \escape\ spectrum of a low-redshift quasar observed through a low-column sightline of the ISM.   The underlying spectrum is assumed to have the spectral slope of the composite 600-1000 \AA\ EUV spectrum \cite{shull12}, normalized to a characteristic observed FUV flux from low-redshift quasars in the HST-COS archive (5~$\times$~10$^{-14}$ erg cm$^{-2}$ s$^{-1}$ \AA$^{-1}$).  The gray data points are the simulated \escape\ spectral data for an exposure time of 100 ks, with the spectrum binned to 8~\AA\ spectral resolution, after including a Milky Way column density of N(HI) = 10$^{20}$ cm$^{-2}$.  
    The quasar continuum is observable with S/N $>$ 3 per 8~\AA\ spectral bins from approximately 80 to 150~\AA; the error bars on the fluxes are smaller than the size of the points.}
    \label{fig:quasar}
\end{figure}

In the absence of an EUV observatory, these short wavelength photons must be redshifted to z $>$ 2.5 to fall into the bands of previous space-borne observatories (e.g., \cite{telfer02,stevans14}). However, the opacity of the intergalactic medium at z $>$ 2 (the Lyman-$\alpha$ forest) often prevents the transmission of these photons. Despite these challenges, \euve\ made photometric detections of a small number of quasars, demonstrating that EUV photons from extragalactic objects (e.g., the 3C 273 system, z = 0.16, $m_{FUV}$~$\approx$~13) can be detected through low-ionization holes in the Galactic ISM~\cite{marshall95}.
With its $>$50$\times$ sensitivity of \euve, \escape\ could empirically determine the high-energy EUV spectral energy distribution of low-z quasars for the first time (Figure~\ref{fig:quasar}).

\section{Conclusions} \label{sec:conclusions}

\escape, a NASA Small Explorer mission concept proposed in 2026, is designed to resolve key uncertainties in the long-term stability of rocky exoplanet atmospheres by observing stellar EUV light and CMEs. Heliophysics, Earth science, and planetary science have long recognized solar EUV and CME observations as essential for understanding orbiting planets, and \escape\ closes this major observational gap for exoplanetary science. \escape\ employs time-resolved EUV and FUV spectroscopy in a two-year survey of over 300 nearby FGKM stars to uncover the EUV irradiance in the habitable zone, how the EUV irradiance evolves in time, and the properties of stellar CMEs through coronal dimming signatures. By combining stellar observations with planetary atmosphere models, \escape\ will enables us to understand other worlds in context; interpreting current observations of all types of exoplanets and providing a roadmap for NASA's habitable exoplanet characterization missions of the future.

The \escape\ instrument comprises a grazing incidence telescope feeding four diffraction gratings that are imaged onto a single MCP detector, providing R$\sim$200 point-source spectroscopy simultaneously covering 80-825 \AA\ and 1250-1650 \AA. The \escape\ mission concept is relatively mature thanks to a NASA-funded Phase A concept study in 2020-2021 \cite{France2022}, NASA-funded MANTIS smallsat prototype mission \cite{indahl2022} launching in 2028, and additional design, analyses, and laboratory tests described in this work.

As a broadly capable EUV observatory with $>$50$\times$ the sensitivity of \euve, \escape\ would answer additional science questions via a general observer program in an extended mission, including solar system science, giant exoplanets, white dwarfs, interacting binaries, the interstellar medium, and AGN.

\subsection*{Disclosures}
The authors declare that there are no financial interests, commercial affiliations, or other potential conflicts of interest that could have influenced the objectivity of this research or the writing of this paper.

\subsection* {Code, Data, and Materials Availability} 
The manuscript makes use of publicly available code resources, which are noted in the manuscript.

\subsection* {Acknowledgments}

The scientific and technical development of the ESCAPE mission concept have benefited from numerous NASA investments, including 80GSFC21C0003, NNX16AG28G, and 80NSSC20K0412 to the University of Colorado. In addition, the authors make use of results from HST/XMM/Chandra programs HST GO13650, GO15071, GO16164, GO16701, and GO17595.  The authors gratefully acknowledge conversations with J. Michael Shull that guided components of this work.


\bibliography{report.bib}   
\bibliographystyle{spiejour}   




\end{spacing}
\end{document}